\documentclass[aps,prd,a4paper,twocolumn]{revtex4}

\usepackage{graphicx}
\usepackage{bm}
\usepackage{amsfonts}
\usepackage{amsmath}

\newcommand{\muas}[0]{\hbox{\rm $\mu$as}}
\newcommand{\ve}[1]{\mbox{\boldmath $#1$}}

\begin{document}

\title{Light propagation in the gravitational field of $N$ arbitrarily moving bodies in 1PN approximation for high-precision astrometry} 

\author{Sven Zschocke}

\affiliation{Institute of Planetary Geodesy - Lohrmann Observatory,  
Dresden Technical University, Helmholtzstrasse 10, D-01069 Dresden, Germany}  

\begin{abstract}
The light-trajectory in the gravitational field of $N$ extended bodies in arbitrary motion  
is determined in the first post-Newtonian approximation.  
According to the theory of reference systems, the gravitational fields of  
these massive bodies are expressed in terms of their intrinsic multipoles,  
allowing for arbitrary shape and inner structure of these bodies.
The results of this investigation aim towards a consistent 
general-relativistic theory of light propagation in the Solar system for 
high-precision astrometry at sub-micro-arcsecond level of accuracy.  
\end{abstract}



\maketitle

\section{Introduction}\label{Section0}  

The primary objective of astrometry is the determination of the positions and motions of celestial objects,  
like stars or Solar system objects, from angular observations, that is to say to trace a lightray detected
by an observer back to the celestial light-source.
Consequently, one fundamental assignment in relativistic astrometry
concerns the precise description of the trajectory of a light-signal, which is emitted by the celestial object
and propagates through the gravitational field of the Solar system towards the observer.
The growing accuracy of observations and new observational techniques have made it
necessary to take subtle relativistic effects into account.
In this respect, a breakthrough in astrometric precision has been achieved by the
space-mission Hipparcos (launch: 8 August 1989) of
European Space Agency (ESA), which has accomplished an astrometric precision
of up to 1 milli-arcsecond (${\rm mas}$) in measuring the positions of stars \cite{Hipparchos1,Hipparchos2}.  
The next milestone in astrometry is established by the ESA astrometry mission Gaia (launch: 19 December 2013), where 
the positions of celestial objects can be determined within an accuracy of several micro-arcseconds (\muas) in the ideal case 
(bright stars) \cite{GAIA}.  

While micro-arcsecond astrometry has been realized both theoretically and
technologically within the Gaia mission, the dawning of
sub-micro-arcsecond (sub-\muas) or even nano-arcsecond (${\rm nas}$) astrometry is going to pass into the  
strategic focus of astronomers. For instance, NEAT \cite{NEAT1,NEAT2} has been proposed to ESA  
as a candidate for one of the M-size missions within the
Cosmic Vision 2015 - 2025, and is intended to reach a precision of about $50\,{\rm nas}$.
To achieve such accuracy, NEAT utilizes a pair of spacecraft that would fly in formation
at a separation of 40 meters. This provides the long focal length necessary to generate
high angular resolution to detect Earth-like planets.
Further space missions like ASTROD \cite{Astrod1,Astrod2}, LATOR \cite{Lator1,Lator2},
ODYSSEY \cite{Odyssey}, SAGAS \cite{Sagas}, or TIPO \cite{TIPO}
are under discussion by ESA which require the knowledge of light propagation
through the Solar system at sub-\muas $\;$ or even at ${\rm nas}$ level of accuracy.
These missions are designed for a highly precise measurement of the spatial distance
between two spacecrafts in order to determine the gravitational field within the
Solar system.
Also feasibility studies of earth-bound telescopes are presently under consideration which aim at an
accuracy of about $10\,{\rm nas}$ \cite{nas_telescopes}.
In view of these technological advancements, a corresponding development in the theory of high-precision astrometry and  
especially in the theory of light propagation is indispensable.

In the limit of geometrical optics the path of a light-signal (photons) is a null geodesic, governed by the geodesic equation  
which is valid in any coordinate system and reads in the exact form \cite{MTW,Brumberg1991}: 
\begin{subequations}
\begin{equation}
\frac{d^2 x^{\alpha}\left(\lambda\right)}{d \lambda^2}
+ \Gamma^{\alpha}_{\mu\nu}\,\frac{d x^{\mu}\left(\lambda\right)}{d \lambda}\,
\frac{d x^{\nu}\left(\lambda\right)}{d \lambda} = 0\,,
\label{Geodetic_Equation} 
\end{equation}
\begin{equation}
g_{\alpha\beta}\,\frac{d x^{\alpha}\left(\lambda\right)}{d \lambda}\,\frac{d x^{\beta}\left(\lambda\right)}{d \lambda} = 0\,, 
\label{Geodetic_Equation1}
\end{equation}
\end{subequations}

\noindent
where (\ref{Geodetic_Equation}) represents the geodesic equation, while the so-called isotropic condition (\ref{Geodetic_Equation1}) must be 
imposed as additional constraint for null geodesics;  
$x^{\alpha}\left(\lambda\right)$ are the four-coordinates of the photon which depend  
on the affine curve parameter $\lambda$, and the Christoffel symbols are functions of the metric of curved space-time,  
\begin{eqnarray}
\Gamma^{\alpha}_{\mu\nu} &=& \frac{1}{2}\,g^{\alpha\beta}
\left(\frac{\partial g_{\beta\mu}}{\partial x^{\nu}}
+ \frac{\partial g_{\beta\nu}}{\partial x^{\mu}}
- \frac{\partial g_{\mu\nu}}{\partial x^{\beta}}\right),
\label{Christoffel_Symbols}
\end{eqnarray}

\noindent
where $g^{\alpha\beta}$ and $g_{\alpha\beta}$ are the contravariant and covariant components of the metric tensor, respectively.  

Facing the fact that in the Solar system the gravitational fields are weak, $\displaystyle \frac{G\,M_A}{c^2\,R_A} \ll 1$,  
and the orbital velocities of the bodies are slow, $\displaystyle \frac{v_A}{c} \ll 1$, one is allowed for utilizing the  
post-Newtonian (PN) approximation for the metric tensor $g_{\alpha\beta}$ which is based on both of these assumptions 
\footnote{A complete list of small parameters characterizing the Solar system and allowing for utilizing the post-Newtonian expansion  
can be found in the introductory section in \cite{KlionerKopeikin1992}.}; 
here $M_A$, $R_A$, and $v_A$ being mass, radius, and velocity, respectively, of some massive Solar system body   
(e.g., A = Sun, planets, moons, planetoids).  
This so-called weak-field slow-motion approximation admits an expansion of the metric of Solar system  
in powers of these small parameters, that means in inverse powers of the speed of light, e.g. \cite{KlionerKopeikin1992}:   
\begin{eqnarray}
g_{\alpha\beta} &=& \eta_{\alpha\beta} + h^{(2)}_{\alpha\beta} + h^{(3)}_{\alpha\beta}
+ h^{(4)}_{\alpha\beta} + {\cal O} \left(c^{-5}\right),
\label{metric_perturbation_2PN_5}
\end{eqnarray}

\noindent
where $\eta_{\alpha\beta} = {\rm diag}\left(-1,+1,+1,+1\right)$ is the metric tensor of flat Minkowski space-time, and  
$h^{(n)}_{\alpha\beta} \ll 1$ are small perturbations of it, which scale as follows, $h^{(n)}_{\alpha\beta} \sim {\cal O} \left(c^{-n}\right)$, while  
their detailed structure will be considered later. 

In doing so one has to bear in mind that such a post-Newtonian expansion  
assumes from the very beginning that all retardations are small. Therefore, the expansion in (\ref{metric_perturbation_2PN_5})  
is only valid inside the so-called near-zone of the Solar system,  
$\left|\ve{x}\right| \ll \lambda_{\rm gr}$, characterized by the length of gravitational waves, $\lambda_{\rm gr}$,
emitted by the Solar system \cite{MTW,Poisson_Will,Kopeikin_Efroimsky_Kaplan,Expansion_2PN}.
To get an idea about the magnitude, one can relate this wavelength to a typical orbital period
$T_{\rm orbit}$ of the Solar system bodies by
$\lambda_{\rm gr} \sim c\,T_{\rm orbit} \sim 10^{17}\,{\rm meter}$, where we have considered
as orbital period one revolution of Jupiter around the Sun. Hence, the boundary of the near-zone,
$\left|\ve{x}\right| \ll 10^{17}\,{\rm meter}$, is still beyond the most outer border
of the Solar system and especially encompasses all Solar system objects.

Since one can define the position of any object only with respect to a concrete reference system, such description necessarily implies
to introduce global coordinates which cover the entire curved space-time and in respect to which the positions of the
massive bodies, celestial objects and photons along their trajectories can be well-defined.
According to the recommendations of the International Astronomical Union (IAU) \cite{IAU_Resolution1,IAU_Resolution2},
the standard global reference system adopted in
modern astrometry is the Barycentric Celestial Reference System (BCRS) with coordinates $\left(ct,\ve{x}\right)$, where $t$
is the coordinate-time and $\ve{x}$ are Cartesian-like spatial coordinates from the origin of the global system
(barycenter of the Solar system) to some field-point.

In BCRS coordinates the exact light-trajectory from the light-source through the Solar system towards the observer, 
that means the exact solution of geodesic equation (\ref{Geodetic_Equation}), can be written as follows,   
\begin{eqnarray}
\ve{x}\left(t\right) &=& \ve{x}_0 + c \left(t-t_0\right) \ve{\sigma} + \Delta \ve{x}\left(t,t_0\right)\,,  
\label{Introduction_5}
\end{eqnarray}

\noindent
where $\ve{x}_0 = \ve{x}\left(t_0\right)$ is the position of the light-source at the moment $t_0$ of emission of the 
light-signal, $\displaystyle \ve{\sigma} = \frac{\dot{\ve{x}}\left(- \infty\right)}{c}$ is the unit-direction  
of the lightray at past-null infinity, and $\Delta \ve{x}$ are gravitational corrections to the 
unperturbed light-trajectory. These corrections are complicated expressions which depend on all  
parameters which characterize the metric of the Solar system.  
According to the post-Newtonian expansion (\ref{metric_perturbation_2PN_5}), the  
gravitational corrections to the unperturbed light-trajectory admit a corresponding expansion,   
\begin{eqnarray}
\Delta \ve{x} &=& \Delta \ve{x}_{\rm 1PN} + \Delta \ve{x}_{\rm 1.5PN} + \Delta \ve{x}_{\rm 2PN} + {\cal O}\left(c^{-5}\right),  
\label{Introduction_6}
\end{eqnarray}

\noindent
where the 1PN terms $\Delta\ve{x}_{\rm 1PN}$, the 1.5PN terms $\Delta\ve{x}_{\rm 1.5PN}$, and the 2PN terms $\Delta\ve{x}_{\rm 2PN}$ 
are of the order ${\cal O}\left(c^{-2}\right)$, ${\cal O}\left(c^{-3}\right)$, and ${\cal O}\left(c^{-4}\right)$, respectively.  

As aforementioned, todays astrometric accuracy has reached a level of a few micro-arcseconds in angular observations, and the  
next scale of precision is the sub-micro-arcsecond level; for a historical survey see \cite{History_Astrometry}.  
In order to analyse such highly precise astrometric data,
a comprehensive and systematic relativistic procedure of data-reduction is required \cite{Brumberg1991,Kovalevsky}.  
Among several aspects of modern astrometry, two specific issues have carefully to be treated:  

{\bf (A)} First, the most fundamental concept in astrometric data reduction concerns the accurate definition of a set of several
reference systems plus the coordinate transformations among them. 
In particular, for the determination of the light-trajectory through the Solar system 
($N$-body system), the following $N$+1 coordinate systems are of primary importance:
one global reference system (BCRS) with coordinates $\left(ct,\ve{x}\right)$ and  
$N$ local coordinate systems with coordinates $\left(cT_A,\ve{X}_A\right)$, one for each massive body ($A=1,...,N$) and 
co-moving with it. These $N$+1 reference systems are fully defined by the form of their metric tensor. Furthermore, it is  
well-known, that the global metric (\ref{metric_perturbation_2PN_5}) of a $N$-body system in the region exterior of the massive bodies  
admits a decomposition into two families of global multipoles, namely  
global mass-multipoles $m_L$ and global spin-multipoles $s_L$ \cite{Thorne,Blanchet_Damour1,Blanchet_Damour2,Multipole_Damour_2,Expansion_2PN}: 
\begin{eqnarray}
h^{(n)}_{\alpha\beta} &=& h^{(n)}_{\alpha\beta} \left(m_L, s_L\right), \quad {\rm for} \quad n=2,3,4\;.  
\label{Expansion_A}
\end{eqnarray}

\noindent
These global multipoles describe the multipole-structure  
of the entire Solar system as a whole. On the other side,  
from the theory of reference systems and in accordance with the IAU resolutions \cite{IAU_Resolution1,IAU_Resolution2},  
it is clear that physically meaningful multipole moments of some massive body A have to be defined in the body's  
local reference system, namely local (also called intrinsic) mass-multipoles $M^A_L$ and local spin-multipoles $S^A_L$.  
These intrinsic multipoles describe the multipole-structure of each individual body separately.  
Consequently, the problem arises about how to express the global metric (\ref{metric_perturbation_2PN_5}) in terms of local multipoles:  
\begin{eqnarray}
h^{(n)}_{\alpha\beta} &=& h^{(n)}_{\alpha\beta} \left(M^A_L, S^A_L\right), \quad {\rm for} \quad n=2,3,4\;. 
\label{Expansion_B}
\end{eqnarray}

\noindent
In this respect, there are two advanced approaches in the relativistic theory of reference systems:  
the {\it Brumberg-Kopeikin} formalism (BK) \cite{Brumberg1991,BK1,Reference_System1,BK2,BK3,Kopeikin_Efroimsky_Kaplan} 
and the {\it Damour-Soffel-Xu} approach (DSX) \cite{DSX1,DSX2,DSX3,DSX4}. Both these approaches coincide for all practical problems  
in celestial mechanics and astrometry \cite{Krivov} and have become a part of the IAU resolutions \cite{IAU_Resolution1,IAU_Resolution2}.  
Thus it appears that the explicit form of the metric perturbations  
$h^{(2)}_{\alpha\beta}$, $h^{(3)}_{\alpha\beta}$, and $h^{(4)}_{00}$ in (\ref{Expansion_B}) are well-established expressions  
in celestial mechanics and modern astrometry, while the spatial components $h^{(4)}_{ij}$ in (\ref{Expansion_B}) deserve special attention 
in case of extended bodies with full multipole structure and is presently an active field of
research \cite{Chinese_Xu_Wu,Xu_Gong_Wu_Soffel_Klioner,MinazzolliChauvineau2009,KS}; note that $h^{(4)}_{0i} = 0$.  

{\bf (B)} Second, the Solar system can be described as an isolated $N$-body system, where the bodies move under the influence of their mutual  
gravitational interaction, therewith associated are orbital motions of the bodies which are highly complicated. One has to be aware that  
the metric (\ref{metric_perturbation_2PN_5}) and, therefore, the light-trajectory (\ref{Introduction_5}) are functions of these complicated  
worldlines $\ve{x}_A\left(t\right)$ of the massive bodies. In order to simplify this problem, one might want to expand the worldline 
of some body A around some time-moment $t_A$ as follows,  
\begin{eqnarray}
\ve{x}_A\left(t\right) &=& \ve{x}_A + \frac{\ve{v}_A}{1!}\left(t - t_A\right) 
+ \frac{\ve{a}_A}{2!}\left(t-t_A\right)^2 + {\cal O}\left(\dot{a}_A\right), 
\nonumber\\
\label{worldline_introduction}
\end{eqnarray}
 
\noindent
where $\ve{x}_A = \ve{x}_A\left(t_A\right)$, $\ve{v}_A = \ve{v}_A\left(t_A\right)$ and $\ve{a}_A = \ve{a}_A\left(t_A\right)$ 
are the position, velocity and acceleration of body A at time-moment $t_A$, respectively, which are constant parameters. 
While terms like $v_A/c$ are beyond 1PN approximation  
in the geodesic equation, one has to realize that the above series expansion is not an expansion in powers over $c$, thus all terms in   
(\ref{worldline_introduction}) will contribute in 1PN approximation to the lightray metric, at least as long as no further assumptions like $a_A \sim G$ are 
asserted; cf. text below Eq.~(\ref{global_metric_perturbation_C}). In principle, the expression  
in (\ref{worldline_introduction}) can be implemented into the metric tensor of the Solar system in (\ref{metric_perturbation_2PN_5}). But  
such an approach leads rapidly to involved integrals when solving the geodesic equation in (\ref{Geodetic_Equation}), 
and implies actually an infinite series of integrals that apparently cannot be summed.  
Also the time-moment $t_A$ is actually an open parameter and remains uncertain without further assumptions.  
Consequently, instead to apply for such an approximative expansion in (\ref{worldline_introduction}), it is much preferable  
to find a solution for the light-trajectory in terms of arbitrary worldlines $\ve{x}_A\left(t\right)$. The actual worldline 
of some massive body can finally be concretized by means of Solar system ephemerides; e.g. the JPL DE421 \cite{JPL}.  
Accordingly, an important point which has carefully to be considered concerns the arbitrary motion of the massive bodies.  

In this investigation we will account for both of these fundamental aspects adressed above: 
issue {\bf (A)} is incorporated by the {\it DSX approach}, while issue {\bf (B)} is accounted for by {\it integration by parts of geodesic equation}  
plus the evidence that the remnants of this procedure represent terms beyond 1PN approximation.  
In this way, a systematic approach is developed in order to determine the light-trajectory in the Solar system 
in (\ref{Introduction_5}) in the first post-Newtonian approximation (1PN approximation),  
\begin{eqnarray}
\ve{x}\left(t\right) &\!=\!& \ve{x}_0 + c \left(t-t_0\right)\ve{\sigma} + \Delta\ve{x}_{\rm 1PN}\left(t,t_0\right)  
+ {\cal O}\left(c^{-3}\right)\!,   
\label{Introduction_8}
\end{eqnarray}

\noindent
where the global metric of the Solar system is described from the very beginning in terms of intrinsic multipoles 
of the extended bodies in arbitrary motion.  
Such a systematic formalism is an imperative prerequisite for   
extending the model to higher-order terms in the post-Newtonian expansion in (\ref{Introduction_6}).  

The article is organized as follows:
In section \ref{Section1} we will motivate the inevitability for an analytical solution of light-trajectory in the field of $N$ arbitrarily moving 
bodies with full multipole structure in post-Newtonian  
order for sub-micro-arcsecond astrometry. In section \ref{Section2} the geodesic equation in 1PN approximation and the  
initial-boundary conditions are introduced which determine an unique solution of the geodesic equation. 
The metric of the Solar system in terms of intrinsic multipoles in accordance with  
the IAU resolutions is given in section \ref{Section4}. In order to simplify the integration procedure, new variables  
for space and time and the corresponding transformation of geodesic equation and metric tensor are presented in  
section \ref{Section3}. In section \ref{First_Integration} the first integration of geodesic equation is performed, while  
in section \ref{Special_Cases_1} some specific cases  
(arbitrarily moving monopoles, dipoles, quadrupoles, and one body at rest with full multipole structure) are considered.  
It will be demonstrated that in the limit of bodies at rest the results are in agreement with known results in the literature.  
In section \ref{Second_Integration} the second integration of geodesic equation is represented, while some specific cases 
(arbitrarily moving monopoles, dipoles, quadrupoles, and one body at rest with full multipole structure)  
are considered in section \ref{Special_Cases_2}. In the limit of bodies at rest an agreement with known results in the literature is shown.  
Expressions for the observable relativistic effects of time-delay and light-deflection are
given in section \ref{Observable_Effects}. A summary and outlook can be found in section \ref{Summary_Outlook}.  

\subsection{Notation of impact vectors:}

It appears to be considerate to introduce the notation in use regarding the impact vectors, while further  
notations are shifted to appendix \ref{Notation}. 

While the exact lightray $\ve{x}\left(t\right)$ in (\ref{Introduction_5}) is a complicated function,   
the unperturbed lightray in flat Minkowski space-time is just given by a straight line:  
\begin{eqnarray}
\ve{x}_{\rm N}\left(t\right) &=& \ve{x}_0 + c\left(t - t_0\right)\ve{\sigma}\,, 
\label{Unperturbed_light_ray_10}
\end{eqnarray}

\noindent 
where the subscript N stands for Newtonian limit. One may introduce the following impact vector:  
\begin{eqnarray}
\ve{\xi} &=& \ve{\sigma} \times \left(\ve{x}_{\rm N}\left(t\right) \times \ve{\sigma}\right)
= \ve{\sigma} \times \left(\ve{x}_0 \times \ve{\sigma}\right),\, d = \left|\ve{\xi}\right|. 
\label{notation_2}
\end{eqnarray}

\noindent
The impact-vector in (\ref{notation_2}) points from the origin of the global system (BCRS) towards the point of closest approach of the 
unperturbed lightray to that origin. The impact vector in (\ref{notation_2}) is time-independent, both in case of massive bodies at rest as well  
as in case of massive bodies in motion.

\subsubsection{Massive bodies at rest:} 

Massive bodies at rest means their positions to be constant with respect to the global system: $\ve{x}_A = {\rm const}$. 
We will make use of the following notation for the vector from the massive body at rest towards the photon propagating along the exact light-trajectory:  
\begin{eqnarray}
\ve{r}_A &=& \ve{x}\left(t\right) - \ve{x}_A \,, 
\label{notation_3a}
\end{eqnarray}

\noindent
with the absolute value $r_A=\left|\ve{r}_A\right|$. 
The vector from the massive body at rest towards the photon along the unperturbed light-trajectory reads:  
\begin{eqnarray}
\ve{r}^{\rm N}_A &=& \ve{x}_{\rm N}\left(t\right) - \ve{x}_A 
\nonumber\\
&=& \ve{x}_0 + c\left(t - t_0\right)\ve{\sigma} - \ve{x}_A\,,  
\label{notation_3b}
\end{eqnarray}
 
\noindent
with the absolute value $r_A^{\rm N}=\left|\ve{r}^{\rm N}_A \right|$, and obviously $\ve{r}_A = \ve{r}^{\rm N}_A + {\cal O}\left(c^{-2}\right)$. 
We also need the vector from the massive body at rest towards the photon at the moment of signal-emission:  
\begin{eqnarray}
\ve{r}^0_A &=& \ve{x}_0 - \ve{x}_A\,, 
\label{notation_3c}
\end{eqnarray}

\noindent
with the absolute value $r_A^0=\left|\ve{r}^0_A \right|$. 
Note that in case of massive bodies at rest there will be no time-argument in $\ve{r}_A$ and $\ve{r}^{\rm N}_A$,
irrespective of the fact that the distance between the photon and the body actually depends on time due to the propagation of the photon.
In case of massive bodies at rest we introduce the following impact-vector:  
\begin{eqnarray}
\ve{d}_A &=& \ve{\sigma} \times \left(\ve{r}^{\rm N}_A \times \ve{\sigma}\right), \quad d_A = \left|\ve{d}_A\right|\,.  
\label{notation_4}
\end{eqnarray}

\noindent
The impact-vector in (\ref{notation_4}) is time-independent, $\ve{\dot{d}}_A = 0$, and points from the origin of local coordinate system 
of massive body A towards the unperturbed lightray at the time of closest approach to that origin,  
defined later by Eq.~(\ref{time_of_closest_approach_t_0}).  
Notice that the term {\it weak gravitational field} implies $\displaystyle d_A \gg \frac{G\,M_A}{c^2}$.

\subsubsection{Massive bodies in motion:} 

In case of massive bodies in motion, their positions become time-dependent: $\ve{x}_A \left(t\right)$.
Then we will make use of the following notation for the vector from the massive body towards the photon propagating along the 
exact light-trajectory: 
\begin{eqnarray}
\ve{r}_A\left(t\right) &=& \ve{x}\left(t\right) - \ve{x}_A\left(t\right),  
\label{notation_5a}
\end{eqnarray}

\noindent
with the absolute value $r_A\left(t\right)=\left|\ve{r}_A\left(t\right)\right|$. 
The vector from the massive body in motion towards the photon along the unperturbed light-trajectory reads:  
\begin{eqnarray}
\ve{r}^{\rm N}_A\left(t\right) &=& \ve{x}_{\rm N}\left(t\right) - \ve{x}_A\left(t\right)
\nonumber\\
&=& \ve{x}_0 + \,c \left(t - t_0\right)\ve{\sigma} - \ve{x}_A\left(t\right)\,, 
\label{notation_5b}
\end{eqnarray}

\noindent
with the absolute value $r_A^{\rm N}\left(t\right)=\left|\ve{r}^{\rm N}_A\left(t\right)\right|$ 
and obviously $\ve{r}_A\left(t\right) = \ve{r}^{\rm N}_A\left(t\right) + {\cal O}\left(c^{-2}\right)$. 
We also will need the vector from the massive body towards the photon at the time-moment of 
emission of the light signal, given by  
\begin{eqnarray}
\ve{r}^{\rm N}_A\left(t_0\right) &=& \ve{x}_0 - \ve{x}_A\left(t_0\right)\,,  
\label{notation_5c}
\end{eqnarray}

\noindent
with the absolute value $r^{\rm N}_A\left(t_0\right) = \left|\ve{r}^{\rm N}_A\left(t_0\right)\right|$.  
In case of massive bodies in motion we introduce the following impact vector:
\begin{eqnarray}
\ve{d}_A\left(t\right) &=& \ve{\sigma} \times \left(\ve{r}^{\rm N}_A \left(t\right) \times \ve{\sigma}\right), \quad  
d_A\left(t\right) = \left|\ve{d}_A\left(t\right)\right|\,. 
\label{notation_6}
\end{eqnarray}

\noindent
The impact-vector in (\ref{notation_6}) is time-dependent, $\ve{\dot{d}}_A \neq 0$, and points from the origin of local  
coordinate system of massive body A towards the unperturbed lightray at the time of closest approach to that origin.   
The time-dependence of the impact-vector in (\ref{notation_6}) is solely caused by the motion of the massive body, that
means a time-derivative of (\ref{notation_6}) is proportional to the orbital velocity of this body,
$\ve{\dot{d}}_A\left(t\right) = \ve{\sigma} \times \left(\ve{\sigma} \times \ve{v}_A\left(t\right)\right)$.  
The term {\it weak gravitational field} implies $\displaystyle d_A\left(t_A^{\ast}\right) \gg \frac{G\,M_A}{c^2}$ 
for the time of closest approach of the lightray to the massive body, which will 
be defined later; see Eq.~(\ref{time_of_closest_approach_t_0}).

\section{Motivation}\label{Section1}

Before representing our approach, it is most appropriate to review in brief the recent advancements in the theory of light propagation in the 
weak gravitational field of $N$ massive bodies.  
It is clear that in a short review like the present, it is impossible to consider all articles written on the subject during  
the last decades, and many important calculations must remain unmentioned. Instead, the brief survey is enforced to be focussed on those results, which are of  
upmost relevance for our considerations.  
 
As mentioned in the introductory section, the BCRS metric of Solar system admits an expansion in terms of multipoles.  
By inserting the decomposition of the metric in terms of global multipoles (\ref{Expansion_A}) into the geodesic equation (\ref{Geodetic_Equation}) one obtains   
a corresponding decomposition of the lightray-perturbation (\ref{Introduction_6}) in terms of global multipoles:
\begin{eqnarray}
\Delta \ve{x} &=& \sum\limits_{l=0}^{\infty} \Delta \ve{x}\left(m_L,s_L\right) + {\cal O}\left(c^{-5}\right).   
\label{Introduction_7a}
\end{eqnarray}

\noindent
Likewise, inserting the decomposition of the metric in terms of local multipoles (\ref{Expansion_B}) into 
the geodesic equation (\ref{Geodetic_Equation}) one obtains a corresponding decomposition of the lightray-perturbation (\ref{Introduction_6}) in 
terms of local multipoles:  
\begin{eqnarray}
\Delta \ve{x} &=& \sum\limits_{l=0}^{\infty} \Delta \ve{x}\left(M^A_L,S^A_L\right) + {\cal O}\left(c^{-5}\right). 
\label{Introduction_7b}
\end{eqnarray}

\noindent
In the subsequent survey it will carefully be distinguished whether a decomposition in terms of global multipoles (\ref{Introduction_7a})  
or in terms of local multipoles (\ref{Introduction_7b}) is meant. Let us gradually consider these individual terms,   
depending on how accurate the astrometric measurements are.

\subsection{Astrometry at milli-arcsecond level of accuracy}  

For astrometry on milli-arcsecond (${\rm mas}$) level of accuracy it is sufficient to approximate all Solar system bodies as spherically symmetric objects.  
In case of $N$ monopoles at rest the corresponding correction-term in (\ref{Introduction_7b}) reads \cite{Brumberg1991}:  
\begin{eqnarray}
&& \Delta \ve{x}_{\rm 1PN}^M \left(t,t_0\right) =  
- \frac{2\,G}{c^2} \sum\limits_{A=1}^N M_A   
\nonumber\\
&& \times \bigg(\frac{\ve{d}_A}{r^{\rm N}_A - \ve{\sigma} \cdot \ve{r}^{\rm N}_A}
- \frac{\ve{d}_A}{r^0_A - \ve{\sigma} \cdot \ve{r}^0_A}
- \ve{\sigma}\,\ln \frac{r^{\rm N}_A - \ve{\sigma} \cdot \ve{r}^{\rm N}_A}
{r^0_A - \ve{\sigma} \cdot \ve{r}^0_A}\bigg),  
\nonumber\\
\label{Introduction_10}
\end{eqnarray}

\noindent
where the sum in (\ref{Introduction_10}) runs over all massive bodies of the Solar system.   
For a comparison of (\ref{Introduction_10}) with \cite{Brumberg1991} it might be useful to recall:
$\ln \frac{r^{\rm N}_A - {\bf \sigma} \cdot {\bf r}^{\rm N}_A}{r^0_A - {\bf \sigma} \cdot {\bf r}^0_A}
= - \ln \frac{r^{\rm N}_A + {\bf \sigma} \cdot {\bf r}^{\rm N}_A}{r^0_A + {\bf \sigma} \cdot {\bf r}^0_A}$.
The magnitude of light-deflection for grazing rays amounts to be: $1.75 \times 10^3$ mas for Sun,
$16.3$ mas for Jupiter, $5.8$ mas for Saturn, $2.1$ mas for Uranus, $2.5$ mas for Neptune \cite{Klioner2003a}.  
 
Since in reality these massive bodies are moving, the question arises about how to implement the time-dependence of the positions of  
these gravitating bodies. This particular issue has thoroughly been solved in \cite{KopeikinSchaefer1999} in  
first post-Minkowskian (1PM) approximation, and will be one aspect in the following section.

\subsection{Astrometry at micro-arcsecond level of accuracy}  

Meanwhile, modern space-based astrometry has accomplished the step from milli-arcsecond level  
to micro-arcsecond level of accuracy \cite{History_Astrometry}.  
In order to determine the light-trajectory on \muas-level of accuracy, besides the monopole-term 
in (\ref{Introduction_10}) some further subtle relativistic effects of light propagation need to be accounted for, as there are:  
\begin{enumerate}
\item[$\bullet$] 1. the quadrupole structure of the massive bodies,  
\item[$\bullet$] 2. the motion of the massive bodies, 
\item[$\bullet$] 3. the post-post-Newtonian monopole-term.   
\end{enumerate}

\noindent
The fundamentals of the corresponding theoretical model of light propagation have been worked out in  
\cite{Brumberg1991,Klioner2003a,Klioner1991,KlionerKopeikin1992}, and later be refined 
in \cite{Klioner_Zschocke,Zschocke_Klioner}.  
The results of these investigations have been adopted as one of two model for the Gaia data reduction 
and which is called GREM (Gaia Relativistic Model).  
Another approach has been developed in \cite{RAMOD1,RAMOD2,RAMOD3,RAMOD4,RAMOD5}, which 
is the second model in use for Gaia data reduction and which is called RAMOD (Relativistic Astrometric Model).  
Both these model are designed for relativistic astrometry at micro-arcsecond level of accuracy and allow for an independent 
check of their results. Let us consider in more detail each of these three subtle effects which are listed above.

\subsubsection{Impact of the quadrupole field on light-trajectory}  

The analytical solution for the light-trajectory in a quadrupole
field of a body at rest and in post-Newtonian approximation has been determined in \cite{Klioner1991},  
where the time-dependence  
of the coordinates of the photon and the solution of the boundary value problem
for the geodesic equations has been obtained at the first time. These results were later confirmed by  
different approaches in \cite{Kopeikin1997,LePoncinLafitteTeyssandier2008,Crosta_Quadrupole}.
The formula for the quadrupole light-deflection in 1PN approximation can be found 
in \cite{Klioner2003a,Klioner1991,KlionerKopeikin1992} and should be given here in its complete form:  
\begin{eqnarray}
&& \Delta \ve{x}_{\rm 1PN}^Q \left(t,t_0\right) =
\frac{G}{c^2} \sum\limits_{A=1}^N \frac{1}{d_A^2} \bigg[ \ve{\alpha}_A \left({\cal U}_A - {\cal U}^0_A\right)
\nonumber\\ 
&& + \ve{\beta}_A \left({\cal V}_A - {\cal V}^0_A\right)   
+ \ve{\gamma}_A \left({\cal F}_A - {\cal F}^0_A\right) 
+ \ve{\delta}_A \left({\cal E}_A - {\cal E}^0_A\right)\bigg].  
\nonumber\\ 
\label{Introduction_12}
\end{eqnarray}

\noindent
The sum in (\ref{Introduction_12}) runs over all massive bodies of the Solar system.  
In \cite{Zschocke_Klioner} it has been shown that all terms in the second line of Eq.~(\ref{Introduction_12}) 
are negligible for \muas--astrometry, but this fact is not of much relevance in our investigation here.  
The time-independent vectorial coefficients in (\ref{Introduction_12}) are given by  
\begin{eqnarray}
\alpha_A^k &=&
- M^A_{i_1 i_2}\,d_A^k\,\sigma^{i_1}\,\sigma^{i_2}
+ 2 M^A_{i_1 k}\,d_A^{i_1}
- 2 M^A_{i_1 i_2}\,d_A^{i_2}\,\sigma^{i_1}\,\sigma^k
\nonumber\\ 
&& - \frac{4}{d_A^2} M^A_{i_1 i_2} d_A^{i_1}\,d_A^{i_2}\,d_A^k\,,  
\label{Introduction_coefficient_alpha}
\\
\nonumber\\
\beta_A^k &=&
+ M^A_{i_1 i_2}\,\sigma^{i_1}\,\sigma^{i_2}\,\sigma^k
- 2 M^A_{i_1 k}\,\sigma^{i_1}
\nonumber\\ 
&& + \frac{4}{d_A^2} M^A_{i_1 i_2} d_A^{i_2}\,d_A^k\,\sigma^{i_1}
- \frac{2}{d_A^2} M^A_{i_1 i_2} d_A^{i_1}\,d_A^{i_2}\,\sigma^k ,
\label{Introduction_coefficient_beta}
\\
\nonumber\\
\gamma_A^k &=&
+ M^A_{i_1 i_2}\,d_A^{i_1}\,d_A^{i_2}\,d_A^k
- M^A_{i_1 i_2}\,d_A^2 \,d_A^k\,\sigma^{i_1}\,\sigma^{i_2}
\nonumber\\
&& + 2\,M^A_{i_1 i_2}\,d_A^{i_2}\,d_A^2\,\sigma^{i_1}\,\sigma^k\,,
\label{Introduction_coefficient_gamma}
\\
\nonumber\\
\delta_A^k &=&
- M^A_{i_1 i_2}\,d_A^{i_1}\,d_A^{i_2}\,\sigma^k
+ M^A_{i_1 i_2}\,d_A^2\,\sigma^{i_1}\,\sigma^{i_2}\,\sigma^k
\nonumber\\
&& + 2\,M^A_{i_1 i_2}\,d_A^{i_2}\,d_A^k\,\sigma^{i_1}\,, 
\label{Introduction_coefficient_delta}
\end{eqnarray}

\noindent
with intrinsic mass-quadrupole-moments $M^A_{i_1 i_2}$.     
The time-dependent scalar functions in (\ref{Introduction_12}) are given by:
\begin{eqnarray}
{\cal U}_A &=& \frac{1}{r^{\rm N}_A}\,
\frac{r^{\rm N}_A+\ve{\sigma}\cdot\ve{r}^{\rm N}_A}{r^{\rm N}_A-\ve{\sigma}\cdot\ve{r}^{\rm N}_A}\,,\; 
{\cal U}^0_A = \frac{1}{r^0_A}\,
\frac{r^0_A+\ve{\sigma}\cdot\ve{r}^0_A}{r^0_A-\ve{\sigma}\cdot\ve{r}^0_A}\,, 
\label{Introduction_function5}
\\
\nonumber\\
{\cal V}_A &=& \frac{\ve{\sigma}\cdot\ve{r}^{\rm N}_A}{r^{\rm N}_A}\,,\; 
{\cal V}^0_A = \frac{\ve{\sigma}\cdot\ve{r}^0_A}{r^0_A}\,,
\label{Introduction_function6}
\\
\nonumber\\
{\cal F}_A &=& \frac{1}{\left(r^{\rm N}_A\right)^3}\,,\; 
{\cal F}^0_A = \frac{1}{\left(r^0_A\right)^3}\,,
\label{Introduction_function7}
\\
\nonumber\\
{\cal E}_A &=& \frac{\ve{\sigma}\cdot\ve{r}^{\rm N}_A}{\left(r^{\rm N}_A\right)^3}\,,\; 
{\cal E}^0_A = \frac{\ve{\sigma}\cdot\ve{r}^0_A}{\left(r^0_A\right)^3}\,.  
\label{Introduction_function8}
\end{eqnarray}

\noindent
The light-deflection for grazing rays at giant planets due to their quadrupole-structure amounts to be:  
$240$ \muas$\;$ for Jupiter, $95$ \muas$\;$ for Saturn, $8$ \muas$\;$ for Uranus, $10$ \muas$\;$ for Neptune \cite{Klioner2003a},  
which clearly indicate that the effect of quadrupole light-deflection must be taken into account 
for astrometry at micro-arcsecond level of accuracy.

\subsubsection{Impact of the motion of massive bodies on light-trajectory}\label{Motion_A}  

One of the most sophisticated challenges in relativistic astrometry concerns the problem of the motion of massive bodies and 
its impact on the light-trajectories.   
While the solutions in (\ref{Introduction_10}) and (\ref{Introduction_12}) are valid for bodies at rest, $\ve{x}_A=0$,  
in reality the global coordinates of the bodies depend on time, $\ve{x}_A \left(t\right)$, which is a highly complicated function 
in a $N$-body system due to the mutual gravitational interaction of the massive bodies.  
These complicated worldlines of the massive bodies in the Solar system can be series-expanded \cite{Klioner2003a,KlionerKopeikin1992}, 
\begin{eqnarray}
\ve{x}_A(t) &=& \ve{x}_A + \ve{v}_A\left(t-t_A\right)
+ {\cal O}\left(a_A\right),
\label{Introduction_15}
\end{eqnarray}

\noindent
where $\ve{x}_A$ and $\ve{v}_A$ can be thought of as the
actual position and velocity of body A taken from an ephemeris for some instant of time $t_A$.   
Let us underline here that the impact of the term $\ve{v}_A\left(t-t_A\right)$ in (\ref{Introduction_15})
on the light-trajectory is of 1PN order, besides the fact that this term is proportional
to the velocity of the body; recall that on the other side terms proportional to $v_A/c$ are of 1.5PN
order in the theory of light propagation; cf. text below Eq.~(\ref{global_metric_perturbation_C}).

An analytical integration of light-trajectory in the field of an uniformly moving
body (\ref{Introduction_15}) has been derived in closed form in 1PN approximation in \cite{Klioner1989} and later 
also in 1PM approximation by means of a suitable Lorentz transformation of the light-trajectory \cite{Klioner_LT}.
As long as one considers uniformly moving bodies, the instant of time $t_A$ in the expansion (\ref{Introduction_15})  
remains an open parameter, but by all means heuristic arguments can be put forward for a meaningful choice for it.  
Perhaps the most fruitful suggestion was that given in \cite{Hellings}, where it was  
supposed to accept that this parameter coincides with the time of  
closest approach of the lightray to the massive body, $t_A^{\ast}$, given by an implicit relation:  
\begin{eqnarray}
t_A^{\ast} &=& t_0 - \frac{\ve{\sigma} \cdot \left(\ve{x}_0 - \ve{x}_A\left(t_A^{\ast}\right)\right)}{c} 
+ {\cal O}\left(c^{-2}\right)\,, 
\label{time_of_closest_approach_t_0}
\\
&=& t_1 - \frac{\ve{\sigma} \cdot \left(\ve{x}_1 - \ve{x}_A\left(t_A^{\ast}\right)\right)}{c}
+ {\cal O}\left(c^{-2}\right)\,,
\label{time_of_closest_approach_t_1}
\end{eqnarray}

\noindent
where $\ve{x}_0 = \ve{x}\left(t_0\right)$ is the global spatial coordinate of the source at the moment of emission of the light-signal 
and $\ve{x}_1 = \ve{x}\left(t_1\right)$ is the global spatial coordinate of the space-based observer at the moment of observation  
of the light-signal; cf. Eq.~(5.13) in \cite{KlionerKopeikin1992}. As a result, in the light propagation formulae (\ref{Introduction_10}) and  
(\ref{Introduction_12}) one would have to insert $\ve{x}_A\left(t_A^{\ast}\right)$.  
That educated guess was triggered by the idea that the biggest influence on the lightray the body exerts when the photon passes nearest to it.  
But an unique justification of this suggestion has not been evidenced at that time. Further arguments have later been put forward  
that partially justify the computation of the parameters of the linear model (\ref{Introduction_15}) to match the real position and velocity 
of the body at the moment of closest approach between the lightray and the real trajectory of the body \cite{KlionerPeip2003}.  

A rigorous solution of the problem of light propagation in the field of arbitrarily moving pointlike monopoles and 
in the first post-Minkowskian approximation has been found in \cite{KopeikinSchaefer1999}, where advanced integration methods 
have been applied that were originally been introduced in \cite{Kopeikin1997} for stationary fields and further developed in
\cite{KopeikinSchaefer1999_Gwinn_Eubanks} for time-dependent fields. According to the solution in \cite{KopeikinSchaefer1999},  
the positions of the bodies have to be computed at the retarded instant of time, $t_A^{\rm ret}$, given by the implicit relation  
\begin{eqnarray}
t_A^{\rm ret} &=& t - \frac{\left| \ve{x}\left(t\right) - \ve{x}_A\left(t_A^{\rm ret}\right)\right|}{c}\,. 
\label{retarded_time}
\end{eqnarray}

\noindent  
The expression (\ref{retarded_time}) is valid for an arbitrary time, e.g. either $t=t_0$ or $t=t_1$.  
With the aid of this rigorous approach in \cite{KopeikinSchaefer1999} it has been shown that  
if the positions and velocities of the bodies are taken at $t_A^{\rm ret}$ then
the effects of acceleration and the effects due to time-dependence of velocity
of the bodies are much smaller than 1 micro-arcsecond. The numerical accuracy of various approaches have been investigated 
in \cite{KlionerPeip2003}, where it was demonstrated for the monopole-term that  
for an accuracy of $1\,$\muas$\;$ it is sufficient to take the 1PN solution of a motionless body in (\ref{Introduction_10}),  
if the position $\ve{x}_A$ of body A is taken at either $t_A^{\ast}$ or $t_A^{\rm ret}$.

\subsubsection{The post-post-Newtonian monopole term}\label{2PN_Term}

Actually, corrections of post-post-Newtonian (2PN) order to the lightray in (\ref{Introduction_6}) will not be on the scope of  
the present investigation, but should briefly be mentioned here for reasons of completeness about \muas--astrometry.

While several post-post-Newtonian effects of light-deflection due to a monopole at rest have been determined a long time ago  
\cite{EpsteinShapiro,FischbachFreeman,RichterMatzner1,RichterMatzner2,RichterMatzner3,Cowling},  
the determination of the explicit time-dependence of the photons coordinate is mandatory in the 
data reduction for highly sophisticated astrometry missions like Gaia.  
In this respect, an important progress has been made in \cite{Brumberg1987}, where a 2PN solution for the light-trajectory in the
Schwarzschild field as function of coordinate-time in a number of coordinate gauges was obtained;
see also \cite{Brumberg1991,KlionerKopeikin1992}. In harmonic gauge, the solution reads  
\begin{eqnarray}
\ve{x}_{\rm 2PN}^M \left(t\right) &=& \ve{x}_0 + c \left(t-t_0\right) \ve{\sigma} 
\nonumber\\
\nonumber\\
&& + \frac{G\,M_A}{c^2}\,\left[\ve{B}_1\left(\ve{r}_A^{\rm 1PN}\right) - \ve{B}_1\left(\ve{r}_0\right)\right] 
\nonumber\\
\nonumber\\
&& + \frac{G^2\,M_A^2}{c^4}\,\left[\ve{B}_2\left(\ve{r}_A^{\rm N}\right) - \ve{B}_2\left(\ve{r}_0\right)\right],  
\label{Introduction_14_A}
\end{eqnarray}

\noindent
where the vectorial functions read (cf. Eqs.~(50) and (51) in \cite{Klioner_Zschocke}):
\begin{widetext}
\begin{eqnarray}
\ve{B}_1(\ve{r}_A^{\rm 1PN}) &=&
- 2 \, \frac{\ve{\sigma}\times(\ve{r}_A^{\rm 1PN} \times \ve{\sigma})}{r_A^{\rm 1PN} - \ve{\sigma}\cdot\ve{r}_A^{\rm 1PN}}
+2\, \ve{\sigma}\,\ln \left(r_A^{\rm 1PN}-\ve{\sigma}\cdot\ve{r}_A^{\rm 1PN}\right)\,,
\label{Appendix_2PN_15}
\\
\nonumber\\
\ve{B}_2(\ve{r}_A^{\rm N}) &=&
+ 4\,\frac{\ve{\sigma}}{r_A^{\rm N} - \ve{\sigma}\cdot \ve{r}_A^{\rm N}} 
+ 4\,\frac{\ve{d}_A}{\left(r_A^{\rm N} - \ve{\sigma}\cdot \ve{r}_A^{\rm N}\right)^2}
+ \frac{1}{4}\,\frac{\ve{r}_A^{\rm N}}{\left(r_A^{\rm N}\right)^2} 
- \frac{15}{4}\,\frac{\ve{\sigma}}{d_A}\,\arctan \left(\frac{\ve{\sigma}\cdot\ve{r}_A^{\rm N}}{d_A}\right)
\nonumber\\
&& - \frac{15}{4}\,\ve{d}_A\,\frac{\ve{\sigma} \cdot \ve{r}_A^{\rm N}}{d_A^3}\,
\left[\frac{\pi}{2} + \arctan \left(\frac{\ve{\sigma}\cdot\ve{r}_A^{\rm N}}{d_A}\right)\right], 
\label{Appendix_2PN_20}
\end{eqnarray}
\end{widetext}

\noindent
while the expressions $\ve{r}_A^{\rm N}$ and $\ve{r}_A^{\rm 1PN}$ are given by:  
\begin{eqnarray}
\ve{r}_A^{\rm N} &=& \ve{x}_0 + c \left(t-t_0\right) \ve{\sigma} - \ve{x}_A\,,
\label{Appendix_2PN_25}
\\
\nonumber\\
\ve{r}_A^{\rm 1PN} &=& \ve{r}_A^{\rm N} 
- 2\,\frac{G\,M_A}{c^2}\,\left(\frac{\ve{d}_A}{r_A^{\rm N} - \ve{\sigma}\cdot \ve{r}_A^{\rm N}} - \frac{\ve{d}_A}{r_A^0 - \ve{\sigma}\cdot \ve{r}_A^0}\right)
\nonumber\\
&& + 2\,\frac{G\,M_A}{c^2}\,\ve{\sigma}\,\ln \frac{r_A^{\rm N} - \ve{\sigma}\cdot \ve{r}_A^{\rm N}}{r_A^0 - \ve{\sigma}\cdot \ve{r}_A^0} \,,
\label{Appendix_2PN_30}
\end{eqnarray}

\noindent
while $\ve{r}_A^0$ is defined by Eq.~(\ref{notation_3c}).
Notice that the expression (\ref{Appendix_2PN_15}) is the source of 1PN and 2PN terms.
Generalizations of that 2PN solution for the case of the parametrized
post-post-Newtonian metric have been given in \cite{Klioner_Zschocke}, where
the numerical magnitudes of
the post-post-Newtonian terms have been estimated and a practical algorithm for
highly-effective computation of the post-post-Newtonian effects has been formulated.

Two alternative approaches to the calculation of propagation-time and direction of the lightrays
have been formulated recently. Both approaches allow one to avoid explicit integration of
the geodesic equations for lightrays.  
The first approach in \cite{LePoncinLafitteLinetTeyssandier2004,TeyssandierLePoncinLafitte2008}
is based on the use of Synge's world function. 
Another approach is based on the eikonal concept and has been developed in \cite{AshbyBertotti2010} in order to investigate
the light propagation in the field of a spherically symmetric body. 

In order to get an idea about the magnitude of 2PN effects, let us recall the well-known fact that the 2PN monopole correction  
for grazing lightrays at the Sun is about $11$ \muas$\;$  
\cite{Brumberg1991,KlionerKopeikin1992,EpsteinShapiro,FischbachFreeman,RichterMatzner1,RichterMatzner2,RichterMatzner3,Cowling}. 
In the concrete case of ESA astrometry-mission Gaia there is a sunshield which is tilted at a $45$ degree angle to the Sun,
so that the telescopes observe a space-region where the post-post-Newtonian effects of the Sun become negligible.
However, while the Gaia mission will not observe close to the Sun, it will observe very close to the surface of
giant planets. A corresponding detailed investigation in \cite{Klioner_Zschocke} has recovered the remarkable fact,
that post-post-Newtonian corrections become relevant for lightrays grazing the surface of the giant planets.
As outlined in \cite{Klioner_Zschocke}, the reason for this fact is the inevitable occurrence of coordinate-dependent 
enhanced terms, because real astrometric measurements incorporate the use of concrete global coordinate systems  
and inherit the choice of coordinate-dependent impact parameters, see also \cite{BodennerWill2003}.

\subsection{Astrometry at sub-micro-arcsecond level of accuracy}  

In order to determine the light-trajectory with an unprecedented accuracy at sub-\muas-level of accuracy, many  
further subtle relativistic effects in the theory of light propagation have to be accounted for. Let us deploy just 
a minimal set of corresponding requirements which need to be considered:  
\begin{enumerate}
\item[$\bullet$] 1. full set of mass-multipoles,   
\item[$\bullet$] 2. spin-dipole,      
\item[$\bullet$] 3. some higher spin-multipoles,       
\item[$\bullet$] 4. motion of arbitrarily moving massive bodies,
\item[$\bullet$] 5. post-post-Newtonian effects.   
\end{enumerate}

\noindent
Of course, what is really necessary to implement into the final relativistic model depends on what is  
actually meant by the term sub-\muas-level of accuracy. For instance, for a model aiming at an accuracy of 
$0.1$ \muas-level there is no need to take into account any higher spin-multipoles in 1.5PN approximation,   
while a model on $0.01$ \muas-level necessitates such terms. 
Let us look at the present situation in the theory of light propagation at sub-\muas-level of accuracy by 
considering each of these five issues mentioned.

\subsubsection{Impact of higher mass-multipoles on light-trajectory}

Keeping the magnitude of quadrupole light-deflection by giant planets in mind, it can easily be foreseen  
that a light propagation model at sub-\muas-level needs to take into account the impact of
higher mass-multipoles beyond the well-known mass-quadrupole term in (\ref{Introduction_12}). 

A systematic approach to the integration of light geodesic equations in the stationary gravitational field
of a localized source at rest, $\ve{x}_A = {\rm const}$, located at the origin of coordinate system and having time-independent
local multipole structure, $M^A_L$ and $S^A_L$, has been worked out in \cite{Kopeikin1997} in 1PN and 1.5PN approximation.
Especially, sophisticated integration methods have been introduced in \cite{Kopeikin1997} allowing for analytical integrations
of geodesic equations in the complex field of multipoles to arbitrary order.

Furthermore, the case of light propagation in the field of a localized source at rest which is  
characterized by time-dependent multipoles has been investigated in \cite{KopeikinKorobkovPolnarev2006,KopeikinKorobkov2005} in 1PM approximation. 
This solution can be interpreted in two different ways: 

(i) Either the localized source is thought of to be  
composed of {\it $N$ arbitrarily moving bodies}, but then the 
time-dependent multipoles have to be interpreted as global multipoles, $m_L\left(t\right)$ and $s_L\left(t\right)$, 
which characterize the entire $N$-body system as a whole.  

(ii) Or the localized source is thought of as being just {\it one body A at rest} with 
intrinsic multipoles, $M^A_L\left(t\right)$ and $S^A_L\left(t\right)$, which characterize that single body.  

But neither of these two interpretations allow one to consider the solution in \cite{KopeikinKorobkovPolnarev2006,KopeikinKorobkov2005} 
to be valid for the case of arbitrarily moving bodies, $\ve{x}_A\left(t\right)$, and with local multipoles $M^A_L\left(t\right)$ and $S^A_L\left(t\right)$  
characterizing each individual body A of the $N$-body system.  

The influence of time-independent intrinsic mass-multipoles of higher-order on a lightray by an isolated
axisymmetric body at rest has also been investigated in \cite{LePoncinLafitteTeyssandier2008}, using a different approach
based on the multipole expansion of time transfer function.
Explicitly, a formula for the bending of light due to any order of multipole moments has been derived and numerical
estimates have been presented. For instance, it has been found in \cite{LePoncinLafitteTeyssandier2008}
that the light-deflection due to mass-octupole structure amounts to be $0.016$ \muas $\;$
and due to mass-hexadecupole structure amounts to be $9.6$ \muas $\;$ for grazing rays at Jupiter.

Recently, in \cite{moving_axisymmetric_body} the light propagation in the field of an uniformly moving axisymmetric body
has been determined in terms of the full multipole structure of the body.
Furthermore, an analytical formula for the time-delay caused by the gravitational field of a body in slow and
uniform motion with arbitrary multipoles has been derived in \cite{Soffel_Han}.

{\bf 1. assessment:} 
According to these investigations in the literature,  
the 1PN solution $\Delta\ve{x}_{\rm 1PN}\left(t,t_0\right)$ in (\ref{Introduction_8}) 
in the gravitational field of $N$ arbitrarily moving bodies, $\ve{x}_A\left(t\right)$, and with time-dependent  
intrinsic mass-multipoles, $M_L^A\left(t\right)$, has not been determined thus far, but appears to be an inevitable requirement  
for sub-\muas--astrometry.

\subsubsection{Light propagation in the field of spin-dipoles}

The next term beyond \muas--astrometry which is certainly required at sub-\muas-level is the impact of  
rotational motion of massive bodies on the light propagation;  
note that such a term is already of 1.5PN order.  
For instance, the light-deflection due to rotational motion of Solar system bodies amounts to be  
$0.7$ \muas $\;$ for grazing ray at Sun, $0.2$ \muas $\;$ for grazing ray at Jupiter, and     
$0.04$ \muas $\;$ for grazing ray at Saturn \cite{Klioner2003a,Klioner1991}.  

The first solution of the light-trajectory $\Delta \ve{x}^S_{\rm 1.5PN}\left(t,t_0\right)$ in the
gravitational field of massive bodies at rest possessing a time-independent intrinsic spin-dipole, $\ve{S}^A$, has
been obtained in \cite{Klioner1991}. This solution provides all the details of
light propagation, especially the time-dependence of the coordinates
of the photon and the solution of the corresponding boundary value problem.

Utilizing advanced integration methods, a solution for the light-trajectory in the field of one body at rest and having  
time-independent local spin-dipole, $\ve{S}^A$, has also been obtained in \cite{Kopeikin1997} in 1.5PN approximation.  
Moreover, an analytical solution in 1PM approximation
for the case of light propagation in the field of an arbitrarily moving pointlike spin-dipole,   
$\ve{s}\left(t\right)$ (expressed in terms of a global spin-tensor)  
has been derived in \cite{KopeikinMashhoon2002}.  

{\bf 2. assessment:} 
In view of these few investigations available in the literature, the task remains to determine the light-trajectory 
$\Delta \ve{x}^S_{\rm 1.5PN}\left(t,t_0\right)$ in the gravitational field of an arbitrarily moving body,  
$\ve{x}_A\left(t\right)$, carrying a time-dependent intrinsic spin-dipole, $\ve{S}^A\left(t\right)$.

\subsubsection{Impact of higher spin-multipoles on light-trajectory}

As mentioned above, a solution for the light-trajectory in the stationary gravitational field of a localized source at rest, 
$\ve{x}_A = {\rm const}$,  
with time-independent local multipoles, $M^A_L$ and $S^A_L$, has been determined in 1.5PN approximation in \cite{Kopeikin1997}.  
Furthermore, the light-trajectory in the field of a localized source with time-dependent global
multipoles, $m_L\left(t\right)$ and $s_L\left(t\right)$, has been obtained
in \cite{KopeikinKorobkovPolnarev2006,KopeikinKorobkov2005} in 1PM approximation.
As it has been noticed already, the results in \cite{KopeikinKorobkovPolnarev2006,KopeikinKorobkov2005} 
can be considered as solution for the light-trajectory in the field of either a system  
of {\it $N$ arbitrarily moving bodies} characterized by global multipoles or in the field of   
{\it one body A at rest} characterized by local multipoles, but not as solution for for the light-trajectory in the field of 
arbitrarily moving bodies characterized by intrinsic multipoles.  

Recent calculations \cite{Jan-Meichsner_Diploma_Thesis} have revealed, that the light-deflection 
due to spin-octupole structure of massive bodies at rest  
amounts to be about $0.015$ \muas$\;$ for Jupiter and about $0.006$ \muas$\;$ for Saturn  
for grazing rays. Therefore, a model at sub-\muas-level has to take into account at least the spin-octupole term 
which is of 1.5PN order in the theory of light propagation. 

{\bf 3. assessment:} 
According to these facts, the 1.5PN solution $\Delta\ve{x}_{\rm 1.5PN}\left(t,t_0\right)$ in (\ref{Introduction_6}) in the 
gravitational field of $N$ arbitrarily moving bodies, $\ve{x}_A\left(t\right)$, and with time-dependent  
intrinsic spin-multipoles, $S_L^A\left(t\right)$, has not been determined so far and remains an unavoidable task
in order to achieve an astrometric accuracy at sub-\muas-level.

\subsubsection{Impact of the motion of the bodies on light-trajectory}

The Solar system bodies are moving along their individual worldlines, $\ve{x}_A\left(t\right)$, which are complicated functions of time due to  
the mutual interaction among the bodies, implying that the metric and the light-trajectory become also complicated functions of time.   
As explicated in section \ref{Motion_A}, for \muas-astrometry this highly sophisticated problem 
can be treated by using the standard 1PN solutions of motionless bodies, $\ve{x}_A = {\rm const}$, as long as the positions of the bodies  
are taken at either their retarded times $t_A^{\rm ret}$ or at their time of closest approach to the lightray $t_A^{\ast}$.  

However, in the investigation \cite{KlionerPeip2003} it has been shown that for an astrometric astrometry better than $0.2$ \muas$\;$ 
one needs to take into account the motion of the bodies. Especially, it is not sufficient to apply for a simple series-expansion of the 
bodies worldline, $\ve{x}_A\left(t\right)=\ve{x}_A + \ve{v}_A\left(t - t_A\right)$, as given by Eq.~(\ref{Introduction_15}).  
Instead, one has to determine the light-trajectory in the field of arbitrarily moving bodies $\ve{x}_A\left(t\right)$.  
For the case of arbitrarily moving monopoles such a solution has been provided in \cite{KopeikinSchaefer1999}, and 
for the case of arbitrarily moving bodies with quadrupole structure such a solution has been found in \cite{KopeikinMakarov2007}.  
But for arbitrarily moving bodies with higher intrinsic multipoles there are no solutions available so far.  

{\bf 4. assessment:}
As a result, for sub-\muas-astrometry the approximative expansion in (\ref{Introduction_15}) is not applicable, instead of that
one has to find a solution for the light-trajectory in terms of arbitrary worldlines $\ve{x}_A\left(t\right)$. The real worldlines
of the massive bodies can finally be implemented into the model by means of Solar system ephemerides \cite{JPL}.

\subsubsection{Post-post-Newtonian effects}\label{2PN_Effects} 

The most intricate issue in the theory of sub-micro-arcsecond astrometry will be the post-post-Newtonian effects  
$\Delta \ve{x}_{\rm 2PN}\left(t,t_0\right)$ in (\ref{Introduction_6}). Such 2PN corrections to the lightray  
will not be on the scope of this investigation, but some remarks should be in order.  

The largest perturbation term is of course the monopole-term, $\Delta \ve{x}^M_{\rm 2PN}\left(t,t_0\right)$, which 
in case of pointlike bodies at rest, $\ve{x}_A={\rm const}$, has been calculated at the first time in \cite{Brumberg1987};  
see also \cite{Brumberg1991,KlionerKopeikin1992}. In reality, the bodies are moving, and one has to 
treat the problem of moving monopoles in post-post-Newtonian approximation where, however,   
only very limited results are available thus far.
Especially, in \cite{Bruegmann2005} the light-deflection
in 2PN approximation in the field of two moving point-like bodies has been determined, using two essential approximations:
(i) both the light-source and the observer are assumed to be located at infinity in an
asymptotically flat space, and  
(ii) the relative separation distance of the bodies is assumed to be much smaller than the impact parameter of incoming lightray. These
approximations are of interest in case of studying light propagation in the field of a binary pulsar,
but they are not applicable for real astrometric observations in the Solar system.

Presently it remains unknown, how large the impact of higher mass-multipoles on light-deflection in post-post-Newtonian order is.   
In order to tackle this problem, an extension of the DSX-metric  
\cite{DSX1,DSX2} towards post-linear order is mandatory; see text below Eq.~(\ref{Expansion_B}).   
There are several preliminary and promising efforts to extend relativistic astrometry to post-post-Newtonian order for lightrays,  
especially to focus on the 2PN gravitational field of
arbitrarily moving bodies endowed with arbitrary intrinsic mass- and spin-multipole moments. 
There have been several attempts to solve this problem \cite{Chinese_Xu_Wu,Xu_Gong_Wu_Soffel_Klioner,MinazzolliChauvineau2009},  
but they are far from being complete.
Problems, that have been ignored in these articles are
related with the internal structure of extended bodies. For a single body at rest
these problems are well understood for both the post-Newtonian \cite{Thorne,Blanchet_Damour1} and the
post-Minkowskian case \cite{Blanchet_Damour2,Multipole_Damour_2}, where
many structure dependent terms appear in intermediate calculations that
cancel exactly in virtue of the local equations of motion or can be
eliminated by corresponding gauge transformations. However, in  
post-post-Newtonian order the situation is still unclear. 
For a spherically symmetric body the complete derivation of the metric in the  
exterior of the massive body (Schwarzschild metric) was recently solved in \cite{KS},  
where it has been shown how such structure-dependent
terms cancel so that one finally ends up with the well-known Schwarzschild solution in harmonic
gauge. This work allows in principle to determine the light-trajectory in the field 
of a spherically symmetric and extended massive body at rest in 2PN approximation.  

{\bf 5. assessment:}
So far, the light-trajectory in 2PN approximation, $\Delta \ve{x}_{\rm 2PN}\left(t,t_0\right)$ in (\ref{Introduction_6}),  
is only known for pointlike monopoles at rest. Moreover, the DSX-metric in post-linear approximation has    
to be determined, in order to ascertain the impact on light-deflection  
of terms in second post-Newtonian order beyond the monopole-term, either numerically or analytically.

\section{Geodesic equation in 1PN approximation}\label{Section2}

The description of the metric of the Solar system becomes more complex the more accurate the astrometric measurements are
and one has to resort on approximation schemes to solve the geodesic equation (\ref{Geodetic_Equation}).  
Since the gravitational fields of Solar system are weak and the motions of the massive bodies are slow, we can utilize the 
so-called post-Newtonian expansion (weak-field slow-motion approximation) for the metric as given by Eq.~(\ref{metric_perturbation_2PN_5}).  
The main objective of this investigation is an analytical solution for the light-trajectory in 1PN approximation, see Eq.~(\ref{Introduction_8}).  
As a result, terms of the order ${\cal O}\left(c^{-2}\right)$ in the metric tensor are required for such an approximation:   
\begin{eqnarray}
g_{\alpha\beta}\left(t,\ve{x}\right) &=& \eta_{\alpha\beta} 
+ h^{(2)}_{\alpha\beta}\left(t,\ve{x}\right)   
+ {\cal O} \left(c^{-3}\right). 
\label{metric_perturbation_pN}
\end{eqnarray}

\noindent
Inserting (\ref{metric_perturbation_pN}) into (\ref{Geodetic_Equation}) with virtue of (\ref{Christoffel_Symbols}) yields  
the geodesic equation in 1PN approximation, which can be rewritten
in terms of global coordinate-time; cf. Refs.~\cite{Brumberg1991,KlionerKopeikin1992,KlionerPeip2003} 
(cf. the first four terms in Eq.~(A.4) in \cite{KlionerPeip2003}):  
\begin{eqnarray}
\frac{\ddot{x}^i \left(t\right)}{c^2} &=& 
+ \frac{1}{2}\,h_{00,i}^{(2)} - h_{00,j}^{(2)} \frac{\dot{x}^i\left(t\right)}{c}\frac{\dot{x}^j\left(t\right)}{c}
- h_{ij,k}^{(2)}\,\frac{\dot{x}^j\left(t\right)}{c}\frac{\dot{x}^k\left(t\right)}{c}  
\nonumber\\
&& + \frac{1}{2}\,h_{jk,i}^{(2)}\,\frac{\dot{x}^j\left(t\right)}{c}\frac{\dot{x}^k\left(t\right)}{c}
+ {\cal O}\left(c^{-3}\right),  
\label{geodesic_equation_1}
\end{eqnarray}

\noindent
where $h^{(2)}_{\alpha\beta,i}=\partial h^{(2)}_{\alpha\beta}/\partial x^i$, while a dot denotes derivative with respect 
to coordinate-time.
In order to find an unique solution of the geodesic equation in (\ref{geodesic_equation_1}), so-called mixed
initial-boundary conditions must be imposed, which have extensively been used in the literature, e.g.
\cite{Brumberg1991,KlionerKopeikin1992,Klioner_Zschocke,Kopeikin1997,KopeikinSchaefer1999_Gwinn_Eubanks,Brumberg1987,KopeikinKorobkovPolnarev2006}: 
\begin{eqnarray}
\ve{x}_0 &=& \ve{x}\left(t_0\right),
\label{Initial_Boundary_Condition_1}
\\
\nonumber\\
\ve{\sigma} &=& \lim_{t \rightarrow - \infty}\, \frac{\dot{\ve{x}}\left(t\right)}{c}\,.
\label{Initial_Boundary_Condition_2}
\end{eqnarray}

\noindent
The first condition (\ref{Initial_Boundary_Condition_1}) defines the spatial coordinates of the photon at the moment
$t_0$ of emission of light. The second condition (\ref{Initial_Boundary_Condition_2}) defines the unit-direction
$\left(\ve{\sigma}\cdot\ve{\sigma} = 1\right)$ of the lightray at past null infinity, that means the
unit-tangent vector along the light-path at infinite distance in the past from the origin of the global coordinate system.

The metric perturbations in (\ref{geodesic_equation_1})
are functions of the coordinates of the global reference system (BCRS).
It is, however, important to realize that in the geodesic equation this coordinate-dependence has always
to be understood as being the coordinates of the photon $\ve{x}\left(t\right)$ at time $t$, that means
\begin{eqnarray}
h_{\alpha \beta}^{(2)} &=& h_{\alpha \beta}^{(2)}\left(t,\ve{x}\right)
\Bigg|_{\ve{x}=\ve{x}\mbox{\normalsize $\left(t\right)$}}\,. 
\label{geodesic_equation_2}
\end{eqnarray}

\noindent
Consequently, the spatial derivatives in (\ref{geodesic_equation_1}) are taken along the lightray:  
\begin{eqnarray}
h_{\alpha \beta, i}^{(2)} &=& \frac{\partial h_{\alpha \beta}^{(2)}\left(t,\ve{x}\right)}{\partial x^i}
\Bigg|_{\ve{x}=\ve{x}\mbox{\normalsize $\left(t\right)$}}\;.
\label{geodesic_equation_3}
\end{eqnarray}

\noindent
The geodesic equation in (\ref{geodesic_equation_1}) has usually been solved by an iteration procedure. In the first iteration the  
right-hand side in (\ref{geodesic_equation_1}) vanishes, $\ddot{x}^i = 0$, and the integration of this differential  
equation yields the unperturbed lightray in Eq.~(\ref{Unperturbed_light_ray_10}).  
The exact light-trajectory $\ve{x}\left(t\right)$ deviates from the Newtonian approximation by terms of
the order ${\cal O}\left(c^{-2}\right)$, that means   
\begin{eqnarray}
\ve{x}\left(t\right) &=& \ve{x}_{\rm N}\left(t\right) + {\cal O}\left(c^{-2}\right).
\label{Unperturbed_light_ray_15}
\end{eqnarray}

\noindent
Solving the geodesic equations (\ref{geodesic_equation_1}) by iteration implies that $\dot{\ve{x}}\left(t\right)$ can be replaced by its Newtonian  
approximation, $\dot{\ve{x}}_{\rm N}\left(t\right) = c\,\ve{\sigma}$, which follows by time-derivative of (\ref{Unperturbed_light_ray_10}), 
so that the geodesic equation in (\ref{geodesic_equation_1}) simplifies as follows:  
\begin{eqnarray}
\frac{\ddot{x}^i \left(t\right)}{c^2} &=&
+ \frac{1}{2}\,h_{00,i}^{(2)}  
- h_{00,j}^{(2)}\,\sigma^i\,\sigma^j - h_{ij,k}^{(2)}\,\sigma^j\,\sigma^k
\nonumber\\
&& + \frac{1}{2}\,h_{jk,i}^{(2)}\,\sigma^j\,\sigma^k
+ {\cal O}\left(c^{-3}\right).  
\label{geodesic_equation_5}
\end{eqnarray}

\noindent
In 1PN approximation, the metric perturbations in (\ref{geodesic_equation_5}) have to be taken at the
spatial coordinates of the unperturbed lightray given by (\ref{Unperturbed_light_ray_10}),
that means   
\begin{eqnarray}
h_{\alpha \beta}^{(2)} &=& h_{\alpha \beta}^{(2)}\left(t,\ve{x}\right)
\Bigg|_{\ve{x}=\ve{x}_{\rm N}\mbox{\normalsize $\left(t\right)$}}\;,
\label{transformed_geodesic_equation_10}
\end{eqnarray}

\noindent
and in (\ref{geodesic_equation_5}) one has first to differentiate with respect to  
spatial coordinates and afterwards one inserts the unperturbed lightray, that means   
\begin{eqnarray}
h_{\alpha \beta, i}^{(2)} &=& \frac{\partial h_{\alpha \beta}^{(2)}\left(t,\ve{x}\right)}{\partial x^i}
\Bigg|_{\ve{x}=\ve{x}_{\rm N}\mbox{\normalsize $\left(t\right)$}}\,.
\label{transformed_geodesic_equation_15}
\end{eqnarray}

\noindent
In our investigation we will solve the geodesic equation (\ref{geodesic_equation_5})  
in 1PN approximation, that means the exact light-trajectory $\ve{x}\left(t\right)$ is
determined up to terms of the order ${\cal O}\left(c^{-3}\right)$:
\begin{eqnarray}
\ve{x}\left(t\right) &=& \ve{x}_{\rm 1PN}\left(t\right) + {\cal O}\left(c^{-3}\right). 
\label{exact_light_ray_PN_1}
\end{eqnarray}

\noindent
The first and second integral of geodesic equations (\ref{geodesic_equation_5}) in 1PN approximation
can formally be written as follows \cite{Brumberg1991}:
\begin{eqnarray}
\dot{\ve{x}}_{\rm 1PN} \left(t\right) &=& c\,\ve{\sigma} + \Delta\dot{\ve{x}}_{\rm 1PN}\left(t\right),
\label{light_trajectory_A}
\\
\nonumber\\
\ve{x}_{\rm 1PN}\left(t\right) &=& \ve{x}\left(t_0\right) + c\,\ve{\sigma}\left(t-t_0\right) +
\Delta \ve{x}_{\rm 1PN} \left(t,t_0\right),
\label{light_trajectory_B}
\end{eqnarray}

\noindent
where $\Delta \ve{x}_{\rm 1PN}$ are small perturbations of the unperturbed light-trajectory,   
and $\Delta\dot{\ve{x}}_{\rm 1PN}$ is the time-derivative of these small perturbations.

\section{The metric of Solar system}\label{Section4}

In order to describe and to interpret observational data in astrometry correctly, a set of 
several reference systems and the transformation laws among their coordinates must be introduced.  
In this respect, two standard reference systems are of fundamental importance, which are adopted by
the IAU resolution B1.3 (2000) \cite{IAU_Resolution1}: the Barycentric Celestial Reference System (BCRS)  
with coordinates $\left(ct,\ve{x}\right)$ and the 
Geocentric Celestial Reference System (GCRS) with coordinates $\left(cT,\ve{X}\right)$.  
Furthermore, for any massive body A of the Solar system a so-called GCRS-like reference system  
with coordinates $\left(cT_A,\ve{X}_A\right)$ can be introduced. In this section we will 
give a summary about how to combine these systems to a global metric tensor in terms of local multipoles, which 
is the physically adequate reference system for modeling of light-trajectories through the Solar system.  

\subsection{BCRS}\label{BCRS} 

The harmonic coordinates of BCRS are denoted by $x^{\mu}=\left(ct,x^i\right)$, where $t={\rm TCB}$ is the BCRS coordinate-time, 
and cover the entire space-time and can therefore be used to model light-trajectories from distant celestial objects to the observer. 
The origin of the BCRS is located at the barycenter of the Solar system, and the IAU Resolution B2 (2006) \cite{IAU_Resolution2} 
recommends the spatial axes of BCRS to be oriented according to the spatial axes of the International Celestial Reference System  
(ICRS) \cite{ICRS}.   
According to IAU resolution B1.3 (2000) \cite{IAU_Resolution1}, the Solar system is assumed to be isolated and the space-time  
is asymptotically flat, that means the BCRS metric $g_{\mu\nu}\left(t,\ve{x}\right)$ at spatial infinity reads:  
\begin{eqnarray}
\lim_{\mbox{\scriptsize $\left|\ve{x}\right| \rightarrow \infty$}}\,
g_{\mu\nu}\left(t,\ve{x}\right) &=& \eta_{\mu\nu}\,.
\label{boundary_condition_global_metric}
\end{eqnarray}

\noindent
The BCRS is completely characterized by the form of its metric tensor, up to order ${\cal O}\left(c^{-3}\right)$ 
given by \cite{IAU_Resolution1}:  
\begin{eqnarray}
g_{00}\left(t,\ve{x}\right) &=& - 1 + \frac{2\,w\left(t,\ve{x}\right)}{c^2} + {\cal O}\left(c^{-4}\right),  
\label{BCRS_1}
\\
\nonumber\\
g_{0i}\left(t,\ve{x}\right) &=& {\cal O}\left(c^{-3}\right),
\label{BCRS_2}
\\
\nonumber\\
g_{ij}\left(t,\ve{x}\right) &=& \left( 1 + \frac{2\,w\left(t,\ve{x}\right)}{c^2}\right) \delta_{ij} + {\cal O}\left(c^{-4}\right).  
\label{BCRS_3}
\end{eqnarray}

\noindent
The scalar gravitational potential in (\ref{BCRS_1}) and (\ref{BCRS_3}) is given by the integral 
\begin{eqnarray}
w\left(t,\ve{x}\right) &=& \frac{G}{c^2} \int d^3 x^{\prime} \;
\frac{t^{00}\left(t,\ve{x}^{\prime}\right)}{\left|\ve{x} - \ve{x}^{\prime}\right|} + {\cal O}\left(c^{-2}\right),  
\label{BCRS_4}
\end{eqnarray}

\noindent
which runs over the entire Solar system, and 
where $t^{00}$ is the time-time-component of the energy-momentum tensor $t^{\mu\nu}$ in global BCRS
coordinates; recall the components of energy-momentum tensor scale as follows:
$t^{00} = {\cal O}\left(c^2\right), t^{0i} = {\cal O}\left(c^1\right), t^{ij} = {\cal O}\left(c^0\right)$. 

The global gravitational potential in (\ref{BCRS_4}) admits an expansion in terms of global STF multipoles, which characterize  
the multipole structure of the Solar system as a whole \cite{Thorne,Blanchet_Damour1,Blanchet_Damour2}:    
\footnote{For the relations between metric tensor and gothic metric tensor in 
\cite{Blanchet_Damour1,Blanchet_Damour2} we refer to \cite{Thorne}; see also Appendix A in \cite{Zschocke_Soffel}.}: 
\begin{eqnarray}
w\left(t,\ve{x}\right) &=& G \sum\limits_{l=0}^{\infty} \frac{\left(-1\right)^l}{l!}\,m_L\left(t\right)\,
\partial_L\,\frac{1}{r} + {\cal O}\left(c^{-2}\right),
\label{global_multipole_expansion}
\end{eqnarray}

\noindent
where $\displaystyle \partial_L=\frac{\partial}{\partial x^{a_1}}\,...\,\frac{\partial}{\partial x^{a_l}}$.
The global mass-multipoles in (\ref{global_multipole_expansion})
are Cartesian symmetric and trace-free (STF) tensors, in Newtonian approximation given by; 
cf. Eq.~(2.34a) in \cite{Blanchet_Damour1}:  
\begin{eqnarray}
m_L\left(t\right) &=& \int d^3x\,\hat{x}_L\, \frac{t^{00}\left(t,\ve{x}\right)}{c^2} + {\cal O} \left(c^{-2}\right), 
\label{global_mass_multipoles}
\end{eqnarray}

\noindent
where the integral runs over the entire Solar system. 
The global mass-monopole, i.e. $l=0$ in Eq.~(\ref{global_mass_multipoles}), is just the total (Newtonian) 
mass, $M = {\rm const}$, of the entire Solar system, while  
the global mass-dipole term vanishes, i.e. $m_i = 0$, because the origin of BCRS is located at the barycenter of the Solar system.  

A further comment should be in order about a possible retarded time argument of the energy-momentum tensor in Eq.~(\ref{BCRS_4});   
cf. text below Eqs.~(17) in \cite{IAU_Resolution1}. One may easily recognize that such retarded time-argument would be 
beyond 1PN approximation for the lightrays. Especially, in terms of multipole-expansion one may demonstrate the following relation:  
\begin{eqnarray}
w\left(t,\ve{x}\right) &=& G \sum\limits_{l=0}^{\infty} \frac{\left(-1\right)^l}{l!}\,m_L\left(t\right)\, 
\partial_L\,\frac{1}{r} + {\cal O}\left(c^{-2}\right)  
\nonumber\\
&=& G \sum\limits_{l=0}^{\infty} \frac{\left(-1\right)^l}{l!}\,  
\partial_L\,\frac{m_L\left(t_{\rm ret}\right)}{r} + {\cal O}\left(c^{-2}\right), 
\label{global_multipole_expansion_retarded}
\end{eqnarray}

\noindent
where the retarded time has been defined by Eq.~(\ref{retarded_time}).  
If one expands the retarded multipoles (second line in Eq.~(\ref{global_multipole_expansion_retarded})) in inverse powers of $c$, then one finds  
that all terms proportional to $1/c$ cancel against each other. This cancellation is important, because terms of odd powers $1/c$  
would violate the time-reversal symmetry; cf. the corresponding statement in the text below Eq.~(17) in the IAU resolutions \cite{IAU_Resolution1}. 
The time-reversal symmetry is violated because of the gravitational radiation emitted by the Solar system which is, however,   
an effect much beyond 1PN approximation.

The expansion in (\ref{global_multipole_expansion}) has two specific characteristics,
which prevent a direct use for our intentions:

(1) As emphasized in \cite{Thorne,Blanchet_Damour1,Blanchet_Damour2,Multipole_Damour_2},
the expansion in (\ref{global_mass_multipoles}) is valid outside a sphere which encloses the
complete $N$-body system, see also \cite{Zschocke_Multipole_Expansion}. However, for a description
of lightrays inside the Solar system (light-trajectories between
the massive bodies) one has to apply a metric tensor which is also valid inside this sphere, i.e. in space-regions
between these massive bodies; cf. text on page 3298 in \cite{DSX1}.

(2) From the theory of relativistic reference systems it is clear that physically meaningful
multipole moments of some body A have to be defined in the body's local reference system $\left(cT_A, \ve{X}_A\right)$.

For these reasons, in our approach we will have to express the gravitational potential in (\ref{global_multipole_expansion})  
by local (intrinsic) mass-multipoles $M^A_L$,  
which are defined in the local coordinate system $\left(cT_A,\ve{X}_A\right)$ of the corresponding massive body. 
This crucial issue will be the subject in what follows.

\subsection{GCRS}\label{GCRS} 

The harmonic coordinates of GCRS are denoted by $X^{\mu}=\left(cT,X^i\right)$, where $T={\rm TCG}$ is the GCRS coordinate-time.  
According to IAU resolution B1.3 (2000) \cite{IAU_Resolution1}, the origin of GCRS is co-moving with the Earth and  
located at the barycenter of the Earth, and is adequate to describe physical processes in the vicinity of the Earth.  
The spatial axes of GCRS are kinematically non-rotating with respect to BCRS, i.e. they are locally non-inertial.  
The GCRS is completely characterized by the form of its metric tensor, up to order ${\cal O}\left(c^{-3}\right)$ given by
\cite{IAU_Resolution1,DSX1,DSX2},
\begin{eqnarray}
G_{00}\left(T,\ve{X}\right) &=& - 1 + \frac{2\,W\left(T,\ve{X}\right)}{c^2} + {\cal O}\left(c^{-4}\right),
\label{GCRS_1}
\\
\nonumber\\
G_{0i}\left(T,\ve{X}\right) &=& {\cal O}\left(c^{-3}\right),
\label{GCRS_2}
\\
\nonumber\\
G_{ij}\left(T,\ve{X}\right) &=& \left(1 + \frac{2\,W\left(T,\ve{X}\right)}{c^2}\right)\delta_{ij} + {\cal O}\left(c^{-4}\right).
\label{GCRS_3}
\end{eqnarray}

\noindent
The scalar gravitational potential in (\ref{GCRS_1}) and (\ref{GCRS_3}) can uniquely be separated into two terms:
a local potential, $W_{\rm loc}$, which originates from the body A itself and an external potential, $W_{\rm ext}$,  
which is associated with inertial effects (due to the accelerated motion of the local system)
and tidal forces (caused by the other bodies of the Solar system) \cite{IAU_Resolution1,DSX1,DSX2}:
\begin{eqnarray}
W\left(T,\ve{X}\right) &=& W_{\rm loc}\left(T,\ve{X}\right) + W_{\rm ext}\left(T,\ve{X}\right).
\label{GCRS_4}
\end{eqnarray}

\noindent
Explicit expressions for the external potential $W_{\rm ext}$ are given in \cite{DSX1,DSX2}, while  
the potential $W_{\rm loc}$ is defined by the following integral,  
\begin{eqnarray}
W_{\rm loc}\left(T,\ve{X}\right) &=& \frac{G}{c^2} \int_{V_E} d^3 X^{\prime}\;  
\frac{T^{00}\left(T,\ve{X}^{\prime}\right)}{\left|\ve{X} - \ve{X}^{\prime}\right|} + {\cal O}\left(c^{-2}\right),
\nonumber\\
\label{GCRS_6}
\end{eqnarray}

\noindent
which runs over the entire volume $V_E$ of the Earth, and
where $T^{00}$ is the time-time-component of the energy-momentum tensor $T^{\mu\nu}$ of the isolated Earth and expressed in GCRS
coordinates; recall the components of energy-momentum tensor scale as follows:
$T^{00} = {\cal O}\left(c^2\right), T^{0i} = {\cal O}\left(c^1\right), T^{ij} = {\cal O}\left(c^0\right)$.
The local potential (\ref{GCRS_6}) is generated by the Earth and can be expanded into a series of local STF multipole moments, which 
characterize the multipole structure of the Earth as an isolated body  
\cite{IAU_Resolution1,Thorne,Blanchet_Damour1,Blanchet_Damour2,Multipole_Damour_2,DSX1}:
\begin{eqnarray}
W_{\rm loc}\left(T,\ve{X}\right) &=& G\,\sum\limits_{l=0}^{\infty} \frac{\left(-1\right)^l}{l!}
M_L\left(T\right)\;{\cal D}_L\,\frac{1}{R} + {\cal O} \left(c^{-2}\right),
\nonumber\\
\label{BCRS_10}
\end{eqnarray}

\noindent
where $\displaystyle {\cal D}_L = \frac{\partial}{\partial X^{a_1}}\,...\,\frac{\partial}{\partial X^{a_l}}$.  

The local mass-monopole, i.e. $l=0$ in Eq.~(\ref{BCRS_10}), is just the (Newtonian) mass of the Earth, $M = {\rm const}$.  
Actually, the origin of the GCRS is assumed to be located at the barycenter of the Earth, hence the dipole-term in 
(\ref{BCRS_10}) vanishes: $M_{i}=0$. But in real measurements of celestial mechanics the center-of-mass of massive 
Solar system bodies can usually not be determined exactly, so it is meaningful to keep  
this term and to assume $M_{i} \neq 0$ in general.   
The STF mass-multipoles $M_L$ in (\ref{BCRS_10}) in Newtonian approximation are given by
\begin{eqnarray}
M_L \left(T\right) &=&
\int_{V_E} d^3 X\;\hat{X}_L\;\frac{T^{00}\left(T,\ve{X}\right)}{c^2} + {\cal O}\left(c^{-2}\right). 
\nonumber\\
\label{local_Newtonian_Mass_Multipole}
\end{eqnarray}

\noindent
According to the theory of reference systems, 
\cite{IAU_Resolution1,Brumberg1991,BK1,Reference_System1,BK2,BK3,DSX1,DSX2,DSX3,DSX4},  
the GCRS is the standard reference system to define local multipoles of the Earth.   
However, as it has been noted in \cite{IAU_Resolution1}, the detailed form of mass-multipoles in (\ref{local_Newtonian_Mass_Multipole})
is not needed for practical astrometry or celestial mechanics, since
these terms are related to observational quantities. That means, the gravitational potentials  
can be expanded in terms of vector spherical harmonics and the coefficients of
such an expansion are equivalent to the local multipoles, see appendix A in \cite{IAU_Resolution1}.

\subsection{Metric of Solar system in terms of intrinsic multipoles in the DSX-framework}\label{Section4_2}

Physically meaningful multipoles of the massive bodies can only be defined in their local reference systems. On these grounds,  
for each massive body A of the Solar system a GCRS-like reference system with coordinates $\left(cT_A,\ve{X}_A\right)$ and co-moving  
with the body A is introduced, to permit the definition of local multipoles of this body. Hence, for a $N$-body system  
there are in total $N$+1 reference systems, one global chart $\left(ct,\ve{x}\right)$ and $N$ local charts $\left(c T_A, \ve{X}_A\right)$,  
which are linked to each other via coordinate-transformations, which allow the construction of one global reference system in terms of  
local multipoles $M^A_L$ of the massive bodies A=1,...,$N$. That reference system is valid in the entire near-zone of the Solar system, and  
combines the advantage of locally defined multipoles and is well-defined in space-regions between the massive bodies; 
cf. text above Eq.~(6.9a) in \cite{DSX1}.  
Such a system is also physically adequate for modeling the light-trajectory from a light-source through the near-zone of
the Solar system towards the observer.
The corresponding framework has been elaborated within the DSX theory \cite{DSX1,DSX2,DSX3,DSX4},
which has originally been established for celestial mechanics and for deriving the equations of motion
of a $N$-body system. This framework has later be reformulated in terms of PPN formalism in \cite{Klioner_Soffel_DSX},
aiming at several tests of relativity in celestial mechanics, e.g. tests of equivalence principle.
One main result of the DSX-formalism are these transformation rules for the coordinates   
$\left(ct, \ve{x}\right) \longleftrightarrow \left(cT_A, \ve{X}_A\right)$ and for the metric potentials  
$w \longleftrightarrow W_A$.  
According to \cite{DSX1,DSX2}, the global coordinates $\left(ct,\ve{x}\right)$ and the local coordinates
$\left(cT_A,\ve{X}_A\right)$ of some body A are related by the following coordinate transformation; 
cf. Eq.~(2.8a) in \cite{DSX1} (for the inverse transformation we refer to \cite{IAU_Resolution1}):   
\begin{eqnarray}
x^{\mu} &=& x^{\mu}_A \left(T_A\right) + e^{\mu}_{a}\left(T_A\right) X^a_A + {\cal O} \left(c^{-2}\right),
\label{coordinate_transformation_1}
\end{eqnarray}

\noindent
where $x^{\mu}_A$ is the worldline of body A in BCRS coordinates (i.e. a selected point associated with body A)
and $e^{\mu}_{a}$ are tetrads along the worldline of this body; cf. Eqs.~(2.16) in \cite{DSX1}:
\begin{eqnarray}
e_a^0\left(T_A\right) &=& \frac{\dot{x}_A^a\left(T_A\right)}{c} + {\cal O}\left(c^{-3}\right),
\label{tetrade_2}
\\
\nonumber\\
e_a^i\left(T_A\right) &=& \delta_{ai} + {\cal O}\left(c^{-2}\right),
\label{tetrade_3}
\end{eqnarray}

\noindent
where in (\ref{tetrade_2}) a dot means derivative with respect to $T_A$; thus $\dot{x}_A^a\left(T_A\right)$ are the spatial  
components of the three-velocity of body A in the global system and given in terms of the body's local coordinate-time $T_A$.
Without going into the details, using the tensorial transformation rule for metric tensors 
in different coordinate systems (cf. Eq.~(4.11) in \cite{DSX1}),
it has been demonstrated in \cite{DSX1} that the global potential  
can be expressed in terms of local (intrinsic) STF multipoles $M_L^A$ as follows  
(for the inverse transformation we refer to \cite{IAU_Resolution1}):
\begin{eqnarray}
w\left(t,\ve{x}\right) &=& \sum\limits_{A=1}^{N}\, w_A\left(t,\ve{x}\right),  
\label{global_metric_potentials_1}
\\
\nonumber\\
w_A\left(t,\ve{x}\right) &=& G \sum\limits_{l = 0}^{\infty} \frac{\left(-1\right)^l}{l!}  
M_L^A \left(T_A\right)\;{\cal D}^A_L \frac{1}{R_A} + {\cal O}\left(c^{-2}\right),
\nonumber\\
\label{global_metric_potentials_2}
\end{eqnarray}

\noindent
where in (\ref{global_metric_potentials_1}) the sum runs over all bodies of the $N$-body system, 
$R_A =\left|\ve{X}_A\right|$ is the spatial distance from the origin of local coordinate system
to some field point located outside the massive body, and 
$\displaystyle {\cal D}^A_L = \frac{\partial}{\partial X^{a_1}_A}\,...\,\frac{\partial}{\partial X_A^{a_l}}$.  
The local STF mass-multipoles $M_L^A$ in (\ref{global_metric_potentials_2})   
in Newtonian approximation are given by 
\begin{eqnarray}
M^A_L \left(T_A\right) &=& 
\int_{V_A} d^3 X_A\;\hat{X}^A_L\;\frac{T^{00}_A\left(T_A,\ve{X}_A\right)}{c^2} + {\cal O}\left(c^{-2}\right), 
\nonumber\\
\label{local_Newtonian_Mass_Multipole_A}
\end{eqnarray}

\noindent
where the integration runs over the volume $V_A$ of the massive body A under consideration,  
and where $T_A^{00}$, is the time-time-component of the energy-momentum tensor $T_A^{\mu\nu}$ of the isolated massive body A  
and expressed in the coordinates of the local reference system of that envisaged body.

In order to complete the transformation, also the partial derivatives in (\ref{global_metric_potentials_2}) have to be transformed, 
which follow from the coordinate transformations (\ref{coordinate_transformation_1}) and read explicitly; 
cf. Eqs.~(2.10) with virtue of Eqs.~(2.16) in \cite{DSX1} :  
\begin{eqnarray}
\frac{\partial}{\partial cT_A} &=& \frac{\partial}{\partial c t}
+ \frac{v_A^a\left(T_A\right)}{c} \frac{\partial}{\partial x^a} + {\cal O}\left(c^{-2}\right),
\label{coordinate_transformation_6}
\\
\nonumber\\
\frac{\partial}{\partial X_A^a} &=& \frac{\partial}{\partial x^a}
+ \frac{v_A^a\left(T_A\right)}{c} \frac{\partial}{\partial ct} + {\cal O}\left(c^{-2}\right). 
\label{coordinate_transformation_7}
\end{eqnarray}

\noindent
Let us note already here, that the second term in (\ref{coordinate_transformation_6}) and  
(\ref{coordinate_transformation_7}) yield terms of the order ${\cal O}\left(c^{-4}\right)$ in the global metric,  
hence these terms do finally not appear in Eq.~(\ref{global_metric_potentials_A}).  
Furthermore, we note that from (\ref{coordinate_transformation_1}) follows  
the relation \cite{IAU_Resolution1,Kopeikin_Efroimsky_Kaplan,DSX1,DSX2}:  
\begin{eqnarray}
R_A &=& \left| \ve{x} - \ve{x}_A\left(t\right)\right| + {\cal O} \left(c^{-2}\right), 
\label{coordinate_relation_5} 
\end{eqnarray}

\noindent
where according to (\ref{transformed_geodesic_equation_10}) the field-point $\ve{x}$ in (\ref{coordinate_relation_5})
will later be replaced by the photons light-trajectory.
The coordinate-time in the global and local systems are related via \cite{IAU_Resolution1,Kopeikin_Efroimsky_Kaplan,DSX1,DSX2}:   
\begin{eqnarray}
T_A &=& t + {\cal O} \left(c^{-2}\right).  
\label{coordinate_relation_10}
\end{eqnarray}

\noindent
Actually, a constant $b^0_A$ could be added on the right-hand side in (\ref{coordinate_relation_10}), 
which would indicate different initial times of the clocks in the global and local systems, cf. Eq.~(4) in \cite{Zschocke_Soffel},  
but has been omitted in favor of simpler notation and could formally be added at any stage of the calculations; about the general  
problem of clock-synchronization in the gravitational field of the Solar system we refer to \cite{Synchronization1}. 
From (\ref{coordinate_relation_10}) we conclude  
\begin{eqnarray}
M_L^A\left(T_A\right) &=& M_L^A\left(t\right) + {\cal O}\left(c^{-2}\right), 
\label{coordinate_relation_15}
\end{eqnarray}

\noindent
where the neglected terms in (\ref{coordinate_relation_15}) are beyond 1PN approximation for lightrays.   
By inserting (\ref{coordinate_transformation_6}) - (\ref{coordinate_relation_15}) into 
(\ref{global_metric_potentials_1}) - (\ref{global_metric_potentials_2}) we arrive at 
the global gravitational potential in terms of local mass-multipoles $M_L^A$:  
\begin{eqnarray}
w_A\left(t,\ve{x}\right) &=& G \sum\limits_{l = 0}^{\infty} \frac{\left(-1\right)^l}{l!}  
M_L^A \left(t\right)\;\partial_L\,\frac{1}{r_A\left(t\right)} + {\cal O}\left(c^{-2}\right),
\nonumber\\
\label{global_metric_potentials_A}
\end{eqnarray}

\noindent
where $r_A\left(t\right) = \left|\ve{x} - \ve{x}_A\left(t\right)\right|$, and  
$\displaystyle \partial_L=\frac{\partial}{\partial x^{i_1}}\,...\,\frac{\partial}{\partial x^{i_l}}$  
are partial derivatives in the global system.   
In summary of this section, the metric perturbation in the near-zone 
of the Solar system and expressed in terms of local multipoles is given by: 
\begin{eqnarray}
h^{\left(2\right)}_{00}\left(t,\ve{x}\right) &=& \sum\limits_{A=1}^{N} h^{\left(2\right) A}_{00}\left(t,\ve{x}\right),  
\label{global_metric_perturbation_A}
\\
\nonumber\\
h^{\left(2\right) A}_{00}\left(t,\ve{x}\right) &=& \frac{2\,G}{c^2}  
\sum\limits_{l = 0}^{\infty} \frac{\left(-1\right)^l}{l!}
M_L^A \left(t\right)\;\partial_L\,\frac{1}{r_A\left(t\right)}\,, 
\label{global_metric_perturbation_B}
\\
\nonumber\\
h^{\left(2\right)}_{ij}\left(t,\ve{x}\right) &=& \delta_{ij}\,h^{\left(2\right)}_{00}\left(t,\ve{x}\right),
\label{global_metric_perturbation_C}
\end{eqnarray}

\noindent
where the sum in (\ref{global_metric_perturbation_A}) runs over all massive bodies of the Solar system and the 
metric perturbation caused by one individual body is given by (\ref{global_metric_perturbation_B}).   
The metric perturbation in (\ref{global_metric_perturbation_A}) - (\ref{global_metric_perturbation_C})  
has to be implemented into the geodesic equation in (\ref{geodesic_equation_5}). 

At this stage let us underline again that an implementation of the infinite series expansion (\ref{worldline_introduction}) into 
(\ref{global_metric_perturbation_B}) via $r_A\left(t\right) = \left|\ve{x} - \ve{x}_A\left(t\right)\right|$, would more explicitly  
elucidate the fact that an arbitrary worldline of the body, $\ve{x}_A\left(t\right)$, implicitly generates terms in the metric  
tensor (\ref{global_metric_perturbation_B}) which are proportional to the velocity and acceleration of the body. However, such  
terms would be proportional either to $v_A\left(t-t_A\right)$ or $a_A\left(t-t_A\right)^2$, but neither to $v_A/c$ nor $a_A/c$,   
hence they would not be beyond 1PN approximation for the lightrays. From this consideration it becomes obvious that 
an arbitrary worldline $\ve{x}_A\left(t\right)$ implies a summation over all terms in the series expansion (\ref{worldline_introduction}) 
and, therefore, a solution of the geodesic equation in terms of arbitrary worldlines $\ve{x}_A\left(t\right)$ is much preferable 
compared to a solution in terms of approximative worldlines (\ref{worldline_introduction}).

\section{Transformation of geodesic equation}\label{Section3}

According to Eqs.~(\ref{transformed_geodesic_equation_10}) - (\ref{transformed_geodesic_equation_15}), the geodesic  
equation in (\ref{geodesic_equation_5}) has to be integrated along the unperturbed light-trajectory (\ref{Unperturbed_light_ray_10}).  
In view of this fact, it is meaningful to express the geodesic equation, i.e. the metric tensor and the derivatives, in terms of 
new parameters which characterize the unperturbed light-trajectory from the very beginning of the integration procedure.  
In this respect, the investigations in \cite{KopeikinSchaefer1999_Gwinn_Eubanks,KopeikinKorobkovPolnarev2006,KopeikinKorobkov2005}  
have recovered the remarkable efficiency of the following two independent variables $\tau$ and $\ve{\xi}$:  
\begin{eqnarray}
c\,\tau &=& \ve{\sigma}\cdot\ve{x}_{\rm N}\left(t\right)\,,\quad  
c\,\tau_0 = \ve{\sigma}\cdot\ve{x}_{\rm N}\left(t_0\right),  
\label{variable_1}
\\
\nonumber\\
\xi^i &=& P^i_j\,x_{\rm N}^j\left(t\right),
\label{variable_2}
\end{eqnarray}

\noindent
where $P^i_j = P_{ij} = P^{ij}$ is the operator of projection onto the plane perpendicular to the vector $\ve{\sigma}$,  
\begin{eqnarray}
P^{ij} &=& \delta_{i j} - \sigma^i\,\sigma^j\,. 
\label{variable_3}
\end{eqnarray}

\noindent
The three-vector 
$\ve{\xi} = \ve{\sigma}\times\left(\ve{x}_{\rm N}\left(t\right)\times\ve{\sigma}\right)=\ve{\sigma}\times\left(\ve{x}_0\times\ve{\sigma}\right)$ 
in (\ref{variable_2}) is the impact vector of the unperturbed lightray, see also Eq.~(\ref{notation_2}).  
Especially, $\ve{\xi}$ is time-independent and directed from the origin of global coordinate system to the point of closest
approach of the unperturbed light-trajectory; its absolute value is denoted by $d = \left|\ve{\xi}\right|$.

\begin{figure}[!ht]
\begin{center}
\includegraphics[scale=0.155]{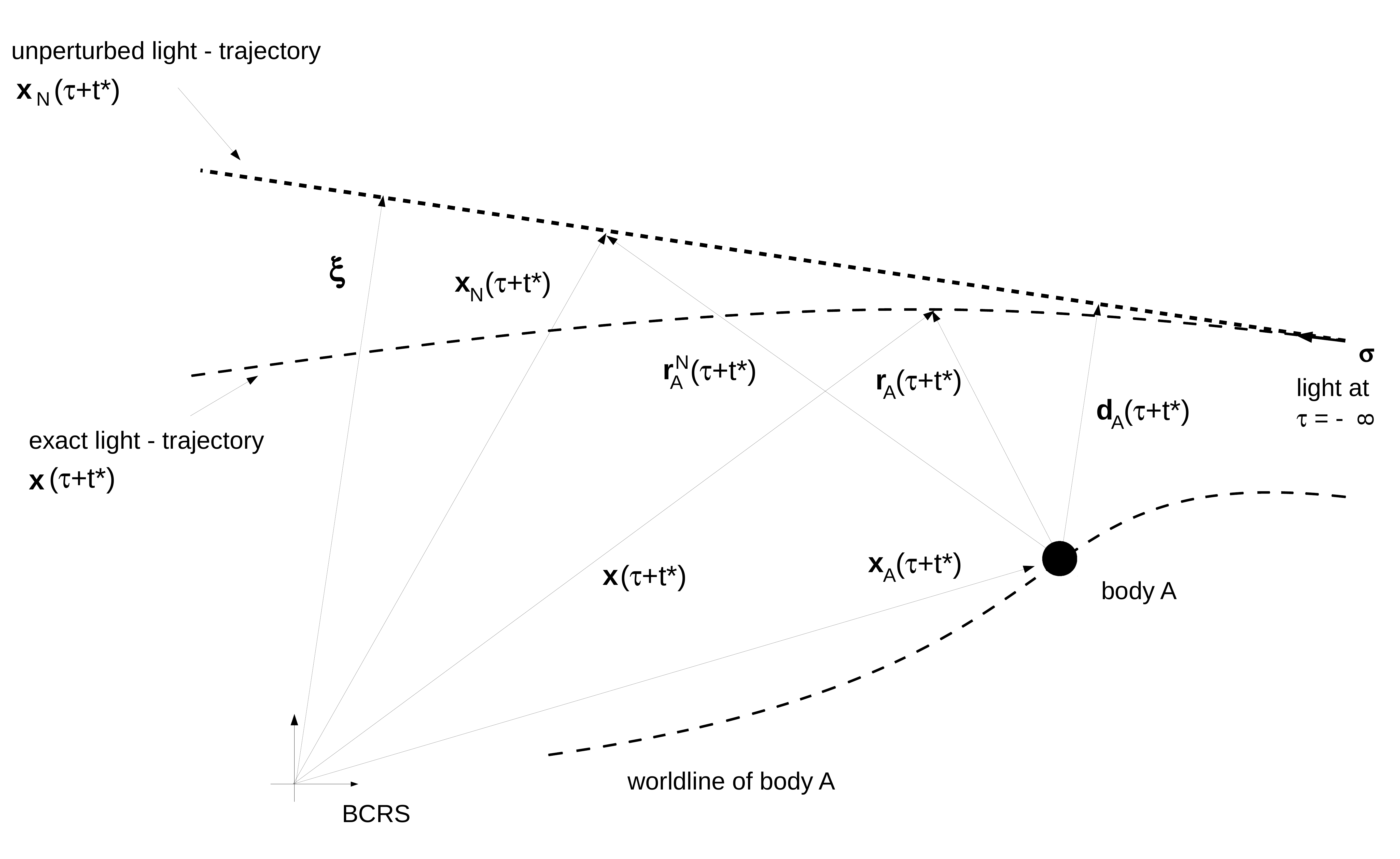}
\end{center}
\caption{A geometrical representation of the light-trajectory through the Solar system in terms of the
new variables $\ve{\xi}$ and $\tau$. The impact vector $\ve{\xi}$ is defined by Eq.~(\ref{variable_2})
and points from the origin of global system to the point of closest approach of the
unperturbed lightray to that origin, and is time-independent. The
impact vector $\ve{d}_A\left(\tau+t^{\ast}\right)$ is defined by Eq.~(\ref{First_Integration_22})
and points from the origin of local system of body A towards the point of closest approach
of unperturbed lightray to that origin, and is time-dependent due to the motion of the body.
Furthermore, $\ve{x} \left(\tau+t^{\ast}\right)$ and $\ve{x}_{\rm N}\left(\tau+t^{\ast}\right)$ are the global spatial positions of the photon
of the exact light-trajectory and unperturbed light-trajectory, respectively. The
worldline of massive body A in the global system is given by $\ve{x}_A\left(\tau+t^{\ast}\right)$, and
$\ve{r}_A\left(\tau+t^{\ast}\right)$ points from the origin of local system towards the exact position of the photon, while
$\ve{r}^{\rm N}_A\left(\tau+t^{\ast}\right)$ points from the origin of local system towards the unperturbed lightray.}
\label{Diagram1}
\end{figure}

While some detailed explanations and geometrical elucidations can be found in \cite{KopeikinSchaefer1999_Gwinn_Eubanks},  
two comments should be in order about these new variables.  

(i) First, one can easily recognize that (\ref{variable_1}) can also be written in the form 
$c \tau = c\left(t - t^{\ast}\right)$ and $c \tau_0 = c \left(t_0 - t^{\ast}\right)$, where  
\begin{eqnarray} 
t^{\ast} &=& t_0 - \frac{\ve{\sigma}\cdot\ve{x}_0}{c}\,,
\label{time_of_closest_approach}  
\end{eqnarray} 

\noindent 
is the time of closest approach of unperturbed lightray to the origin of the global 
coordinate system; note that (\ref{time_of_closest_approach}) differs from (\ref{time_of_closest_approach_t_0}) which is the 
time of closest approach of the lightray to the origin of the local coordinate system of some massive body A.  
With the aid of these new variables $\ve{\xi}$ and $\tau$, the mixed initial-boundary conditions 
(\ref{Initial_Boundary_Condition_1}) and (\ref{Initial_Boundary_Condition_2}) take the form  
\begin{eqnarray}
\ve{x}_0 &=& \ve{x}\left(\tau_0 + t^{\ast}\right),
\label{Transformed_Initial_Boundary_Condition_1}
\\
\nonumber\\
\ve{\sigma} &=& \lim_{\tau \rightarrow - \infty}\, \frac{\dot{\ve{x}}\left(\tau + t^{\ast}\right)}{c}\,, 
\label{Transformed_Initial_Boundary_Condition_2}
\end{eqnarray}

\noindent
where a dot means derivative with respect to variable $\tau$. In terms of the new variables the interpretation of these initial-boundary   
conditions remains the same: the first condition (\ref{Transformed_Initial_Boundary_Condition_1}) defines the spatial coordinates  
of the photon at the moment of emission of light, while the second condition (\ref{Transformed_Initial_Boundary_Condition_2})  
defines the unit-direction ($\ve{\sigma}\cdot\ve{\sigma} = 1$) at infinite past and infinite distance from the origin of 
global coordinate system, that means at the so-called past null infinity.  

(ii) Second, it is important to mention that with the aid  
of the new variable (\ref{variable_1}) and (\ref{variable_2}), the unperturbed lightray 
in (\ref{Unperturbed_light_ray_10}) transforms as follows 
\footnote{The unperturbed lightray in Eq.~(\ref{Unperturbed_light_ray_10}) for $t=\tau+t^{\ast}$ reads 
$\ve{x}_{\rm N}\left(\tau+t^{\ast}\right)=\ve{x}_0 + c\left(\tau+t^{\ast}-t_0\right)\ve{\sigma}$. By means of  
the above mentioned relation $\displaystyle t^{\ast} = t_0 - \frac{\ve{\sigma}\cdot\ve{x}_0}{c}$ one obtains 
$\ve{x}_{\rm N}\left(\tau+t^{\ast}\right)=\ve{x}_0 - \ve{\sigma} \left(\ve{\sigma}\cdot\ve{x}_0\right) + c\,\tau\,\ve{\sigma}$.  
Using the definition (\ref{variable_2}) one obtains $\ve{x}_{\rm N}\left(\tau+t^{\ast}\right)=\ve{\xi} + c\,\tau\,\ve{\sigma}$, which is 
just relation (\ref{variable_5}).}: 
\begin{eqnarray}
\ve{x}_{\rm N}\left(\tau+t^{\ast}\right) &=& \ve{\xi} + c\,\tau\,\ve{\sigma}\,.  
\label{variable_5}
\end{eqnarray}

\noindent
In these new variables, the vector from the arbitrarily moving body and the light-trajectory in (\ref{notation_5a}) 
transforms as follows:
\begin{eqnarray}
\ve{r}_A\left(\tau + t^{\ast}\right) &=& \ve{x}\left(\tau+t^{\ast}\right) - \ve{x}_A\left(\tau + t^{\ast}\right), 
\label{notation_14}
\end{eqnarray}

\noindent
with the absolute value $r_A\left(\tau + t^{\ast}\right)=\left|\ve{r}_A\left(\tau + t^{\ast}\right)\right|$, 
while the distance between the unperturbed lightray and the arbitrarily moving body in (\ref{notation_5b}) reads now:  
\begin{eqnarray}
\ve{r}^{\rm N}_A\left(\tau + t^{\ast}\right) &=& \ve{x}_{\rm N}\left(\tau + t^{\ast}\right) - \ve{x}_A\left(\tau + t^{\ast}\right) 
\nonumber\\
\nonumber\\
&=& \ve{\xi} + c\,\tau\,\ve{\sigma} - \ve{x}_A\left(\tau + t^{\ast}\right),   
\label{notation_15}
\end{eqnarray}

\noindent 
with the absolute value $r^{\rm N}_A\left(\tau + t^{\ast}\right)=\left|\ve{r}^{\rm N}_A\left(\tau + t^{\ast}\right)\right|$, 
and we note $\ve{r}_A\left(\tau + t^{\ast}\right) = \ve{r}^{\rm N}_A\left(\tau + t^{\ast}\right) + {\cal O}\left(c^{-2}\right)$.  
The impact parameter in (\ref{notation_6}) for arbitrarily moving bodies in these new variables reads:  
\begin{eqnarray}
\ve{d}_A \left(\tau + t^{\ast}\right) &=& \ve{\sigma} \times \left(\ve{r}^{\rm N}_A\left(\tau + t^{\ast}\right) \times \ve{\sigma}\right),
\label{First_Integration_22}
\end{eqnarray}

\noindent 
with the absolute value $d_A\left(\tau + t^{\ast}\right)=\left|\ve{d}_A\left(\tau + t^{\ast}\right)\right|$. 
For an illustration of the expressions in Eqs.~(\ref{variable_2}) and (\ref{variable_5}) - (\ref{First_Integration_22}) see Fig.~\ref{Diagram1}.  

Furthermore, it has been outlined in \cite{KopeikinSchaefer1999_Gwinn_Eubanks,KopeikinMashhoon2002} that  
by means of the new variables (\ref{variable_1}) and (\ref{variable_2}), the following relation is valid for 
a smooth function $F\left(t,\ve{x}\right)$;    
cf. Eq.~(33) in \cite{KopeikinSchaefer1999_Gwinn_Eubanks} or Eq.~(C4) in \cite{KopeikinMashhoon2002}: 
\begin{eqnarray}
&& \hspace{-1.0cm} \left(\frac{\partial}{\partial x^i} + \sigma^i\,\frac{\partial}{\partial ct}\right)\,F\left(t,\ve{x}\right) 
\Bigg|_{\ve{x}=\ve{x}_{\rm N}\mbox{\normalsize $\left(t\right)$}} 
\nonumber\\
&& \hspace{-1.0cm} = \left(P^{ij}\,\frac{\partial}{\partial \xi^j} + \sigma^i\,\frac{\partial}{\partial c \tau}\right) 
F\left(\tau + t^{\ast},\ve{\xi} + c\,\tau\,\ve{\sigma}\right).
\label{Relation_A} 
\end{eqnarray}

\noindent
It is important to realize that on the left-hand side in (\ref{Relation_A}) one has first to differentiate with respect to the  
fieldpoint $\ve{x}$ and global coordinate-time $t$ and afterwards one has to substitute the unperturbed lightray  
$\ve{x}_{\rm N}\left(t\right) = \ve{x}_0 + c \left(t-t_0\right) \ve{\sigma}$,  
while on the right-hand side in (\ref{Relation_A}) one has first to substitute $\tau + t^{\ast}$ and
$\ve{x}_{\rm N}\left(\tau + t^{\ast}\right) = \ve{\xi} + c\,\tau\,\ve{\sigma}$ and afterwards to perform the differentiation with respect  
to $\ve{\xi}$ and $\tau$.  

From now on, the smooth function $F\left(t,\ve{x}\right)$ in relation (\ref{Relation_A}) is considered to be one of the components of 
the metric perturbation  
$h_{\alpha \beta}^{(2)}\left(t,\ve{x}\right)$. Then, the derivatives with respect to variable $c t$ on the left-hand side of 
relation (\ref{Relation_A}) yield only terms of higher-order beyond 1PN approximation,  
\begin{eqnarray}
\frac{\partial h_{\alpha \beta}^{(2)}\left(t,\ve{x}\right)}{\partial ct} \Bigg|_{\ve{x}=\ve{x}_{\rm N}\mbox{\normalsize $\left(t\right)$}}  
&=& {\cal O}\left(c^{-3}\right),   
\label{Transformation_Derivative_2A} 
\end{eqnarray}
 
\noindent
because they are proportional to either $\dot{M}^A_L/c$ or $\ve{v}_A/c$; for actually the same reason there is no time-derivative 
in the geodesic equation either, see (\ref{geodesic_equation_1}) or (\ref{geodesic_equation_5}).
However, one has to keep the differentiation with respect to variable $c \tau$ in the right-hand side of
relation (\ref{Relation_A}), because that derivative does not only act on the
multipoles $M_L^A\left(\tau + t^{\ast}\right)$ and spatial coordinates of the massive bodies $\ve{x}_A\left(\tau + t^{\ast}\right)$,
but also on the unperturbed lightray $\ve{\xi} + c \tau \ve{\sigma}$. 
Therefore, in 1PN approximation the relation (\ref{Relation_A}) simplifies as follows:  
\begin{eqnarray}
&& \frac{\partial h_{\alpha \beta}^{(2)}\left(t,\ve{x}\right)}{\partial x^i}
\Bigg|_{\ve{x}=\ve{x}_{\rm N}\mbox{\normalsize $\left(t\right)$}} 
\nonumber\\
&& \hspace{-0.3cm} = \left(P^{ij} \frac{\partial}{\partial \xi^j} + \sigma^i \frac{\partial}{\partial c \tau}\right) 
h_{\alpha \beta}^{(2)}\left(\tau + t^{\ast},\ve{\xi} + c \tau \ve{\sigma}\right) + {\cal O}\left(c^{-3}\right).
\nonumber\\
\label{Transformation_Derivative_2}
\end{eqnarray}

\noindent
If the derivative with respect to variable $c \tau$ in (\ref{Transformation_Derivative_2}) acts 
on the multipoles or spatial coordinates of the massive bodies, then terms will be generated which are beyond 1PN approximation, 
namely terms proportional to either $\dot{M}^A_L/c$ or $\ve{v}_A/c$, respectively, which, however, can easily be identified.  

By means of relation (\ref{Transformation_Derivative_2}), the geodesic equation in 1PN approximation 
in (\ref{geodesic_equation_5}) transforms as follows:  
\begin{eqnarray}
\frac{\ddot{x}^i \left(\tau+t^{\ast}\right)}{c^2} &=&  
+ \frac{1}{2}\,P^{ij}\,\frac{\partial}{\partial \xi^j}\,h_{00}^{(2)} 
- \frac{1}{2}\,\sigma^i\,\frac{\partial}{\partial c \tau}\,h_{00}^{(2)} 
\nonumber\\
\nonumber\\
&& + \frac{1}{2}\,\sigma^k \sigma^l P^{ij} \frac{\partial}{\partial \xi^j} h_{kl}^{(2)} 
+ \frac{1}{2} \sigma^i \sigma^j \sigma^k \frac{\partial}{\partial c \tau} h_{jk}^{(2)} 
\nonumber\\
&& - \sigma^j\,\frac{\partial}{\partial c \tau}\,h_{ij}^{(2)}  
+ {\cal O}\left(c^{-3}\right)\,,  
\label{transformed_geodesic_equation_A}
\end{eqnarray}

\noindent
where the double-dot on the left-hand side in (\ref{transformed_geodesic_equation_A}) means twice of the total derivative   
with respect to the new variable $\tau$. 
By taking into account (\ref{global_metric_perturbation_C}), the geodesic equation further simplifies:  
\begin{eqnarray}
\frac{\ddot{x}^i \left(\tau+t^{\ast}\right)}{c^2} &=&
P^{ij}\,\frac{\partial}{\partial \xi^j}\,h_{00}^{(2)}
- \sigma^i\,\frac{\partial}{\partial c \tau}\,h_{00}^{(2)}
+ {\cal O}\left(c^{-3}\right). 
\nonumber\\
\label{transformed_geodesic_equation_B}
\end{eqnarray}

\noindent
As next step, the metric perturbations in (\ref{global_metric_perturbation_A}) - (\ref{global_metric_perturbation_C}) have to be transformed  
in terms of these new variables $\ve{\xi}$ and $\tau$. Since the metric perturbations in (\ref{global_metric_perturbation_B}) contain spatial  
derivatives, $\partial_L\,r^{-1}_A\left(t\right)$, we will have to transform these differential operators in terms of these new variables.  
For that one might want to use relation (\ref{Relation_A}), which is valid for any smooth function, 
but a possible time-derivative on the left-hand side of (\ref{Relation_A})  
generates only terms beyond 1PN approximation, 
\begin{eqnarray}
\frac{\partial}{\partial c t}\,\frac{1}{r_A\left(t\right)}\Bigg|_{\ve{x}=\ve{x}_{\rm N}\mbox{\normalsize $\left(t\right)$}} 
&=& {\cal O}\left(c^{-1}\right). 
\label{Beyon_1PN_A}
\end{eqnarray}

\noindent
Therefore, like in (\ref{Transformation_Derivative_2}), we may use the simpler relation, 
\begin{eqnarray}
\frac{\partial}{\partial x^i}\,\frac{1}{r_A\left(t\right)}\Bigg|_{\ve{x}=\ve{x}_{\rm N}\mbox{\normalsize $\left(t\right)$}} &=& 
\left(P^{ij}\,\frac{\partial}{\partial \xi^j} + \sigma^i\,\frac{\partial}{\partial c \tau}\right) \frac{1}{r^{\rm N}_A\left(\tau + t^{\ast}\right)} 
\nonumber\\
&& + {\cal O}\left(c^{-1}\right),  
\label{Beyon_1PN_B}
\end{eqnarray}

\noindent
where we have taken into account that the derivative with respect to $c \tau$ in the right-hand side of (\ref{Beyon_1PN_B}) 
must be kept because of; cf. relation (\ref{dipole_12}):  
\begin{eqnarray}
\frac{\partial}{\partial c \tau} \frac{1}{r^{\rm N}_A\left(\tau + t^{\ast}\right)} &=& 
- \frac{\ve{\sigma} \cdot \ve{r}_A^{\rm N}\left(\tau+t^{\ast}\right)}{\left(r_A^{\rm N}\left(\tau+t^{\ast}\right)\right)^3} 
+ {\cal O}\left(\frac{v_A}{c}\right).
\nonumber\\
\label{Beyon_1PN_C}
\end{eqnarray}

\noindent 
The outcome of (\ref{Beyon_1PN_B}) and (\ref{Beyon_1PN_C}) is, that the metric perturbation in (\ref{global_metric_perturbation_B}) for one massive body A and  
in terms of these new variables $\ve{\xi}$ and $\tau$ is given by:  
\begin{eqnarray}
h^{\left(2\right)}_{00}\left(\tau,\ve{\xi}\right) &=& \sum\limits_{A=1}^{N} h^{\left(2\right) A}_{00}\left(\tau,\ve{\xi}\right),
\label{Transformation_Derivative_4}
\\
\nonumber\\
h^{\left(2\right) A}_{00}\left(\tau,\ve{\xi}\right) &=& \frac{2 G}{c^2} 
\sum\limits_{l = 0}^{\infty} \frac{\left(-1\right)^l}{l!} M_L^A\left(\tau+t^{\ast}\right) 
\partial_L \frac{1}{r^{\rm N}_A\left(\tau+t^{\ast}\right)},  
\nonumber\\
\label{Transformation_Derivative_5}
\end{eqnarray}

\noindent
where, by means of binomial theorem, the spatial derivatives in (\ref{Transformation_Derivative_5}) in terms of new variables can be 
written in the following form (cf. Eq.~(24) in \cite{Kopeikin1997}):  
\begin{eqnarray}
\partial_L &=& \sum\limits_{p=0}^{l} \frac{l!}{\left(l-p\right)!\;p!}\;
\sigma^{i_1}\,...\,\sigma^{i_p}\;
P^{i_{p+1}\,j_{p+1}}\;...\;P^{i_l\,j_l}\;
\nonumber\\
&& \times \frac{\partial}{\partial \xi^{j_{p+1}}}\;...\;
\frac{\partial}{\partial \xi^{j_{l}}}\;\left(\frac{\partial}{\partial c\tau}\right)^p\,.
\label{Transformation_Derivative_3}
\end{eqnarray}

\noindent
Here, we recall that $M^A_L$ are STF multipoles, therefore for a smooth function $F$ we have  
$M^A_L\,\partial_L\,F = M^A_L\,\underset{i_1 ... i_l}{\rm STF}\,\partial_L\,F$, 
so that the expression in (\ref{Transformation_Derivative_3})  
must be interpreted in combination with $M^A_L$.  
The insertion of metric perturbation (\ref{Transformation_Derivative_4}) - (\ref{Transformation_Derivative_5}) 
into the geodesic equation (\ref{transformed_geodesic_equation_B}) finally yields the geodesic equation for lightrays   
which propagate in the gravitational field of one arbitrarily moving body A:  
\begin{eqnarray}
&& \hspace{-0.5cm} \frac{\ddot{x}^{i\,A} \left(\tau+t^{\ast}\right)}{c^2} 
\nonumber\\
&=& + \frac{2\,G}{c^2}\,  
P^{ij}\,\frac{\partial}{\partial \xi^j}\,
\sum\limits_{l = 0}^{\infty} \frac{\left(-1\right)^l}{l!}\;M_L^A\left(\tau+t^{\ast}\right) 
\partial_L\;\frac{1}{r^{\rm N}_A\left(\tau+t^{\ast}\right)} 
\nonumber\\
\nonumber\\
&& - \frac{2\,G}{c^2}\,\sigma^i\,\frac{\partial}{\partial c \tau}\, 
\sum\limits_{l = 0}^{\infty} \frac{\left(-1\right)^l}{l!}\;M_L^A\left(\tau+t^{\ast}\right) 
\partial_L\;\frac{1}{r^{\rm N}_A\left(\tau+t^{\ast}\right)}  
\nonumber\\
&& + {\cal O}\left(c^{-3}\right)\,,
\label{transformed_geodesic_equation_C}
\end{eqnarray}

\noindent
where the derivative operator $\partial_L$ is given by (\ref{Transformation_Derivative_3}).  
Eq.~(\ref{transformed_geodesic_equation_C}) completes the transformation of geodesic equation in 1PN approximation 
and for the case of one arbitrarily moving massive body having arbitrary shape and structure. 
Due to the linearity of post-Newtonian equations, the case of $N$ arbitrarily moving bodies is easily obtained by  
a summation over all massive bodies $A=1,2,...,N$.  
 
In the limit of: (i) one massive body at rest, (ii) time-independent multipoles, and (iii) assuming that the 
center-of-mass is located at the 
origin of the global coordinate-system, the geodesic equation (\ref{transformed_geodesic_equation_C})  
agrees with the geodesic equation given in \cite{Kopeikin1997}; recall that there are no spin-multipole terms 
in (\ref{transformed_geodesic_equation_C}) because they contribute to the order ${\cal O}\left(c^{-3}\right)$.

\section{First integration of geodesic equation}\label{First_Integration}  

The first integral determines the coordinate-velocity of the photon and, due to the linearity of geodesic equation in 1PN approximation, 
can be written as follows; cf. Eq.~(\ref{light_trajectory_A}):  
\begin{eqnarray}
\dot{\ve{x}}_{\rm 1PN} \left(\tau + t^{\ast}\right) &=&
c\,\ve{\sigma} + \sum\limits_{A=1}^N \Delta \dot{\ve{x}}_{\rm 1PN}^A\left(\tau + t^{\ast}\right),
\label{first_integral_geodesic_equation}
\end{eqnarray}

\noindent
where the contribution of one body A reads:  
\begin{eqnarray}
\frac{\Delta \dot{\ve{x}}_{\rm 1PN}^A\left(\tau + t^{\ast}\right)}{c} &=& \int\limits_{-\infty}^{\tau}\,d c\tau^{\prime}\,
\frac{\Delta \ddot{\ve{x}}_{\rm 1PN}^A\left(\tau^{\prime} + t^{\ast}\right)}{c^2}\,, 
\label{Additional_Equation_A}
\end{eqnarray}

\noindent
where the integrand is given by Eq.~(\ref{transformed_geodesic_equation_C}).  
Accordingly, one obtains: 
\begin{eqnarray}
\frac{\Delta \dot{x}_{\rm 1PN}^{i\,A} \left(\tau+t^{\ast}\right)}{c} &=& + \frac{2 G}{c^2} P^{ij}\,\frac{\partial}{\partial \xi^j}  
\sum\limits_{l = 0}^{\infty} \frac{\left(-1\right)^l}{l!} {\cal I}_A\left(\tau+t^{\ast},\ve{\xi}\right)  
\nonumber\\ 
&& \hspace{-0.3cm} - \frac{2 G}{c^2} \sigma^i  
\sum\limits_{l = 0}^{\infty} \frac{\left(-1\right)^l}{l!} {\cal I}_B\left(\tau+t^{\ast},\ve{\xi}\right).  
\label{First_Integration_5} 
\end{eqnarray}

\noindent
The integrals in (\ref{First_Integration_5}) are defined by (the arguments of the integrals are omitted)  
\begin{eqnarray}
{\cal I}_A &=& \int\limits_{-\infty}^{\tau}\,d c\tau^{\prime}\, 
M_L^A\left(\tau^{\prime}+t^{\ast}\right) \partial^{\prime}_L\;
\frac{1}{r^{\rm N}_A\left(\tau^{\prime}+t^{\ast}\right)}\,, 
\label{Integration_A}
\\
\nonumber\\
{\cal I}_B &=& 
\int\limits_{-\infty}^{\tau}\,d c\tau^{\prime}\,\frac{\partial}{\partial c \tau^{\prime}}\,
M_L^A\left(\tau^{\prime}+t^{\ast}\right) 
\partial^{\prime}_L\;
\frac{1}{r^{\rm N}_A\left(\tau^{\prime}+t^{\ast}\right)}\,, 
\nonumber\\
\label{Integration_B}
\end{eqnarray}

\noindent
where the differential operator $\partial^{\prime}_L$ in (\ref{Integration_A}) and (\ref{Integration_B})
is given by (cf. Eq.~(\ref{Transformation_Derivative_3}):  
\begin{eqnarray}
\partial^{\prime}_L &=& \sum\limits_{p=0}^{l} \frac{l!}{\left(l-p\right)!\;p!}\;
\sigma^{i_1}\,...\,\sigma^{i_p}\;
P^{i_{p+1}\,j_{p+1}}\;...\;P^{i_l\,j_l}\;
\nonumber\\ 
&& \times \frac{\partial}{\partial \xi^{j_{p+1}}}\;...\;
\frac{\partial}{\partial \xi^{j_{l}}}\;\left(\frac{\partial}{\partial c\tau^{\prime}}\right)^p\,. 
\label{Transformation_Derivative}
\end{eqnarray}

\noindent
Here, we recall again that $M^A_L$ are STF multipoles, 
therefore for a smooth function $F$ we have $M_L^A\,\partial^{\prime}_L\,F = M_L^A\,\underset{i_1 ... i_l}{\rm STF}\,\partial^{\prime}_L\,F$, 
so that relation (\ref{Transformation_Derivative}) must be interpreted in combination with $M_L^A$.   
In (\ref{First_Integration_5}) we have taken into account that $dt = d \tau$ for the total
differentials because $t^{\ast}={\rm const}$ is a constant for each individual lightray.
Also the following integration rule ($\tau$ and $\ve{\xi}$ are independent variables)
for indefinite integrals along the unperturbed lightray has been used; cf. Eq.~(4.10) in \cite{KopeikinKorobkovPolnarev2006}:
\begin{eqnarray}
\int\limits_{-\infty}^{\tau} d c\tau^{\prime}\;\frac{\partial}{\partial \xi^i}\,f\left(\tau^{\prime},\ve{\xi}\right) &=&
\frac{\partial}{\partial \xi^i} \int\limits_{-\infty}^{\tau} d c\tau^{\prime}\;f\left(\tau^{\prime},\ve{\xi}\right).
\label{Integration_Rule1}
\end{eqnarray}

\noindent
The integral in (\ref{Integration_A}) runs over the unknown worldline $\ve{x}_A\left(t\right)$ of the massive body A and, 
therefore, can only be integrated by parts. Such strategy intrinsically inherits to demonstrate that the non-integrated terms of the integration procedure  
involve terms which are beyond 1PN approximation, that means it elaborates on the fact that the non-integrated terms imply an additional factor $c^{-1}$.  
In this way, the integral ${\cal I}_A$ is determined by Eqs.~(\ref{Appendix_First_Integration_10}) - (\ref{Integral_A_1}) in appendix \ref{Appendix_Integral_A},  
while the integral ${\cal I}_B$ can immediately be calculated without integration by parts:  
\begin{eqnarray}
{\cal I}_B\left(\tau+t^{\ast},\ve{\xi}\right) &=& M_L^A\left(\tau+t^{\ast}\right) \partial_L\;
\frac{1}{r^{\rm N}_A\left(\tau+t^{\ast}\right)}\,. 
\label{Integration_C}
\end{eqnarray}

\noindent
Altogether one obtains for the first integral of geodesic equation (\ref{transformed_geodesic_equation_C}): 
\begin{widetext}
\begin{eqnarray}
&& \frac{\Delta \dot{\ve{x}}_{\rm 1PN}^A \left(\tau+t^{\ast}\right)}{c} 
\nonumber\\ 
&=& - \frac{2\,G}{c^2}\,
\sum\limits_{l=1}^{\infty} \sum\limits_{p=1}^{l}  
\frac{\left(-1\right)^l}{\left(l-p\right)!\,p!}\;M^A_L \left(\tau+t^{\ast}\right)
\sigma^{i_1}\,...\,\sigma^{i_p}\,
P^{i_{p+1}\,j_{p+1}}\,...\,P^{i_l\,j_l}\;
\frac{\partial}{\partial \xi^{j_{p+1}}}\,...\,\frac{\partial}{\partial \xi^{j_{l}}}\,
\left(\frac{\partial}{\partial c\tau}\right)^{p-1}
\frac{\ve{d}_A\left(\tau+t^{\ast}\right)}
{\left(r^{\rm N}_A\left(\tau+t^{\ast}\right)\right)^3}\;
\nonumber\\
\nonumber\\
\nonumber\\
&& -\,\frac{2\,G}{c^2}\,
\sum\limits_{l=0}^{\infty} \frac{\left(-1\right)^l}{l!}\;M^A_L \left(\tau+t^{\ast}\right)  
P^{i_1j_1}\,...\,P^{i_l j_l}\;
\frac{\partial}{\partial \xi^{j_1}}\,...\,\frac{\partial}{\partial \xi^{j_l}}\,
\frac{\ve{d}_A\left(\tau+t^{\ast}\right)}
{r^{\rm N}_A\left(\tau + t^{\ast}\right) - \ve{\sigma}\cdot\ve{r}^{\rm N}_A\left(\tau + t^{\ast}\right)}
\frac{1}{r^{\rm N}_A\left(\tau + t^{\ast}\right)}  
\nonumber\\
\nonumber\\
\nonumber\\
&& -\,\frac{2\,G}{c^2}\,\ve{\sigma}\; 
\sum\limits_{l=0}^{\infty} 
\sum\limits_{p=0}^{l} \frac{\left(-1\right)^l}{\left(l-p\right)!\,p!}\; M^A_L \left(\tau+t^{\ast}\right)  
\sigma^{i_1}\,...\,\sigma^{i_p}\, 
P^{i_{p+1}\,j_{p+1}}\,...\,P^{i_l\,j_l}\;
\frac{\partial}{\partial \xi^{j_{p+1}}}\,...\,
\frac{\partial}{\partial \xi^{j_{l}}}\,\left(\frac{\partial}{\partial c\tau}\right)^p\,  
\frac{1}{r^{\rm N}_A\left(\tau+t^{\ast}\right)}\,,  
\label{First_Integration_21}
\end{eqnarray}
\end{widetext}

\noindent
where we recall the notation $M^A_L = M^A_{i_1 ... i_l}$. The expression in (\ref{First_Integration_21}) represents the solution  
for the first integration of geodesic equation in 1PN approximation in (\ref{transformed_geodesic_equation_C}),  
in the gravitational field of one arbitrarily moving body A to any order of its intrinsic mass-multipoles.  
It should be underlined, that after performing of the differentiations in (\ref{First_Integration_21})  
one can replace $\tau+t^{\ast}$ by the global coordinate-time $t$. Let us also note that the following relations have been used 
in order to obtain (\ref{First_Integration_21}):  
\begin{eqnarray}
P^{ij}\,\frac{\partial}{\partial \xi^j}\,
\frac{1}{r^{\rm N}_A\left(\tau+t^{\ast}\right)}
&=& - \frac{d_A^i\left(\tau + t^{\ast}\right)}{\left(r^{\rm N}_A\left(\tau + t^{\ast}\right)\right)^3}\,, 
\label{First_Integration_35}
\end{eqnarray}

\noindent
and 
\begin{eqnarray}
&& \hspace{-0.5cm} P^{ij}\,\frac{\partial}{\partial \xi^j}\,
\ln \left[r^{\rm N}_A\left(\tau+t^{\ast}\right) - \ve{\sigma}\cdot\ve{r}^{\rm N}_A\left(\tau+t^{\ast}\right)\right]
\nonumber\\ 
&& \hspace{1.0cm} = \frac{d_A^i\left(\tau + t^{\ast}\right)}{r^{\rm N}_A\left(\tau + t^{\ast}\right)}  
\frac{1}{r^{\rm N}_A\left(\tau + t^{\ast}\right) - \ve{\sigma}\cdot\ve{r}^{\rm N}_A\left(\tau + t^{\ast}\right)}\,,   
\nonumber\\
\label{First_Integration_20}
\end{eqnarray}

\noindent
and the relation $P^{ij} \left(\xi^j - x_A^j\left(\tau + t^{\ast}\right)\right)=d_A^i\left(\tau + t^{\ast}\right)$.

\section{Some special cases of first integration}\label{Special_Cases_1}  

Modern computer algebra systems allow for highly-efficient computation of partial differentiations which 
occur in the first integral (\ref{First_Integration_21}) of geodesic equation.  
Here, the first few terms of (\ref{First_Integration_21}) as instructive examples  
are considered and compared with known results in the literature, namely: arbitrarily moving 
monopoles, dipoles, quadrupoles, and the case of one massive body at rest with full mass-multipole structure.  
These examples can also serve as further elucidation about how the formula in (\ref{First_Integration_21}) works. 

\subsection{Monopoles in arbitrary motion} 

For the case of light propagation in the gravitational field of $N$ extended mass-monopoles in arbitrary motion 
we have to consider the term $l=0$ in (\ref{First_Integration_21}), which reads:  
\begin{eqnarray}
\frac{\Delta \dot{\ve{x}}_M\left(t\right)}{c} &=& - \frac{2\,G}{c^2} \sum\limits_{A=1}^{N}  
\frac{M_A}{r^{\rm N}_A\left(t\right)}  
\left(\frac{\ve{d}_A\left(t\right)}
{r^{\rm N}_A\left(t\right) - \ve{\sigma}\cdot\ve{r}^{\rm N}_A\left(t\right)} + \ve{\sigma}\right),   
\nonumber\\
\label{Comparison_10}
\end{eqnarray}
  
\noindent
where $\tau+t^{\ast}$ has finally been replaced by the global coordinate-time $t$. 
We recall that $\ve{r}^{\rm N}_A\left(t\right) = \ve{x}_{\rm N}\left(t\right) - \ve{x}_A\left(t\right)$, with   
$\ve{x}_{\rm N}\left(t\right)$ being the spatial position of the unperturbed light-signal and  
$\ve{x}_A\left(t\right)$ is the spatial position of the arbitrarily moving massive monopole.  

By taking the limit of monopoles at rest $\ve{x}_A = {\rm const}$ in (\ref{Comparison_10}), 
one may easily recognize an agreement of (\ref{Comparison_10}) with Eq.~(3.2.14) in \cite{Brumberg1991}  
and with Eq.~(28) in \cite{Klioner2003a}, where the mass-monopoles are 
displaced by some constant vector $\ve{x}_A$ from the origin of the global coordinate-system.  

In \cite{KopeikinSchaefer1999} the light-trajectory in the field of $N$ arbitrarily moving pointlike monopoles has been determined  
in 1PM approximation. The 1PM approximation is a weak-field approximation, that means the pointlike 
monopoles could even be in ultra-relativistic motion, while (\ref{Comparison_10}) is for extended monopoles but 
in 1PN approximation, which is a weak-field slow-motion approximation.  
By expansion of the 1PM solution (Eqs.~(32) and (34) in \cite{KopeikinSchaefer1999})  
in powers of $v_A/c$, one may show an agreement with our solution in (\ref{Comparison_10})  
up to terms of the order ${\cal O}\left(v_A/c\right)$.  

\subsection{Dipoles in arbitrary motion} 

Let us consider the dipole-term, given by the term $l=1$ in (\ref{First_Integration_21}). 
Inserting the derivatives given by Eqs.~(\ref{dipole_10}) - (\ref{dipole_12}) in appendix \ref{Appendix_Derivatives_1}, we obtain  
\begin{widetext}
\begin{eqnarray}
\frac{\Delta \dot{\ve{x}}_D\left(t\right)}{c} &=& + \frac{2 G}{c^2} \sum\limits_{A=1}^{N}  
\frac{\ve{M}_A\left(t\right)}{r^{\rm N}_A\left(t\right)}
\frac{1}{r^{\rm N}_A\left(t\right) - \ve{\sigma}\cdot\ve{r}^{\rm N}_A\left(t\right)} 
\nonumber\\
\nonumber\\
&& + \frac{2 G}{c^2}\,\sum\limits_{A=1}^{N}\frac{\ve{\sigma} \cdot \ve{M_A}\left(t\right)}{r^{\rm N}_A\left(t\right)}
\left(\frac{\ve{d}_A\left(t\right)}{\left(r^{\rm N}_A\left(t\right)\right)^2} 
- \frac{\ve{\sigma}}{r^{\rm N}_A\left(t\right) - \ve{\sigma}\cdot\ve{r}^{\rm N}_A\left(t\right)} 
- \frac{\ve{\sigma} \left(\ve{\sigma}\cdot\ve{r}^{\rm N}_A\left(t\right)\right)}{\left(r^{\rm N}_A\left(t\right)\right)^2}\right)  
\nonumber\\
\nonumber\\
&& - \frac{2 G}{c^2} \sum\limits_{A=1}^{N} \frac{\ve{d}_A\left(t\right) \cdot \ve{M}_A\left(t\right)}{\left(r^{\rm N}_A\left(t\right)\right)^2}\,  
\left(\frac{\ve{\sigma}}{r^{\rm N}_A\left(t\right)} + \frac{\ve{d}_A\left(t\right)}  
{r^{\rm N}_A\left(t\right)\left(r^{\rm N}_A\left(t\right) - \ve{\sigma}\cdot\ve{r}^{\rm N}_A\left(t\right)\right)} 
+ \frac{\ve{d}_A\left(t\right)}{\left(r^{\rm N}_A\left(t\right) - \ve{\sigma}\cdot\ve{r}^{\rm N}_A\left(t\right)\right)^2}\right).   
\label{dipole_5}
\end{eqnarray}
\end{widetext} 

\noindent
If the origin of the local reference system $\left(cT_A, \ve{X}_A\right)$ is located exactly at the center-of-mass 
of the massive body A, then the dipole moment of this body vanishes, $\ve{M}_A=0$. However, in real high-precision astrometry  
the center-of-mass of, for instance, a planet like Jupiter cannot be determined precisely. Therefore, for real astrometric measurements 
$\ve{M}_A \neq 0$, hence the light-deflection caused by the dipole moment of a massive body has to be taken into account, which is  
purely a coordinate effect; see also \cite{Kopeikin_Efroimsky_Kaplan,KopeikinMakarov2007}.

\subsection{Quadrupoles in arbitrary motion} 

As further instructive example we consider the case of light propagation in the gravitational field of   
N arbitrarily moving quadrupoles, given by $l=2$ in (\ref{First_Integration_21}), which reads:  
\begin{widetext} 
\begin{eqnarray}
\frac{\Delta \dot{\ve{x}}_Q\left(\tau+t^{\ast}\right)}{c} &=& - \frac{2\,G}{c^2}\,\sum\limits_{A=1}^N\,
M^A_{i_1 i_2}\;\sigma^{i_1}\,P^{i_2 j_2}\; 
\frac{\partial}{\partial \xi^{j_2}}\,
\frac{\ve{d}_A}{\left(r^{\rm N}_A\right)^3}\,  
- \frac{G}{c^2}\,\sum\limits_{A=1}^N\,M^A_{i_1 i_2}\;\sigma^{i_1}\,\sigma^{i_2}\;
\frac{\partial}{\partial c\,\tau}\,
\frac{\ve{d}_A}{\left(r^{\rm N}_A\right)^3}\,  
\nonumber\\
\nonumber\\
&& - \frac{G}{c^2}\,\sum\limits_{A=1}^N\,M^A_{i_1 i_2}\;P^{i_1 j_1}\,P^{i_2 j_2}\,\frac{\partial}{\partial \xi^{j_1}}\,
\frac{\partial}{\partial \xi^{j_2}}\,\frac{\ve{d}_A}{r^{\rm N}_A - \ve{\sigma}\cdot\ve{r}^{\rm N}_A}\,  
\frac{1}{r^{\rm N}_A}  
- \frac{G}{c^2}\,\ve{\sigma} \sum\limits_{A=1}^N\,M^A_{i_1 i_2}\,\sigma^{i_1}\,\sigma^{i_2}\;
\frac{\partial}{\partial c\,\tau}\,\frac{\partial}{\partial c\tau}\;
\frac{1}{r^{\rm N}_A} 
\nonumber\\
\nonumber\\
&& - \frac{2\,G}{c^2}\,\ve{\sigma}\sum\limits_{A=1}^N\,M^A_{i_1 i_2}\,\sigma^{i_1}\;
P^{i_2 j_2}\,\frac{\partial}{\partial \xi^{j_2}}\,\frac{\partial}{\partial c\tau}\,
\frac{1}{r^{\rm N}_A}
- \frac{G}{c^2}\,\ve{\sigma} \sum\limits_{A=1}^N\,M^A_{i_1 i_2}\;
\,P^{i_1 j_1}\,P^{i_2 j_2}\,\frac{\partial}{\partial \xi^{j_1}}\,
\frac{\partial}{\partial \xi^{j_2}}\,\frac{1}{r^{\rm N}_A}\,,  
\label{Comparison_15}
\end{eqnarray}
\end{widetext}

\noindent
where here for simpler notation the time-arguments have been omitted, i.e. 
$r^{\rm N}_A = r^{\rm N}_A\left(\tau+t^{\ast}\right)$, $\ve{r}^{\rm N}_A = \ve{r}^{\rm N}_A\left(\tau+t^{\ast}\right)$, 
$\ve{d}_A = \ve{d}_A\left(\tau+t^{\ast}\right)$, and $M^A_{i_1 i_2} = M^A_{i_1 i_2}\left(\tau+t^{\ast}\right)$.  
The derivatives in (\ref{Comparison_15}) are given in appendix \ref{Appendix_Derivatives_1}, and by  
inserting (\ref{Comparison_20}) - (\ref{Comparison_35}) into (\ref{Comparison_15}) one obtains  
the first integral of geodesic equation in the field of $N$ arbitrarily moving quadrupoles:  
\begin{widetext}
\begin{eqnarray}
\frac{\Delta \dot{\ve{x}}_Q\left(t\right)}{c} &=& \frac{G}{c^2} \sum\limits_{A=1}^N \frac{1}{d_A^2\left(t\right)} 
\bigg[\ve{\alpha}_A\left(t\right) \frac{\dot{\cal U}_A\left(t\right)}{c}  
+ \ve{\beta}_A\left(t\right) \frac{\dot{\cal V}_A\left(t\right)}{c}  
+ \ve{\gamma}_A\left(t\right) \frac{\dot{\cal F}_A\left(t\right)}{c}  
+ \ve{\delta}_A\left(t\right) \frac{\dot{\cal E}_A\left(t\right)}{c}\bigg],  
\label{Comparison_Quadrupole_5}
\end{eqnarray}
\end{widetext}

\noindent
where $\tau+t^{\ast}$ has finally been replaced by coordinate-time $t$ in (\ref{Comparison_Quadrupole_5}), i.e. after 
performance of all differentiations in (\ref{Comparison_15}).  
Adopting similar notation as used in \cite{Klioner1991},  
the vectorial coefficients in (\ref{Comparison_Quadrupole_5}) are given by:  
\begin{widetext}
\begin{eqnarray}
\alpha_A^k\left(t\right) &=& 
- M^A_{i_1 i_2}\left(t\right)\,d_A^k\left(t\right)\,\sigma^{i_1}\,\sigma^{i_2} 
+ 2\,M^A_{i_1 k}\left(t\right)\,d_A^{i_1}\left(t\right) 
- 2\,M^A_{i_1 i_2}\left(t\right)\,d_A^{i_2}\left(t\right)\,\sigma^{i_1}\,\sigma^k 
- \frac{4}{d_A^2\left(t\right)}\,M^A_{i_1 i_2}\left(t\right)\,d_A^{i_1}\left(t\right)\,d_A^{i_2}\left(t\right)\,
d_A^k\left(t\right)\,, 
\nonumber\\
\label{vectorial_coefficient_1} 
\\
\nonumber\\
\beta_A^k\left(t\right) &=& 
+ M^A_{i_1 i_2}\left(t\right)\,\sigma^{i_1}\,\sigma^{i_2}\,\sigma^k 
- 2\,M^A_{i_1 k}\left(t\right)\,\sigma^{i_1} 
+ \frac{4}{d_A^2\left(t\right)}\,M^A_{i_1 i_2}\left(t\right)\,d_A^{i_2}\left(t\right)\,d_A^k\left(t\right)\,\sigma^{i_1}  
- \frac{2}{d_A^2\left(t\right)}\,M^A_{i_1 i_2}\left(t\right)\,d_A^{i_1}\left(t\right)\,d_A^{i_2}\left(t\right)\,\sigma^k\,, 
\nonumber\\
\label{vectorial_coefficient_2}  
\\
\nonumber\\
\gamma_A^k\left(t\right) &=& 
+ M^A_{i_1 i_2}\left(t\right)\,d_A^{i_1}\left(t\right)\,d_A^{i_2}\left(t\right)\,d_A^k\left(t\right) 
- M^A_{i_1 i_2}\left(t\right)\,d_A^k\left(t\right)\,d_A^2\left(t\right)\,\sigma^{i_1}\,\sigma^{i_2}  
+ 2\,M^A_{i_1 i_2}\left(t\right)\,d_A^{i_2}\left(t\right)\,d_A^2\left(t\right)\,\sigma^{i_1}\,\sigma^k\,, 
\label{vectorial_coefficient_3}
\\
\nonumber\\
\delta_A^k\left(t\right) &=& 
- M^A_{i_1 i_2}\left(t\right)\,d_A^{i_1}\left(t\right)\,d_A^{i_2}\left(t\right)\,\sigma^k 
+ M^A_{i_1 i_2}\left(t\right)\,d_A^2\left(t\right)\,\sigma^{i_1}\,\sigma^{i_2}\,\sigma^k 
+ 2\,M^A_{i_1 i_2}\left(t\right)\,d_A^{i_2}\left(t\right)\,d_A^k\left(t\right)\,\sigma^{i_1}\,.  
\label{vectorial_coefficient_4}
\end{eqnarray}
\end{widetext}
 
\noindent
The scalar functions in (\ref{Comparison_Quadrupole_5}) are given by:  
\begin{eqnarray}
\frac{\dot{\cal U}_A\left(t\right)}{c}  
&=& \frac{d_A^2\left(t\right)}{\left(r^{\rm N}_A\left(t\right)\right)^2}\,
\frac{1}{r^{\rm N}_A\left(t\right) - \ve{\sigma}\cdot\ve{r}^{\rm N}_A\left(t\right)} 
\nonumber\\ 
&& \hspace{-1.0cm} \times \left(\frac{1}{r^{\rm N}_A\left(t\right)} + \frac{1}{r^{\rm N}_A\left(t\right) - \ve{\sigma}\cdot\ve{r}^{\rm N}_A\left(t\right)}\right),
\label{function1}
\end{eqnarray}

\begin{eqnarray}
\frac{\dot{\cal V}_A\left(t\right)}{c} &=& \frac{d_A^2\left(t\right)}{\left(r^{\rm N}_A\left(t\right)\right)^3}\,, 
\label{function2}
\end{eqnarray}

\begin{eqnarray}
\frac{\dot{\cal F}_A\left(t\right)}{c} &=& -3\,\frac{\ve{\sigma}\cdot\ve{r}^{\rm N}_A\left(t\right)}{\left(r^{\rm N}_A\left(t\right)\right)^5}\,,
\label{function3}
\end{eqnarray}

\begin{eqnarray}
\frac{\dot{\cal E}_A\left(t\right)}{c} &=& \frac{1}{\left(r^{\rm N}_A\left(t\right)\right)^3}  
- 3\,\frac{\left(\ve{\sigma}\cdot\ve{r}^{\rm N}_A\left(t\right)\right)^2}{\left(r^{\rm N}_A\left(t\right)\right)^5}\,.
\label{function4}
\end{eqnarray}

\noindent  
In the limit of quadrupoles at rest, $\ve{x}_A = {\rm const}$, and time-independent quadrupole-moments,  
$M_{i_1 i_2}^A = {\rm const}$, the expression in (\ref{Comparison_Quadrupole_5}) - (\ref{function4}) coincides with   
the corresponding results in \cite{Klioner2003a,Klioner1991,KlionerKopeikin1992}.

\subsection{Body at rest with full mass-multipole structure} 

The light-trajectory in the gravitational field of one massive body A at rest and located at the origin 
of coordinate system, $\ve{x}_A=0$, has been determined in \cite{Kopeikin1997} in post-Newtonian approximation  
for the case of time-independent multipoles. In such situation, we have to make the following replacements:  
$\ve{d}_A\left(\tau+t^{\ast}\right) \rightarrow \ve{\xi}$, 
$d_A\left(\tau+t^{\ast}\right) \rightarrow d$, 
$\ve{r}^{\rm N}_A \rightarrow \ve{r} = \ve{\xi} + c\,\tau\,\ve{\sigma}$,  
$r^{\rm N}_A\left(\tau+t^{\ast}\right) \rightarrow r = \sqrt{d^2 + c^2\tau^2}$,  
and $M^A_L\left(\tau+t^{\ast}\right) \rightarrow M^A_L$.    
Then, our solution in (\ref{First_Integration_21}) simplifies as follows (we omit the monopole- and the dipole-term, because 
the former one has already been considered above, while the latter one is not determined in \cite{Kopeikin1997}):  
\begin{widetext}
\begin{eqnarray}
\frac{\Delta \dot{\ve{x}}_{v_A=0}^A \left(\tau+t^{\ast}\right)}{c} &=& 
-\,\frac{2\,G}{c^2}\,
\sum\limits_{l=2}^{\infty} \frac{\left(-1\right)^l}{l!}\,M^A_L\,
P^{i_1j_1}\,...\,P^{i_l j_l}\,  
\frac{\partial}{\partial \xi^{j_1}}\,...\,\frac{\partial}{\partial \xi^{j_l}}\,
\left[\frac{\ve{\xi}}{d^2}\,\left(1 + \frac{c\,\tau}{r}\right) + \frac{\ve{\sigma}}{r}\right]
\nonumber\\
\nonumber\\
&& \hspace{-2.0cm} - \frac{2\,G}{c^2}
\sum\limits_{l=2}^{\infty} \sum\limits_{p=1}^{l}
\frac{\left(-1\right)^l}{\left(l-p\right)! p!}\,M^A_L\,
\sigma^{i_1}\,...\,\sigma^{i_p}\,
P^{i_{p+1}\,j_{p+1}}\,...\,P^{i_l\,j_l}\,
\frac{\partial}{\partial \xi^{j_{p+1}}}\,...\,\frac{\partial}{\partial \xi^{j_{l}}}\,
\left(\frac{\partial}{\partial c\tau}\right)^{p-1}
\frac{\ve{\xi} - c \tau\,\ve{\sigma}}{r^3}\,, 
\label{Comparison_body_at_rest_1}
\end{eqnarray}
\end{widetext}

\noindent
where we have used $\displaystyle \frac{1}{r - c\,\tau} = \frac{r + c\,\tau}{d^2}$ and 
$\displaystyle \left(\frac{\partial}{\partial c\tau}\right)^p \frac{1}{r} 
= - \left(\frac{\partial}{\partial c\tau}\right)^{p-1} \frac{c\,\tau}{r^3}$.  
The expression in (\ref{Comparison_body_at_rest_1}) agrees with the  
time-derivative of Eq.~(36) in \cite{Kopeikin1997}.  

Needless to say that one cannot deduce the general expression in (\ref{First_Integration_21}) from the specific  
solution in (\ref{Comparison_body_at_rest_1}) by some kind of an inverse replacement procedure, because such an approach would  
not be unique. For instance, the above replacement $d_A \rightarrow d$ is unique, but the inverse procedure is not unique,  
because it could either be $d \rightarrow \left|\ve{\xi}\right|$ or $d \rightarrow \left|\ve{d}_A\right|$.  
Similar ambiguities would appear in inverse replacements regarding variables $\ve{\xi}$ or $c\,\tau$.  
In other words: one cannot deduce the general expression in (\ref{First_Integration_21}) from the specific solution
given by Eq.~(34) in \cite{Kopeikin1997}.

\section{Second integration of geodesic equation}\label{Second_Integration}

The second integral determines the trajectory of the  
photon and can be written as follows; cf. Eq.~(\ref{light_trajectory_B}):  
\begin{eqnarray}
\ve{x}_{\rm 1PN} \left(\tau + t^{\ast}\right) &=& \ve{\xi} + c\,\tau\,\ve{\sigma} +
\sum\limits_{A=1}^N \Delta \ve{x}^A_{\rm 1PN}\left(\tau+t^{\ast}, \tau_0+t^{\ast}\right), 
\nonumber\\
\label{second_integral_geodesic_equation}
\end{eqnarray}

\noindent 
where the contribution of one body A is given by: 
\begin{eqnarray}
\Delta \ve{x}_{\rm 1PN}^A\left(\tau+t^{\ast},\tau_0+t^{\ast}\right) &=& 
\int\limits_{\tau_0}^{\tau}\,d c\tau^{\prime}\,\frac{\Delta \dot{\ve{x}}^A_{\rm 1PN}\left(\tau^{\prime}+t^{\ast}\right)}{c}\,,
\nonumber\\
\label{Second_Integration_5}
\end{eqnarray}

\noindent  
where the integrand is given by Eq.~(\ref{First_Integration_21}). 
How one goes about performing the second integration is
not much different in principle from the first integration represented in section \ref{First_Integration}.  
Using relations (\ref{First_Integration_35}) and (\ref{First_Integration_20}) we obtain the following expression  
for the second integration of geodesic equation for the light-trajectory in the gravitational field of one extended body A in arbitrary motion:
\begin{widetext}
\begin{eqnarray}
\Delta x_{\rm 1PN}^{i\;A}\left(\tau+t^{\ast},\tau_0+t^{\ast}\right) &=& + \frac{2\,G}{c^2}\,P^{ij}\,\frac{\partial}{\partial \xi^j}  
\sum\limits_{l=1}^{\infty} \frac{\left(-1\right)^l}{\left(l-1\right)!}\, \sigma^{i_1}\, P^{i_2\,j_2}\,...\,P^{i_l\,j_l}\,  
\frac{\partial}{\partial \xi^{j_2}}\,...\,\frac{\partial}{\partial \xi^{j_l}}\,{\cal I}_{C}
\nonumber\\
\nonumber\\
&& \hspace{-3.0cm} + \frac{2\,G}{c^2}\,P^{ij}\,\frac{\partial}{\partial \xi^j}
\sum\limits_{l=2}^{\infty}\sum\limits_{p=2}^{l} \frac{\left(-1\right)^l}{\left(l-p\right)! p!}\, \sigma^{i_1}\,...\,\sigma^{i_p}\;
P^{i_{p+1}\,j_{p+1}}\,...\,P^{i_l\,j_l}\,
\frac{\partial}{\partial \xi^{j_{p+1}}}\,...\,\frac{\partial}{\partial \xi^{j_l}}\,{\cal I}_{D}
\nonumber\\
\nonumber\\
&& \hspace{-3.0cm} - \frac{2\,G}{c^2}\,P^{ij}\,\frac{\partial}{\partial \xi^j}  
\sum\limits_{l=0}^{\infty} \frac{\left(-1\right)^l}{l!}\;
 P^{i_1j_1}\,...\,P^{i_l j_l}
\frac{\partial}{\partial \xi^{j_1}}\,...\,\frac{\partial}{\partial \xi^{j_l}}\,{\cal I}_{E} 
-\,\frac{2\,G}{c^2}\,\sigma^i\;\sum\limits_{l=0}^{\infty} \frac{\left(-1\right)^l}{l!}\,  
P^{i_1\,j_1}\,...\,P^{i_l\,j_l}\,  
\frac{\partial}{\partial \xi^{j_1}}\,...\,\frac{\partial}{\partial \xi^{j_l}}\,{\cal I}_{C}  
\nonumber\\
\nonumber\\
&& \hspace{-3.0cm} -\,\frac{2\,G}{c^2}\,\sigma^i\;\sum\limits_{l=1}^{\infty} \sum\limits_{p=1}^{l} \frac{\left(-1\right)^l}{\left(l-p\right)!\,p!}\, 
\sigma^{i_1}\,...\,\sigma^{i_p}\,P^{i_{p+1}\,j_{p+1}}\,...\,P^{i_l\,j_l}\, 
\frac{\partial}{\partial \xi^{j_{p+1}}}\,...\,\frac{\partial}{\partial \xi^{j_{l}}}\,{\cal I}_{F}\,. 
\label{Second_Integration_10}
\end{eqnarray}
\end{widetext}

\noindent
In order to obtain the form of the first two terms and of the last two terms in (\ref{Second_Integration_10}), 
the summation over $l,p$ has been separated as follows:  
\begin{eqnarray}
\sum\limits_{l=1}^{\infty} \sum\limits_{p=1}^{l} F\left(l,p\right) &=& \sum\limits_{l=1}^{\infty} F\left(l,p=1\right)
+ \sum\limits_{l=2}^{\infty} \sum\limits_{p=2}^{l} F\left(l,p\right), 
\nonumber\\
\label{Sumamtion_1}
\\
\nonumber\\
\sum\limits_{l=0}^{\infty} \sum\limits_{p=0}^{l} F\left(l,p\right) &=& \sum\limits_{l=0}^{\infty} F\left(l,p=0\right)
+ \sum\limits_{l=1}^{\infty} \sum\limits_{p=1}^{l} F\left(l,p\right).
\nonumber\\
\label{Summation_2}
\end{eqnarray}

\noindent
In (\ref{Second_Integration_10}) we encounter four kind of integrals:  
\begin{eqnarray}
{\cal I}_{C} &=& \int\limits_{\tau_0}^{\tau}\,d c\tau^{\prime}\,
\frac{M_L^A\left(\tau^{\prime}+t^{\ast}\right)}{r^{\rm N}_A\left(\tau^{\prime}+t^{\ast}\right)}\,, 
\label{Integral_I_C}
\\
\nonumber\\
{\cal I}_{D} &=& \int\limits_{\tau_0}^{\tau}\,d c\tau^{\prime}\,
M_L^A\left(\tau^{\prime}+t^{\ast}\right) \left(\frac{\partial}{\partial c\tau^{\prime}}\right)^{p-1}
\frac{1}{r^{\rm N}_A\left(\tau^{\prime}+t^{\ast}\right)}\,, 
\nonumber\\
\label{Integral_I_D}
\\
\nonumber\\
{\cal I}_{E} &=& \int\limits_{\tau_0}^{\tau}\,d c\tau^{\prime}\,
M_L^A\left(\tau^{\prime}+t^{\ast}\right)\,
\nonumber\\
&& \times \ln \left[r^{\rm N}_A\left(\tau^{\prime}+t^{\ast}\right) - \ve{\sigma}\cdot\ve{r}^{\rm N}_A\left(\tau^{\prime}+t^{\ast}\right)\right]\,, 
\label{Integral_I_E}
\\
\nonumber\\
{\cal I}_{F} &=& \int\limits_{\tau_0}^{\tau}\,d c\tau^{\prime}\,
M_L^A\left(\tau^{\prime}+t^{\ast}\right) \left(\frac{\partial}{\partial c\tau^{\prime}}\right)^{p}
\frac{1}{r^{\rm N}_A\left(\tau^{\prime}+t^{\ast}\right)}\,,  
\nonumber\\
\label{Integral_I_F}
\end{eqnarray}

\noindent
which are determined in appendix \ref{Appendix_Integral_C}. 
These integrals run over the unknown worldline $\ve{x}_A\left(t\right)$ of massive body A, and can also be integrated by parts, that means the procedure it 
essentially based upon the fact that the non-integrated remnants are beyond 1PN approximation, because they imply an additional factor $c^{-1}$.  

Then, inserting the solutions of these four integrals, 
given by Eqs.~(\ref{Appendix_Integral_C_A}), (\ref{Appendix_Integral_C_D}), (\ref{Appendix_Integral_C_15}) 
and (\ref{Appendix_Integral_C_20}), into Eq.~(\ref{Second_Integration_10}) 
and performing the differentiations with respect to $\displaystyle P^{ij}\,\frac{\partial}{\partial \xi^j}$, 
the second integration of geodesic equation for the light-trajectory in the field of one body A is given by  
\begin{widetext}
\begin{eqnarray}
\Delta \ve{x}^A_{\rm 1PN}\left(\tau+t^{\ast},\tau_0+t^{\ast}\right)  
 = \Delta \ve{x}^A_{\rm 1PN}\left(\tau+t^{\ast}\right) - \Delta \ve{x}^A_{\rm 1PN}\left(\tau_0+t^{\ast}\right), 
\label{Second_Integration_15}
\end{eqnarray}
\end{widetext}

\noindent
where    
\begin{widetext}
\begin{eqnarray}
&& \hspace{-1.0cm} \Delta \ve{x}_{\rm 1PN}^A\left(\tau+t^{\ast}\right) 
\nonumber\\
&=& - \frac{2\,G}{c^2}\, \sum\limits_{l=1}^{\infty} \frac{\left(-1\right)^l}{\left(l-1\right)!}\;
M^A_L\left(\tau+t^{\ast}\right) \sigma^{i_1}\,
P^{i_2\,j_2}\,...\,P^{i_l\,j_l}\,
\frac{\partial}{\partial \xi^{j_2}}\,...\,\frac{\partial}{\partial \xi^{j_{l}}}\,
\,\frac{\ve{d}_A\left(\tau + t^{\ast}\right)}{r^{\rm N}_A\left(\tau + t^{\ast}\right)}\,  
\frac{1}{r^{\rm N}_A\left(\tau + t^{\ast}\right) - \ve{\sigma}\cdot\ve{r}^{\rm N}_A\left(\tau + t^{\ast}\right)}
\nonumber\\
\nonumber\\
&& - \frac{2\,G}{c^2} \sum\limits_{l=2}^{\infty}  
\sum\limits_{p=2}^{l} \frac{\left(-1\right)^l}{\left(l-p\right)!\,p!}\;M^A_L\left(\tau+t^{\ast}\right) 
\sigma^{i_1}\,...\,\sigma^{i_p}\,
P^{i_{p+1}\,j_{p+1}}\,...\,P^{i_l\,j_l}\,
\frac{\partial}{\partial \xi^{j_{p+1}}}\,...\,\frac{\partial}{\partial \xi^{j_{l}}}\,
\left(\frac{\partial}{\partial c\tau}\right)^{p-2}\,
\frac{\ve{d}_A\left(\tau+t^{\ast}\right)}{\left(r^{\rm N}_A\left(\tau + t^{\ast}\right)\right)^3}
\nonumber\\
\nonumber\\
&& - \frac{2\,G}{c^2} \sum\limits_{l=0}^{\infty}
\frac{\left(-1\right)^l}{l!}\;M^A_L\left(\tau+t^{\ast}\right)  
P^{i_1\,j_1}\,...\,P^{i_l\,j_l}\,
\frac{\partial}{\partial \xi^{j_1}}\,...\,\frac{\partial}{\partial \xi^{j_l}}\,
\frac{\ve{d}_A\left(\tau+t^{\ast}\right)}
{r^{\rm N}_A\left(\tau + t^{\ast}\right) - \ve{\sigma}\cdot\ve{r}^{\rm N}_A\left(\tau + t^{\ast}\right)}
\nonumber\\
\nonumber\\
&& + \frac{2\,G}{c^2}\,\ve{\sigma} \sum\limits_{l=0}^{\infty}
\frac{\left(-1\right)^l}{l!}\;M_L^A\left(\tau + t^{\ast}\right)  
P^{i_1\,j_1}\,...\,P^{i_l\,j_l}\,
\frac{\partial}{\partial \xi^{j_1}}\,...\,\frac{\partial}{\partial \xi^{j_l}}\,
\ln \left[r^{\rm N}_A\left(\tau + t^{\ast}\right) - \ve{\sigma}\cdot\ve{r}^{\rm N}_A\left(\tau + t^{\ast}\right)\right] 
\nonumber\\
\nonumber\\
&& - \frac{2\,G}{c^2}\,\ve{\sigma} \sum\limits_{l=1}^{\infty} \sum\limits_{p=1}^{l}  
\frac{\left(-1\right)^l}{\left(l-p\right)!\,p!}\;M^A_L\left(\tau+t^{\ast}\right)  
\sigma^{i_1}\,...\,\sigma^{i_p}\, 
P^{i_{p+1}\,j_{p+1}}\,...\,P^{i_l\,j_l}\,
\frac{\partial}{\partial \xi^{j_{p+1}}}\,...\,\frac{\partial}{\partial \xi^{j_l}}\,
\left(\frac{\partial}{\partial c \tau}\right)^{p-1} 
\frac{1}{r^{\rm N}_A\left(\tau + t^{\ast}\right)}\,,   
\nonumber\\
\label{Second_Integration_20}
\end{eqnarray}
\end{widetext}

\noindent
and we recall the notation $M^A_L = M^A_{i_1 ... i_l}$.
In order to obtain (\ref{Second_Integration_20}), the relations (\ref{First_Integration_35}) and (\ref{First_Integration_20}) and 
\begin{widetext}
\begin{eqnarray}
P^{ij}\,\frac{\partial}{\partial \xi^j}\,
\bigg(r^{\rm N}_A\left(\tau+t^{\ast}\right) + \ve{\sigma}\cdot\ve{r}_A^{\rm N}\left(\tau+t^{\ast}\right) 
\ln \left[r^{\rm N}_A\left(\tau+t^{\ast}\right) - \ve{\sigma}\cdot\ve{r}^{\rm N}_A\left(\tau+t^{\ast}\right)\right]\bigg) 
&=& \frac{d_A^i\left(\tau + t^{\ast}\right)}{r^{\rm N}_A\left(\tau + t^{\ast}\right) - \ve{\sigma}\cdot\ve{r}^{\rm N}_A\left(\tau + t^{\ast}\right)}\,,
\label{Second_Integration_Relation_A}
\end{eqnarray}
\end{widetext}

\noindent
have also been used. The expression in (\ref{Second_Integration_20}) represents the solution  
for the second integration of geodesic equation in 1PN approximation in (\ref{transformed_geodesic_equation_C}),  
in the field of one arbitrarily moving body A and to any order of its intrinsic mass-multipoles.
Like in the first integral in (\ref{First_Integration_21}), after 
the differentiations in (\ref{Second_Integration_20}) the replacement of   
$\tau+t^{\ast}$ by the global coordinate-time $t$ can be performed. One may easily check that the time-differentiation 
of (\ref{Second_Integration_20}) yields immediately the first integral in (\ref{First_Integration_21}) 
up to terms of higher-order beyond 1PN approximation. So the solution in (\ref{Second_Integration_20}) is consistent   
with the solution in (\ref{First_Integration_21}).

\section{Some special cases of second integration}\label{Special_Cases_2} 

Like in case of first integration, let us consider the very few first terms  
of (\ref{Second_Integration_20}) as instructive examples, and compare them with research findings in the literature, namely:   
arbitrarily moving monopoles, dipoles, quadrupoles, and the case of one massive body at rest with full mass-multipole structure.

\subsection{Monopoles in arbitrary motion} 

For the monopole-term $\left(l=0\right)$ we obtain from (\ref{Second_Integration_15}) and (\ref{Second_Integration_20}):  
\begin{widetext}
\begin{eqnarray}
\Delta \ve{x}_M \left(t,t_0\right) &=& - \frac{2\,G}{c^2} \sum\limits_{A=1}^N M_A\,  
\left(\frac{\ve{d}_A\left(t\right)}{r^{\rm N}_A\left(t\right) - \ve{\sigma} \cdot \ve{r}^{\rm N}_A\left(t\right)} 
- \frac{\ve{d}_A\left(t_0\right)}{r^{\rm N}_A\left(t_0\right) - \ve{\sigma} \cdot \ve{r}^{\rm N}_A\left(t_0\right)}\right)  
+ \frac{2\,G}{c^2}\,\ve{\sigma} \sum\limits_{A=1}^N \; M_A\, 
\ln \frac{r^{\rm N}_A\left(t\right) - \ve{\sigma} \cdot \ve{r}^{\rm N}_A\left(t\right)}
{r^{\rm N}_A\left(t_0\right) - \ve{\sigma} \cdot \ve{r}^{\rm N}_A\left(t_0\right)}\,, 
\nonumber\\
\label{monopole_5}
\end{eqnarray}
\end{widetext}

\noindent
where in the final expression we have replaced $\tau+t^{\ast} = t$ and $\tau_0 + t^{\ast} = t_0$; recall  
$\ve{r}^{\rm N}_A\left(t\right) = \ve{x}_{\rm N}\left(t\right) - \ve{x}_A\left(t\right)$ and  
$\ve{r}^{\rm N}_A\left(t_0\right) = \ve{x}_0 - \ve{x}_A\left(t_0\right)$.  
The time-derivative of (\ref{monopole_5}) yields immediately (\ref{Comparison_10});  
up to terms of order ${\cal O}\left(v_A/c\right)$  
\footnote{Terms $\sim v_A/c$ originate from $\displaystyle \frac{\partial \ve{x}_A}{\partial c t} = \frac{\ve{v}_A}{c}$, 
and given by $\displaystyle - \frac{2\,G\,M_A}{c^2}\frac{\ve{\dot{d}}_A}{c}\,\frac{1}{r_A^N-\ve{\sigma}\cdot\ve{r}_A^N}$  
where $\ve{\dot{d}}_A = \ve{\sigma} \times \left(\ve{\sigma} \times \ve{v}_A\right)$ and  
$\displaystyle  + \frac{2\,G\,M_A}{c^2\,r_A^N}\left(\frac{\ve{d}_A}{r_A^N-\ve{\sigma}\cdot\ve{r}_A^N} + \ve{\sigma}\right)\frac{\ve{v}_A}{c} \cdot 
\frac{\ve{\sigma}\,r_A^N - \ve{r}_A^N}{r_A^N-\ve{\sigma}\cdot\ve{r}_A^N}$.   
About the magnitude of these terms in the limit $\ve{\sigma}\cdot\ve{r}_A^N \rightarrow r_A^N$ see the comment below Eq.~(\ref{Proof1_G}). Let us 
also recall that $\displaystyle \frac{\ve{v}_A}{c}\cdot\frac{\ve{\sigma}\,r_A^N - \ve{r}_A^N}{r_A^N-\ve{\sigma}\cdot\ve{r}_A^N} 
= \frac{\ve{\sigma}\cdot\ve{v}_A}{c} - \frac{\ve{v}_A}{c}\cdot \frac{\ve{d}_A}{r_A^N-\ve{\sigma}\cdot\ve{r}_A^N}$, hence there are no terms 
proportional to $v_A\left(t-t_0\right)$ but only terms proportional to $v_A/c$, a statement which is consistent with the formalism presented.}.  

In the limit of massive bodies at rest, the expression (\ref{monopole_5}) coincides with Eq.~(3.2.13) in \cite{Brumberg1991}  
and with Eq.~(22) in \cite{Klioner2003a}, where the mass-monopoles are not located at the  
origin of the coordinate-system but displaced by some constant vector $\ve{x}_A$; cf. Eq.~(\ref{Introduction_10}).  

In \cite{KopeikinSchaefer1999} the light-trajectory in the field of $N$ arbitrarily moving pointlike monopoles has been determined
in first post-Minkowskian approximation (1PM), that means where the pointlike
monopoles could even move with ultra-relativistic speed, while (\ref{monopole_5}) is for extended monopoles in 1PN approximation.  
By expansion of the 1PM solution (Eqs.~(33) and (35) in \cite{KopeikinSchaefer1999})
in powers of $v_A/c$, one may show an agreement with our solution in (\ref{monopole_5})
up to terms of the order ${\cal O}\left(v_A/c\right)$.

\subsection{Dipoles in arbitrary motion} 

From (\ref{Second_Integration_20}) we obtain for the dipole-term $\left(l=1\right)$: 
\begin{widetext}
\begin{eqnarray}
\Delta \ve{x}_D \left(t,t_0\right) &=& \Delta \ve{x}_D \left(t\right) - \Delta \ve{x}_D \left(t_0\right),
\nonumber\\
\nonumber\\
\Delta \ve{x}_D \left(t\right) &=& + \frac{2\,G}{c^2}\,\sum\limits_{A=1}^N\,
\frac{\ve{M}_A\left(t\right)}{r^{\rm N}_A\left(t\right) - \ve{\sigma}\cdot\ve{r}^{\rm N}_A\left(t\right)}
+ \frac{2\,G}{c^2}\,\sum\limits_{A=1}^N\;\frac{\ve{\sigma} \cdot \ve{M}_A\left(t\right)}{r^{\rm N}_A\left(t\right)}\;
\left(\frac{\ve{d}_A\left(t\right)}{r^{\rm N}_A\left(t\right) - \ve{\sigma}\cdot\ve{r}^{\rm N}_A\left(t\right)} 
- \frac{\ve{\sigma} \left(\ve{\sigma}\cdot\ve{r}^{\rm N}_A\left(t\right)\right)}
{r^{\rm N}_A\left(t\right) - \ve{\sigma}\cdot\ve{r}^{\rm N}_A\left(t\right)}\right)
\nonumber\\
\nonumber\\
&& - \frac{2\,G}{c^2}\,\sum\limits_{A=1}^N\,\frac{\ve{d}_A\left(t\right) \cdot \ve{M}_A\left(t\right)}{r^{\rm N}_A\left(t\right)}\;
\left(\frac{\ve{\sigma}}{r^{\rm N}_A\left(t\right) - \ve{\sigma}\cdot\ve{r}^{\rm N}_A\left(t\right)}  
+ \frac{\ve{d}_A\left(t\right)}{\left(r^{\rm N}_A\left(t\right) - \ve{\sigma}\cdot\ve{r}^{\rm N}_A\left(t\right)\right)^2}\right), 
\label{dipole_20}
\end{eqnarray}
\end{widetext}

\noindent
where we have used the derivatives given in appendix \ref{Appendix_Derivatives_2}. 
The time-derivative of (\ref{dipole_20}) yields immediately (\ref{dipole_5}) up to terms of order ${\cal O}\left(v_A/c\right)$. 
As mentioned above, if the origin of the local reference system $\left(cT_A, \ve{X}_A\right)$ is located exactly at the  
center-of-mass of the massive body A, then the dipole moment of this body vanishes, $\ve{M}_A=0$,  and there would be no dipole-term. 
But in reality one cannot determine precisely the center-of-mass of a massive body (e.g. giant planets) so that 
$\ve{M}_A \neq 0$ and one has carefully to take into account the change in the light-trajectory caused by the dipole-term, 
which is purely a coordinate effect; cf. Ref.~\cite{Kopeikin_Efroimsky_Kaplan,KopeikinMakarov2007}.

\subsection{Quadrupoles in arbitrary motion} 

Now we consider the light-trajectory in the gravitational field of
N arbitrarily moving quadrupoles, given by the term $l=2$ in (\ref{Second_Integration_20}), which reads:
\begin{eqnarray}
\Delta \ve{x}_Q\left(\tau+t^{\ast},\tau_0+t^{\ast}\right) &=& 
\Delta \ve{x}_Q\left(\tau+t^{\ast}\right) - \Delta \ve{x}_Q\left(\tau_0+t^{\ast}\right), 
\nonumber\\
\label{Trajectory_Quadrupole_5}
\end{eqnarray}

\noindent
with
\begin{widetext} 
\begin{eqnarray}
\Delta \ve{x}_Q\left(\tau+t^{\ast}\right) &=& - \frac{2\,G}{c^2}\,\sum\limits_{A=1}^N\;M^A_{i_1 i_2}\; 
\sigma^{i_1}\,P^{i_2 j_2}\,
\frac{\partial}{\partial \xi^{j_2}}\,
\frac{\ve{d}_A}{r^{\rm N}_A}\,\frac{1}{r^{\rm N}_A - \ve{\sigma}\cdot\ve{r}^{\rm N}_A}
\nonumber\\
\nonumber\\
&& \hspace{-3.0cm} - \frac{G}{c^2}\,\sum\limits_{A=1}^N\,M^A_{i_1 i_2}\;\sigma^{i_1}\,\sigma^{i_2}\,
\frac{\ve{d}_A}{\left(r^{\rm N}_A\right)^2}\,\frac{1}{r^{\rm N}_A}
- \frac{G}{c^2}\,\sum\limits_{A=1}^N\,M^A_{i_1 i_2}\,P^{i_1 j_1}\,P^{i_2 j_2}\,\frac{\partial}{\partial \xi^{j_1}}\,
\frac{\partial}{\partial \xi^{j_2}}\,\frac{\ve{d}_A}{r^{\rm N}_A - \ve{\sigma}\cdot\ve{r}^{\rm N}_A}\,
\nonumber\\
\nonumber\\
&& \hspace{-3.0cm} + \frac{G}{c^2}\,\ve{\sigma} \sum\limits_{A=1}^N\,M^A_{i_1 i_2}\; 
P^{i_1 j_1}\,P^{i_2 j_2}\,\frac{\partial}{\partial \xi^{j_1}}\, 
\frac{\partial}{\partial \xi^{j_2}}\,\ln \left(r^{\rm N}_A - \ve{\sigma}\cdot\ve{r}^{\rm N}_A\right)  
\nonumber\\
\nonumber\\
&& \hspace{-3.0cm} - \frac{2\,G}{c^2}\,\ve{\sigma}\,\sum\limits_{A=1}^N\,M^A_{i_1 i_2}\;\sigma^{i_1}\,
P^{i_2 j_2}\,\frac{\partial}{\partial \xi^{j_2}}\,
\frac{1}{r^{\rm N}_A}
- \frac{G}{c^2}\,\ve{\sigma}\,\sum\limits_{A=1}^N\;M^A_{i_1 i_2}\; 
\sigma^{i_1}\,\sigma^{i_2}\,\frac{\partial}{\partial c\tau}\, 
\frac{1}{r^{\rm N}_A}\,, 
\label{Trajectory_Quadrupole_6}
\end{eqnarray}
\end{widetext} 

\noindent
where here for simpler notation the time-arguments have been omitted, i.e.
$r^{\rm N}_A = r^{\rm N}_A\left(\tau+t^{\ast}\right)$, $\ve{r}^{\rm N}_A = \ve{r}^{\rm N}_A\left(\tau+t^{\ast}\right)$,
$\ve{d}_A = \ve{d}_A\left(\tau+t^{\ast}\right)$, and $M^A_{i_1 i_2} = M^A_{i_1 i_2}\left(\tau+t^{\ast}\right)$.
The derivatives in the first, fifth, and sixth term in (\ref{Trajectory_Quadrupole_6}) were already  
given in appendix \ref{Appendix_Derivatives_1}, while the derivatives of the third and fourth term in (\ref{Trajectory_Quadrupole_6})  
were already given in appendix \ref{Appendix_Derivatives_2}. After performing these derivatives the replacements have to be performed: 
$\tau + t^{\ast} = t$ and $\tau_0 + t^{\ast} = t_0$.  
By inserting relations (\ref{dipole_10}), (\ref{dipole_12}), (\ref{quadrupole_40}), (\ref{quadrupole_45}) into  
(\ref{Trajectory_Quadrupole_6}) one obtains the light-trajectory in the field of $N$ arbitrarily moving quadrupoles:  
\begin{eqnarray}
\Delta \ve{x}_Q\left(t,t_0\right) &=& \Delta \ve{x}_Q\left(t\right) - \Delta \ve{x}_Q\left(t_0\right), 
\label{Trajectory_Quadrupole_10}
\end{eqnarray}

\noindent 
with   
\begin{widetext} 
\begin{eqnarray}
\Delta \ve{x}_Q\left(t\right) &=& \frac{G}{c^2} \sum\limits_{A=1}^N \frac{1}{d_A^2\left(t\right)}
\bigg[\ve{\alpha}_A\left(t\right) {\cal U}_A\left(t\right) + \ve{\beta}_A\left(t\right) {\cal V}_A\left(t\right) 
+ \ve{\gamma}_A\left(t\right) {\cal F}_A\left(t\right) + \ve{\delta}_A\left(t\right) {\cal E}_A\left(t\right)\bigg].  
\label{Trajectory_Quadrupole_15}
\end{eqnarray}
\end{widetext}

\noindent
The vectorial coefficients in (\ref{Trajectory_Quadrupole_15}) were given by  
Eqs.~(\ref{vectorial_coefficient_1}) - (\ref{vectorial_coefficient_4})  
and the scalar functions in (\ref{Trajectory_Quadrupole_15}) are given by:
\begin{eqnarray}
{\cal U}_A\left(t\right) &=& \frac{1}{r^{\rm N}_A\left(t\right)}\,
\frac{r^{\rm N}_A\left(t\right)+\ve{\sigma}\cdot\ve{r}^{\rm N}_A\left(t\right)}{r^{\rm N}_A\left(t\right)-\ve{\sigma}\cdot\ve{r}^{\rm N}_A\left(t\right)}\,, 
\label{function5}
\\
\nonumber\\
{\cal V}_A\left(t\right) &=& \frac{\ve{\sigma}\cdot\ve{r}^{\rm N}_A\left(t\right)}{r^{\rm N}_A\left(t\right)} + 1\,, 
\label{function6}
\\
\nonumber\\
{\cal F}_A\left(t\right) &=& \frac{1}{\left(r^{\rm N}_A\left(t\right)\right)^3}\,,
\label{function7}
\\
\nonumber\\
{\cal E}_A\left(t\right) &=& \frac{\ve{\sigma}\cdot\ve{r}^{\rm N}_A\left(t\right)}{\left(r^{\rm N}_A\left(t\right)\right)^3}\,. 
\label{function8}
\end{eqnarray}

\noindent 
The time-derivative of (\ref{Trajectory_Quadrupole_15}) yields (\ref{Comparison_Quadrupole_5}),  
up to terms of higher-order, i.e. ${\cal O}\left(v_A/c\right)$ or ${\cal O}\left(\dot{M}_{i_1 i_2}^A/c\right)$.  

In the limit of bodies at rest, $\ve{x}_A = {\rm const}$, and time-independent quadrupole-moments,
$M_{i_1 i_2}^A = {\rm const}$, the expression in (\ref{Trajectory_Quadrupole_15}) - (\ref{function8}) coincides with
with the corresponding results in \cite{Klioner2003a,Klioner1991,KlionerKopeikin1992},  
cf. Eqs.~(\ref{Introduction_12}) - (\ref{Introduction_function8}).   

One should keep in view that a series expansion of the 
vectorial coefficients (\ref{vectorial_coefficient_1}) - (\ref{vectorial_coefficient_4})  
does not necessarily create terms 
beyond 1PN approximation. For instance, a series expansion of the vectorial coefficients around some time-moment $t_0$ 
implies a corresponding series expansion of the impact vector and quadrupole moment,  
\begin{eqnarray}
\ve{d}_A\left(t\right) &=& \ve{d}_A\left(t_0\right) + \ve{\sigma} \times \left(\ve{\sigma} \times \ve{v}_A\left(t_0\right)\right) 
\left(t - t_0\right) 
\nonumber\\
&& + {\cal O} \left(a_A\right), 
\label{Series_Expansion_A}
\\
\nonumber\\
M^A_{i_1 i_2}\left(t\right) &=& M^A_{i_1 i_2}\left(t_0\right) + \dot{M}^A_{i_1 i_2}\left(t_0\right)\left(t - t_0\right) 
+ {\cal O}\left(\ddot{M}^A_{i_1 i_2}\right),  
\nonumber\\
\label{Series_Expansion_B}
\end{eqnarray}
 
\noindent
which are proportional either to $v_A\left(t - t_0\right) $ or $\dot{M}^A_{i_1 i_2}\left(t - t_0\right)$, 
but neither to $v_A/c$ nor $\dot{M}^A_{i_1 i_2}/c$. Consequently, the individual terms in a series expansion 
of vectorial coefficients are not necessarily beyond 1PN approximation. 

Results for the light-trajectory in the field of quadrupoles in uniform motion, $\ve{v}_A = {\rm const}$, were represented 
in \cite{RAMOD5}. In the limit of uniform motion the expression in (\ref{Trajectory_Quadrupole_10}) - (\ref{function8}) should coincide  
with the results in \cite{RAMOD5}. For such a comparison the series-expansion in (\ref{Introduction_15}) would have to be
inserted into the solution (\ref{Trajectory_Quadrupole_10}) - (\ref{function8}), which leads rapidly to cumbersome expressions.
Consequently, such a comparison constitutes a rather ambitious assignment of a task and spoils the intention of the investigation.

\subsection{Body at rest with full mass-multipole structure} 

As it has been mentioned above, the light-trajectory in the gravitational field of one massive body at rest and located at 
the origin of coordinate system, $\ve{x}_A = 0$, has been determined in \cite{Kopeikin1997} in post-Newtonian approximation  
and for the case of time-independent multipoles. In such situation, we have to make the following replacements:  
$\ve{d}_A\left(\tau+t^{\ast}\right) \rightarrow \ve{\xi}$,
$d_A\left(\tau+t^{\ast}\right) \rightarrow d$,
$\ve{r}^{\rm N}_A \rightarrow \ve{r} = \ve{\xi} + c\,\tau\,\ve{\sigma}$,
$r^{\rm N}_A\left(\tau+t^{\ast}\right) \rightarrow r = \sqrt{d^2 + c^2\tau^2}$,
and $M^A_L\left(\tau+t^{\ast}\right) \rightarrow M^A_L$. Then our solution  
in (\ref{Second_Integration_20}) simplifies as follows (without monopole- and dipole-term):  
\begin{widetext}
\begin{eqnarray}
\Delta \ve{x}_{v_A=0}^{A}\left(\tau+t^{\ast}\right) &=&
- \frac{2\,G}{c^2}\,\sum\limits_{l=2}^{\infty} \frac{\left(-1\right)^l}{\left(l-1\right)!}\;
M^A_L\,\sigma^{i_1}\,
P^{i_{p+1}\,j_{p+1}}\,...\,P^{i_l\,j_l}\,
\frac{\partial}{\partial \xi^{j_{p+1}}}\,...\,\frac{\partial}{\partial \xi^{j_{l}}}\,
\left[\frac{\ve{\xi}}{d^2}\,\frac{c\,\tau}{r} + \frac{\ve{\sigma}}{r}\right]  
\nonumber\\
\nonumber\\
&& - \frac{2\,G}{c^2} \sum\limits_{l=2}^{\infty}
\sum\limits_{p=2}^{l} \frac{\left(-1\right)^l}{\left(l-p\right)!\,p!}\,
M^A_L\,\sigma^{i_1}\,...\,\sigma^{i_p}\,
P^{i_{p+1}\,j_{p+1}}\,...\,P^{i_l\,j_l}
\frac{\partial}{\partial \xi^{j_{p+1}}}\,...\,\frac{\partial}{\partial \xi^{j_{l}}}\,
\left(\frac{\partial}{\partial c\tau}\right)^{p-2}\,
\left[\frac{\ve{\xi} - c\,\tau\,\ve{\sigma}}{r^3}\right]\,
\nonumber\\
\nonumber\\
&& - \frac{2\,G}{c^2} \sum\limits_{l=2}^{\infty}
\frac{\left(-1\right)^l}{l!}\,M^A_L\,  
P^{i_1\,j_1}\,...\,P^{i_l\,j_l}\,
\frac{\partial}{\partial \xi^{j_1}}\,...\,\frac{\partial}{\partial \xi^{j_l}}\,
\left[\frac{\ve{\xi}}{d^2}\left(r + c\,\tau\right) + \ve{\sigma}\,\ln \left(r + c\,\tau\right)\right],  
\label{Comparison_body_at_rest_2}
\end{eqnarray}
\end{widetext}

\noindent
which is in agreement with Eq.~(36) in \cite{Kopeikin1997}, and the  
time-derivative of (\ref{Comparison_body_at_rest_2}) yields (\ref{Comparison_body_at_rest_1}).  
Let us note, that expression (\ref{Comparison_body_at_rest_2}) has to be understood in combination with 
(\ref{Second_Integration_15}), that means in the first line the term  
$\displaystyle \frac{\ve{\xi}}{d^2}\,\frac{r + c\,\tau}{r}$ has been replaced by 
$\displaystyle \frac{\ve{\xi}}{d^2}\,\frac{c\,\tau}{r}$, and also    
$\displaystyle \ln \frac{r - c\,\tau}{r_0 - c\,\tau_0} = - \ln \frac{r + c\,\tau}{r_0 + c\,\tau_0}$ has been used.

It is of course impossible to deduce the general solution in (\ref{Second_Integration_20}) from the specific solution  
in (\ref{Comparison_body_at_rest_2}), by reason that an inverse replacement procedure would not be unique, because it could either 
be $d \rightarrow \left|\ve{\xi}\right|$ or $d \rightarrow \left|\ve{d}_A\right|$; similar problems concern the variables 
$\ve{\xi}$ or $c \tau$. Stated somewhat differently: one cannot deduce the general expression in (\ref{Second_Integration_20})  
from the specific solution given by Eq.~(36) in \cite{Kopeikin1997}; cf. text below Eq.~(\ref{Comparison_body_at_rest_1}).

\section{Observable relativistic effects}\label{Observable_Effects}  

Let us consider two observable effects which are of decisive importance in relativistic astrometry: 
the time-delay and the deflection of photons propagating through the Solar system.  
We shall assume that the observer and the celestial lightsource are at rest with respect to the global system.  

\subsection{Time-delay} 

The classical relativistic effect of time-delay when a light-signal propagates 
through the static gravitational field of a spherically symmetric massive body (monopole) has been predicted 
by {\it Shapiro} in 1963 \cite{Shapiro1} and were detected soon afterwards \cite{Shapiro2,Shapiro3}.  
The results of these experiments have been confirmed with increasing accuracy, and  
the todays most accurate measurement of Shapiro-delay was achieved in 2003 \cite{Shapiro4} using Cassini spacecraft.  
The solution in (\ref{second_integral_geodesic_equation}) allows to determine 
the time-delay of light-signals propagating through the gravitational field of a system of $N$ arbitrarily moving massive bodies.  

Let $\ve{x}_1 = \ve{x}\left(t_1\right)$ be the global spatial coordinate of the space-based observer 
at the moment of observation $t_1$ and $\ve{x}_0 = \ve{x}\left(t_0\right)$ be the global spatial coordinate of the 
source at the moment of emission $t_0$ of the light-signal which is observed at $\ve{x}\left(t_1\right)$.  
In terms of the new variables $\ve{\xi}$ and $\tau$, both of these spatial coordinates are given by 
$\ve{x}_1 = \ve{x}\left(\tau_1+t^{\ast}\right)$ and $\ve{x}_0 = \ve{x}\left(\tau_0+t^{\ast}\right)$.  
Furthermore, we introduce the following vectors:
\begin{eqnarray}
\ve{R} &=& \ve{x}\left(\tau_1+t^{\ast}\right) - \ve{x}\left(\tau_0+t^{\ast}\right),  
\label{Vector_R}
\\
\nonumber\\
\ve{k} &=& \frac{\ve{R}}{R}\,, 
\label{Vector_k}
\end{eqnarray}

\noindent
where $R = \left|\ve{R}\right|$ with $\ve{R}$ being the vector from the source (at the moment of emission) 
to the observer (at the moment of observation) and $\ve{k}$ is the corresponding unit direction.  
Then, using the same procedure as described in \cite{KopeikinSchaefer1999_Gwinn_Eubanks}, one obtains from  
Eq.~(\ref{second_integral_geodesic_equation}) the following expression for the relativistic time-delay, cf. \cite{Klioner2003a}:  
\begin{eqnarray}
&& c \left(\tau_1 - \tau_0 \right)_{\rm 1PN}  
\nonumber\\
&& = R - \sum\limits_{A=1}^N \ve{k}\cdot 
\left(\Delta \ve{x}^A_{\rm 1PN}\left(\tau_1 + t^{\ast}, \tau_0 + t^{\ast}\right)\right),  
\label{Time_Delay}
\end{eqnarray}

\noindent
where the perturbation terms $\Delta \ve{x}^A_{\rm 1PN}$ are given by Eq.~(\ref{Second_Integration_15}) with (\ref{Second_Integration_20}); note that   
the below standing relation (\ref{Light_Deflection_15}) has also been used.  
In case of $N$ arbitrarily moving monopoles Eq.~(\ref{Time_Delay}) agrees with formula (51) in \cite{KopeikinSchaefer1999} 
up to order ${\cal O}\left(v_A/c\right)$,  
and in case of $N$ quadrupoles at rest Eq.~(\ref{Time_Delay}) agrees with formula (23) in \cite{Zschocke_Klioner}.  

The result in (\ref{Time_Delay}) is valid for $N$ slowly-moving bodies with full mass-multipole structure.  
But even for future highly-precise astrometry missions aiming to determine relativity within the Solar system 
(e.g. ASTROD \cite{Astrod1,Astrod2}, LATOR \cite{Lator1,Lator2}, ODYSSEY \cite{Odyssey}, SAGAS \cite{Sagas}, TIPO \cite{TIPO})  
only the impact of the very few first multipoles could be detected.  
However, the exact determination of these relevant parts of the perturbation terms in (\ref{Time_Delay}) implies 
some remarkable effort, see for instance \cite{Zschocke_Klioner} for the efficient computation of 
the quadrupole-term, and is beyond the scope of our present investigation.

\subsection{Light-deflection} 

Assume, that the observer and the celestial light-source are at rest with respect to the global coordinate system.  
The light-deflection at observers position, $\ve{x}_1 = \ve{x}\left(t_1\right)$, which is assumed to be at rest with respect  
to the global coordinate system, is defined by the unit tangent-vector of the lightray at observers position:  
\begin{eqnarray}
\ve{n}_{\rm 1PN}\left(\tau_1+t^{\ast}\right) &=& \frac{\dot{\ve{x}}_{\rm 1PN}\left(\tau_1+t^{\ast}\right)}
{\left|\dot{\ve{x}}_{\rm 1PN}\left(\tau_1+t^{\ast}\right)\right|}\,. 
\label{Light_Deflection_5}
\end{eqnarray}

\noindent
Using (\ref{first_integral_geodesic_equation}), one obtains 
\begin{eqnarray}
\ve{n}_{\rm 1PN}\left(\tau_1+t^{\ast}\right) \!\!&=&\!\! \ve{\sigma} 
+ \sum\limits_{A=1}^N\ve{\sigma} \times \left(\frac{\Delta \dot{\ve{x}}^A_{\rm 1PN}\left(\tau_1+t^{\ast}\right)}{c} \times \ve{\sigma}\right), 
\nonumber\\
\label{Light_Deflection_10}
\end{eqnarray}
 
\noindent
where the perturbation terms $\Delta \dot{\ve{x}}^A_{\rm 1PN}$ are given by (\ref{First_Integration_21}). 
In case of $N$ arbitrarily moving monopoles our result in (\ref{Light_Deflection_10}) agrees with
Eq.~(69) in \cite{KopeikinSchaefer1999}, and in case of $N$ quadrupoles at rest our result in (\ref{Light_Deflection_10})  
agrees with Eq.~(7) in \cite{Zschocke_Klioner}. One has to bear in mind that for astrometry   
within the near-zone of the Solar system, where the light-sources are at finite distance, one needs to determine the  
light-deflection as function of $\ve{k}$ instead of $\ve{\sigma}$, both of which are related by, cf. \cite{Klioner2003a}:    
\begin{eqnarray}
\ve{\sigma} &=& \ve{k} - \frac{1}{R} \sum\limits_{A=1}^N
\left[\ve{k}\times \bigg(\Delta\ve{x}^A_{\rm 1PN}\left(\tau_1 + t^{\ast}, \tau_0 + t^{\ast}\right)\times\ve{k}\bigg)\right], 
\nonumber\\
\label{Light_Deflection_15}
\end{eqnarray}

\noindent
which follows from (\ref{second_integral_geodesic_equation}) and the definition in (\ref{Vector_R}) and (\ref{Vector_k}).
Inserting (\ref{Light_Deflection_15}) into (\ref{Light_Deflection_10}) yields the expression for the light-deflection,
cf. \cite{Klioner2003a}:
\begin{eqnarray}
\ve{n}_{\rm 1PN}\left(\tau_1+t^{\ast}\right) &=& 
\ve{k} + \sum\limits_{A=1}^N \ve{k} \times \left(\frac{\Delta \dot{\ve{x}}^A_{\rm 1PN}\left(\tau_1+t^{\ast}\right)}{c} \times \ve{k}\right) 
\nonumber\\
\nonumber\\
&& \hspace{-2.5cm} - \frac{1}{R} \sum\limits_{A=1}^N
\left[\ve{k} \times \bigg( \Delta \ve{x}^A_{\rm 1PN} \left(\tau_1 + t^{\ast}, \tau_0 + t^{\ast}\right) \times \ve{k}\bigg)\right]. 
\label{Light_Deflection_20}
\end{eqnarray}

\noindent
In case of quadrupoles at rest our result in (\ref{Light_Deflection_20}) agrees with Eq.~(14) in \cite{Zschocke_Klioner}.  
Let us notice here that in order to determine the unit tangent vector of the lightray at observers position,  
one needs to ascertain both the term $\Delta \dot{\ve{x}}^A_{\rm 1PN}$ as well as $\Delta \ve{x}^A_{\rm 1PN}$,  
which are given by (\ref{First_Integration_21}) and (\ref{Second_Integration_15}) with (\ref{Second_Integration_20}), respectively.  

The formulae in (\ref{Light_Deflection_10}) and (\ref{Light_Deflection_20})
determine the light-deflection in the field of $N$ arbitrarily moving
massive bodies with full mass-multipole structure.
Like in case of time-delay, only the very few first  
multipoles in (\ref{Light_Deflection_10}) or (\ref{Light_Deflection_20}) have to be taken into account for  
sub-micro-arcsecond astrometry. But such an exact determination of the relevant multipoles 
implies some considerable amount of effort, see for instance \cite{Zschocke_Klioner} for the quadrupole part,  
and will therefore not be on the scope of the present investigation.

\section{Summary and Outlook}\label{Summary_Outlook}

While the precision of astrometric measurements has made an advance from milli-arcsecond  
to micro-arcsecond in the angle-determination of celestial objects, prospective developments  
in nearest future aim at sub-micro-arcsecond or even nano-arcsecond level of accuracy. 
It is clear that such extremely high accuracy implies the precise determination of the light-trajectory $\ve{x}\left(t\right)$  
from the celestial object through the Solar system towards the observer.   
In respect thereof two aspects are of specific importance:  

{\bf (A)} In the region exterior of the massive bodies, the global metric of the Solar system (BCRS coordinates: $ct, \ve{x}$) can be expressed 
in terms of two families of global multipoles \cite{Thorne,Blanchet_Damour1,Blanchet_Damour2,Multipole_Damour_2}: 
global mass-multipoles $m_L$ and global spin-multipoles $s_L$, which define the multipole structure of the Solar system as a whole.  
On the other side, from the theory of relativistic reference systems 
follows that the multipole structure of the gravitational field of some massive body A can
only be defined in a physically meaningful way within the local reference system (GCRS-like coordinates: $cT_A, \ve{X}_A$) co-moving with 
that body. 
In accordance with these requirements, highly precise astrometric measurements appeal for the use of a 
global metric expressed in terms of intrinsic mass-multipoles $M^A_L$ and intrinsic spin-multipoles $S^A_L$ of each individual body.  
Such a metric is provided by the {\it Brumberg-Kopeikin} (BK) formalism \cite{Brumberg1991,BK1,Reference_System1,BK2,BK3,Kopeikin_Efroimsky_Kaplan} 
as well as by the {\it Damour-Soffel-Xu} (DSX) approach \cite{DSX1,DSX2,DSX3,DSX4}, 
originally been introduced for celestial mechanics, and
which have become a part of the IAU resolutions B1.3 (2000) \cite{IAU_Resolution1}.  

{\bf (B)} Another aspect in the theory of light propagation concerns the fact that the massive bodies of the Solar system 
are moving along their worldline $\ve{x}_A\left(t\right)$, which is a highly complicated function because of the 
mutual interaction of the massive bodies. Formally, the worldline of some massive body A can be series-expanded  
around some time-moment $t_A$,
\begin{eqnarray}
\ve{x}_A\left(t\right) &=& \ve{x}_A + \frac{\ve{v}_A}{1!}\left(t - t_A\right)
+ \frac{\ve{a}_A}{2!}\left(t-t_A\right)^2 + {\cal O}\left(\dot{a}_A\right),
\nonumber\\
\label{worldline_summary}
\end{eqnarray}

\noindent
where $\ve{x}_A$, $\ve{v}_A$ and $\ve{a}_A$ are the position, velocity and acceleration of body A at time-moment $t_A$, respectively.  

The expansion (\ref{worldline_summary}) has some drawbacks: 

(i) It implies to introduce an instant of time $t_A$, which remains an open parameter, 
as long as no additional arguments are put forward to identify that parameter with the 
time of closest approach (\ref{time_of_closest_approach_t_0}) or with the retarded time (\ref{retarded_time}). 
But so far, an unique justification of that suggestion exists only for pointlike bodies in arbitrary motion,  
but not for extended bodies in arbitrary motion and expressed in terms of intrinsic multipoles. 

(ii) If the expansion (\ref{worldline_summary}) is implemented into the metric, 
it leads to rather cumbersome expressions when integrating the geodesic equation.
 
(iii) One has also to realize that (\ref{worldline_summary}) is not an expansion in inverse powers of
the speed of light, hence these terms are not necessarily beyond 1PN approximation of geodesic equation.  

These facts make it much preferable to determine the light-trajectory as function of arbitrary worldlines $\ve{x}_A\left(t\right)$,  
that means to determine the light-trajectory in the field of arbitrarily moving massive bodies.
The actual worldline of the massive bodies can finally be concretized and implemented by some Solar system ephemerides; 
e.g. the JPL DE421 \cite{JPL}.  

As outlined in some detail by a brief survey of recent advancements in the theory of light propagation, so far there was no solution derived  
for the light-trajectory in the gravitational field of arbitrarily shaped bodies in arbitrary  
motion and described in terms of their local multipoles. 
According to the IAU recommendations \cite{IAU_Resolution1}, in this investigation the DSX-metric 
has been employed in order to determine the light-trajectory in 1PN approximation in the gravitational  
field of $N$ arbitrarily moving massive bodies with full mass-multipole structure:
\begin{eqnarray}
\ve{x}\left(t\right) &=& \ve{x}_0 + c \left(t-t_0\right) \ve{\sigma} + \Delta \ve{x}_{\rm 1PN}\left(t,t_0\right)
+ {\cal O}\left(c^{-3}\right).
\nonumber\\
\label{Summary_1}
\end{eqnarray}

\noindent
The main results of this investigation are given by Eq.~(\ref{First_Integration_21}) and Eq.~(\ref{Second_Integration_20}). 
These solutions have taken into account both of these issues {\bf (A)} and {\bf (B)} outlined above:    
expression (\ref{First_Integration_21}) represents the first integration of geodesic equation, while expression  
(\ref{Second_Integration_20}) represents the second integration of geodesic equation, that means the light-trajectory in the gravitational field 
of $N$ arbitrarily moving and extended massive bodies and expressed in terms of their intrinsic multipoles.  
Furthermore, it has been shown that the results presented agree in special cases with well-established results in the literature, namely   
monopoles, quadrupoles, and arbitrarily shaped bodies at rest as well as monopoles in arbitrary motion.  

It is clear, that a comprehensive model of light propagation on sub-\muas$\:$ or even nas-level    
of accuracy requires at least the solution of light-trajectory in 1.5PN approximation as well:
\begin{eqnarray}
\ve{x}\left(t\right) &=& \ve{x}_0 + c \left(t-t_0\right) \ve{\sigma} + \Delta \ve{x}_{\rm 1PN}\left(t,t_0\right)
\nonumber\\
&& + \Delta \ve{x}_{\rm 1.5PN}\left(t,t_0\right) + {\cal O}\left(c^{-4}\right).
\label{Summary_5}
\end{eqnarray}

\noindent
For instance, the light-deflection of a grazing ray at Jupiter amounts to be about 
$n_{\rm 1PN}^Q \sim 240\,\mu{\rm as}$ \cite{Klioner2003a,Klioner1991}. Such terms are already implemented in the 1PN solution.  
On the other side, a typical term of 1.5PN approximation would be $n_{\rm 1.5PN}^Q \sim n_{\rm 1PN}^Q\,v_A/c$, which 
in case of Jupiter ($v_A/c \sim 4.5\times 10^{-5}$) yields a light-deflection of about $n_{\rm 1.5PN}^Q \sim 0.01\,\mu{\rm as}$. 
Another typical term of 1.5PN approximation is the light-deflection due to the spin of the massive bodies, which 
have been determined to be about $n_{\rm 1.5PN}^S \sim 0.7\,\mu{\rm as}, 0.2\,\mu{\rm as}$, and $0.04\,\mu{\rm as}$   
for grazing lightrays at Sun, Jupiter, and Saturn, respectively \cite{Klioner2003a,Klioner1991}.  
Moreover, recent investigations \cite{Jan-Meichsner_Diploma_Thesis} have recovered, that the light-deflection
due to the spin-octupole-structure of massive bodies
amounts to be about $0.015$ \muas$\;$ for Jupiter and about $0.006$ \muas$\;$ for Saturn
for grazing rays. Therefore, a model at sub-\muas-level has also to account for higher spin-multipole terms 
which are of 1.5PN order.

Clearly, the post-Newtonian approach allows for astrometry within the boundary of the near-zone of the Solar system,  
$\left|\ve{x}\right| \ll \lambda_{\rm gr} \sim 3\,{\rm parsec}$, while lightrays which originate from sources lying far outside 
of the Solar system are subject of the far-zone astrometry. The perturbations of the light-trajectory in the far-zone of Solar system  
are extremely weak (less than $1$ \muas$\,$ in the light-deflection), but  
might be of relevance for sub-micro-arcsecond astrometry. These effects can be investigated by means of
a matching procedure of two asymptotic solutions (near-zone and far-zone solution) proposed in \cite{KlionerKopeikin1992} and
further elaborated in \cite{Will_2003}, and will be on the scope of a further investigation \cite{Zschocke_in_preparation}.

A further problem concerns the retardation effect due to the finite speed at which gravitational action travels.
It has, however, been elucidated by Eq.~(\ref{global_multipole_expansion_retarded}) that the effect of retardation cannot be taken into account
within 1PN approximation for the lightrays. For this fact, the solution for the light-trajectory in 1PN approximation, 
Eqs.~(\ref{First_Integration_21}) and (\ref{Second_Integration_20}), are functions of the instantaneous distance between the photon and  
massive body, as given by Eq.~(\ref{notation_5b}) or (\ref{notation_15}).  

Furthermore, the light-trajectory in 2PN approximation reads formally  
\begin{eqnarray}
\ve{x}\left(t\right) &=& \ve{x}_0 + c \left(t-t_0\right) \ve{\sigma} + \Delta \ve{x}_{\rm 1PN}\left(t,t_0\right)
\nonumber\\
&& \hspace{-1.2cm} + \Delta \ve{x}_{\rm 1.5PN}\left(t,t_0\right) + \Delta \ve{x}_{\rm 2PN}\left(t,t_0\right) + {\cal O}\left(c^{-5}\right).
\label{Summary_10}
\end{eqnarray}

\noindent
The most dominant post-post-Newtonian correction is the monopole-term, $\Delta \ve{x}^M_{\rm 2PN}$, which 
is well-known for bodies at rest. Following a suggestion in \cite{Klioner_LT}, 
for the case of uniformly moving bodies this term can be obtained by 
an appropriate Lorentz transformation, while for the case of arbitrarily moving bodies the solution  
might be acquired with the aid of sophisticated integration methods mentioned in this article.  
It might even be that some very few terms in 2PN approximation beyond the monopole-term are required for
nano-arcsecond accuracy. Such terms will rapidly decrease with increasing impact parameter $d_A$ of the lightray   
and might only be of relevance for grazing rays, i.e. where $d_A$ equals the radius $R_A$ of the body. 
But for all that, the final level of ambition must include  
a rigorous estimation of such terms, implicating a clear understanding about whether or not some 2PN terms beyond the monopole-term become  
relevant for astrometry on nano-arcsecond level.

\section{Acknowledgment}
This work was supported by the Deutsche Forschungsgemeinschaft (DFG).  


\appendix

\section{Notations}\label{Notation}

Throughout the article the following notations are in use:

\begin{itemize}

\item $G$ is the Newtonian constant of gravitation.

\item $c$ is the vacuum speed of light in flat Minkowski space.

\item Lower case Latin indices $a$, $b$, \dots, $i$, $j$, \dots take values 1,2,3.

\item Lower case Greek indices $\alpha$, $\beta$, \dots, $\mu$, $\nu$, \dots take values 0,1,2,3.

\item $\delta_{ij} = \delta^{ij} = {\rm diag} \left(+1,+1,+1\right)$ is Kronecker delta.

\item The three-dimensional coordinate quantities (''three-vectors'') referred to
the spatial axes of the corresponding reference system are set in
boldface: $\ve{a}$.

\item The contravariant components of ''three-vectors'' are $a^{i} = \left(a^1,a^2,a^3\right)$.

\item The contravariant components of ''four-vectors'' are $a^{\mu} = \left(a^0,a^1,a^2,a^3\right)$.

\item Repeated indices imply the Einstein's summation irrespective of
their positions (e.g. $a^i\,b^i=a^1\,b^1+a^2\,b^2+a^3\,b^3$ and
$a^\alpha\,b^\alpha=a^0\,b^0+a^1\,b^1+a^2\,b^2+a^3\,b^3$).

\item The absolute value (Euclidean norm) of a ''three-vector'' $\ve{a}$ is
denoted as $|\ve{a}|$ or, simply, $a$ and can be computed as
$a=|\ve{a}|=(a^1\,a^1+a^2\,a^2+a^3\,a^3)^{1/2}$.

\item The scalar product of any two ''three-vectors'' $\ve{a}$ and $\ve{b}$
with respect to the Euclidean metric $\delta_{ij}$ is denoted by
$\ve{a}\,\cdot\,\ve{b}$ and can be computed as
$\ve{a}\,\cdot\,\ve{b}=\delta_{ij}\,a^i\,b^j=a^i\,b^i$.

\item The vector product of any two ''three-vectors'' $\ve{a}$ and $\ve{b}$
is designated by $\ve{a}\times\ve{b}$ and can be computed as
$\left(\ve{a}\times\ve{b}\right)^i=\varepsilon_{ijk}\,a^j\,b^k$, where
$\varepsilon_{ijk}=(i-j)(j-k)(k-i)/2$ is the fully antisymmetric
Levi-Civita symbol.

\item The global coordinate system is denoted by lower-case letters: $\left(c t, \ve{x}\right)$.

\item The local coordinate system of a massive body A is denoted by upper-case letters: $\left(c T_A, \ve{X}_A\right)$.

\item The photon trajectory is denoted by $\ve{x}\left(t\right)$.
In order to distinguish the photon's spatial coordinate $\ve{x}\left(t\right)$
from the spatial coordinate $\ve{x}$ of the global system,
the time-dependence of photon's spatial coordinate will
everywhere be shown explicitly throughout the article.

\item The worldline of massive body A is denoted by $\ve{x}_A\left(t\right)$ 
or $\ve{x}_A\left(T_A\right)$. 

\item Partial derivatives in the global coordinate system:   
$\displaystyle \partial_{\mu} = \frac{\partial}{\partial x^{\mu}}$ or
$\displaystyle \partial_{i} = \frac{\partial}{\partial x^{i}}$.

\item Partial derivatives in the local coordinate system of body A:  
$\displaystyle {\cal D}^A_{\alpha} = \frac{\partial}{\partial X_A^{\alpha}}$ or
$\displaystyle {\cal D}^A_{a} = \frac{\partial}{\partial X_A^{a}}$.

\item $n! = n \left(n-1\right)\left(n-2\right)\cdot\cdot\cdot 2 \cdot 1$ is the faculty for positive integer;
$0! = 1$.

\item $L=i_1 i_2 ...i_l$ is a Cartesian multi-index of a given tensor $T$, that means
$T_L \equiv T_{i_1 i_2 \,.\,.\,.\,i_l}$, and each index $i_1,i_2,...,i_l$ runs from $1$ to $3$  
(i.e. over the Cartesian coordinate label).

\item Two identical multi-indices imply summation, e.g.:
$\partial_L\,T_L \equiv \sum\limits_{i_1\,.\,.\,.\,i_l}\,\partial_{i_1\,.\,.\,.\,i_l}\,T_{i_1\,.\,.\,.\,i_l}$.
\item The symmetric part of a Cartesian tensor $T_L$ is, cf. Eq.~(2.1) in \cite{Thorne}:
\begin{eqnarray}
T_{\left(L\right)} &=& T_{\left(i_1 ... i_l \right)} = \frac{1}{l!} \sum\limits_{\sigma}
A_{i_{\sigma\left(1\right)} ... i_{\sigma\left(l\right)}}\,,
\end{eqnarray}

\noindent
where $\sigma$ is running over all permutations of $\left(1,2,...,l\right)$.

\item The symmetric tracefree (STF) part of a Cartesian tensor $T_L$ (notation:
$\hat{T}_L \equiv \underset{L}{\rm STF}\,T_L$) is, cf. Eq.~(2.2) in \cite{Thorne}:
\begin{eqnarray}
&& \hspace{0.3cm} \hat{T}_L = \sum_{k=0}^{\left[l/2\right]} a_{l k}\,\delta_{(i_1 i_2 ...} \delta_{i_{2k-1} i_{2k}}\,
S_{i_{2k+1 ... i_l) \,a_1 a_1 ... a_k a_k}}\,,
\nonumber\\ 
\label{anti_symmetric_1}
\end{eqnarray}

\noindent
where $\left[l/2\right]$ means the largest integer less than or equal to $l/2$, and $S_L \equiv T_{\left(L\right)}$
abbreviates the symmetric part of tensor $T_L$. For instance, $T_L^{\alpha\beta}$ means STF with respect to indices $L$
but not with respect to indices $\alpha,\beta$. The coefficient in (\ref{anti_symmetric_1}) is given by
\begin{eqnarray}
a_{l k} &=& \left(-1\right)^k \frac{l!}{\left(l - 2 k\right)!}\,
\frac{\left(2 l - 2 k - 1\right)!!}{\left(2 l - 1\right)!! \left(2k\right)!!}\,.
\label{coefficient_anti_symmetric}
\end{eqnarray}

\noindent
As instructive examples of (\ref{anti_symmetric_1}) let us consider the cases $l=2$ and $l=3$:
\begin{eqnarray}
&& \hspace{0.3cm} \hat{T}_{ij} = T_{\left(ij\right)} - \frac{1}{3}\,\delta_{ij}\,T_{ss}\,,
\label{anti_symmetric_2}
\\
\nonumber\\
&& \hspace{0.3cm} \hat{T}_{ijk} = T_{\left(ijk\right)} - \frac{1}{5}
\left(\delta_{ij} T_{\left(kss\right)} + \delta_{jk} T_{\left(iss\right)} + \delta_{ki} T_{\left(jss\right)}\right).
\nonumber\\
\label{anti_symmetric_3}
\end{eqnarray}

\noindent
Throughout the article, the 'hat' will be omitted for the multipoles, $M^A_L \equiv \hat{M}^A_L$, $m_L \equiv \hat{m}_L$,  
$S^A_L \equiv \hat{S}^A_L$, $s_L \equiv \hat{s}_L$, but kept 
for spatial-coordinates $\hat{x}_L$. 

\end{itemize}

\section{Integral ${\cal I}_A$}\label{Appendix_Integral_A}

The integral ${\cal I}_A$ in (\ref{Integration_A}) reads: 
\begin{eqnarray}
{\cal I}_A\left(\tau+t^{\ast},\ve{\xi}\right) &=& \int\limits_{-\infty}^{\tau}\,d c\tau^{\prime}\, 
M_L^A\left(\tau^{\prime}+t^{\ast}\right)\;\partial^{\prime}_L\,
\frac{1}{r^{\rm N}_A \left(\tau^{\prime}+t^{\ast}\right)}\,.   
\nonumber\\
\label{Appendix_First_Integration_1}
\end{eqnarray}

\noindent
In order to determine the integral ${\cal I}_A$, it is useful to incorporate the operator $\displaystyle P^{ij}\,\frac{\partial}{\partial \xi^j}$  
which stands in front of this integral according to Eq.~(\ref{First_Integration_5}).  
Furthermore, using expression in (\ref{Transformation_Derivative}) for the differential operator $\partial^{\prime}_L$, 
the integral ${\cal I}_A$ can be separated into two kind of integrals: 
integral ${\cal I}_1$ which contains differentiations with respect to time-variable (i.e. $p \ge 1$) and 
integral ${\cal I}_2$ which does not contain such differentiations (i.e. $p=0$), that means:  
\begin{eqnarray}
P^{ij}\,\frac{\partial}{\partial \xi^j}\,{\cal I}_A\left(\tau+t^{\ast},\ve{\xi}\right) &=& 
\sum\limits_{p=1}^{l} \frac{l!}{\left(l-p\right)!\,p!}\,
\sigma^{i_1}\,...\,\sigma^{i_p}\,
\nonumber\\
\nonumber\\
&& \hspace{-4.0cm} \times P^{i_{p+1}\,j_{p+1}}\,...\,P^{i_l\,j_l}\,
\frac{\partial}{\partial \xi^{j_{p+1}}}\,...\,\frac{\partial}{\partial \xi^{j_{l}}}\,
P^{ij}\,\frac{\partial}{\partial \xi^j}\,{\cal I}_1 \left(\tau+t^{\ast},\ve{\xi}\right) 
\nonumber\\
\nonumber\\
&& \hspace{-4.0cm} + P^{i_1j_1}\,...\,P^{i_l j_l}  
\frac{\partial}{\partial \xi^{j_1}}\,...\,\frac{\partial}{\partial \xi^{j_l}}\, 
P^{ij}\,\frac{\partial}{\partial \xi^j}\,{\cal I}_2\left(\tau+t^{\ast},\ve{\xi}\right).  
\label{Appendix_First_Integration_10}
\end{eqnarray}
 
\noindent
The integral ${\cal I}_1$, with the differential operation $\displaystyle P^{ij}\frac{\partial}{\partial \xi^j}$ in front, is given by 
\begin{eqnarray}
&& P^{ij}\,\frac{\partial}{\partial \xi^j}\,{\cal I}_1\left(\tau+t^{\ast},\ve{\xi}\right)  
\nonumber\\
&=& P^{ij}\,\frac{\partial}{\partial \xi^j}\,\int\limits_{-\infty}^{\tau}\,d c\tau^{\prime}\;M^A_L \left(\tau^{\prime}+t^{\ast}\right)
\left(\frac{\partial}{\partial c\tau^{\prime}}\right)^p
\frac{1}{r^{\rm N}_A \left(\tau^{\prime}+t^{\ast}\right)}
\nonumber\\
\nonumber\\
&=& P^{ij}\,\frac{\partial}{\partial \xi^j}\,M^A_L \left(\tau+t^{\ast}\right)\,\left(\frac{\partial}{\partial c\tau}\right)^{p-1}
\frac{1}{r^{\rm N}_A\left(\tau + t^{\ast}\right)}
\nonumber\\
\nonumber\\
&& + {\cal O}\left(\frac{\dot{M}_L^A}{c}\right)\,. 
\label{Integral_B_1}
\end{eqnarray}

\noindent
The integral in (\ref{Integral_B_1}) has been integrated by part, using   
the integration rule for integrals along the unperturbed lightray as given  
by Eq. (4.9) in \cite{KopeikinKorobkovPolnarev2006}. It is important to note that the neglected terms are of the  
order $\dot{M}_L^A/c$ and, therefore, they are beyond 1PN approximation because they imply an additional factor $c^{-1}$.  
In view of (\ref{Integral_A_1}) the proof of this fact is rather simple.  

The integral ${\cal I}_2$,with the differential operation $\displaystyle P^{ij}\frac{\partial}{\partial \xi^j}$ in front, is given by
\begin{eqnarray}
&& P^{ij}\frac{\partial}{\partial \xi^j}\,{\cal I}_2\left(\tau+t^{\ast},\ve{\xi}\right) = 
P^{ij}\frac{\partial}{\partial \xi^j} \int\limits_{-\infty}^{\tau} d c\tau^{\prime} 
\frac{M^A_L\left(\tau^{\prime}+t^{\ast}\right)}{r^{\rm N}_A\left(\tau^{\prime}+t^{\ast}\right)}
\nonumber\\
\nonumber\\
&=& - M^A_L\left(\tau+t^{\ast}\right)
P^{ij}\,\frac{\partial \ln \left[r^{\rm N}_A\left(\tau+t^{\ast}\right) - \ve{\sigma}\cdot\ve{r}^{\rm N}_A\left(\tau+t^{\ast}\right)\right]}
{\partial \xi^j} 
\nonumber\\
\nonumber\\
&& + {\cal O} \left(\frac{\dot{M}^A_L}{c}\right) + {\cal O}\left(\frac{v_A}{c}\right), 
\label{Integral_A_1}
\end{eqnarray}

\noindent
where for the lower integration limit we have used, 
\begin{eqnarray}
\lim_{\tau\rightarrow-\infty} P^{ij}\,\frac{\partial}{\partial \xi^j}
\ln \left[r^{\rm N}_A\left(\tau+t^{\ast}\right) - \ve{\sigma}\cdot\ve{r}^{\rm N}_A\left(\tau+t^{\ast}\right)\right] &=& 0\,.
\nonumber\\
\end{eqnarray}

\noindent
Let us note that the physical dimension of a length in the argument of the logarithm in (\ref{Integral_A_1}) is not a problem at all 
and has to be treated according to Eq.~(\ref{First_Integration_20}).
The integral in (\ref{Integral_A_1}) has been integrated by parts, using:
\begin{eqnarray}
\frac{1}{r^{\rm N}_A\left(\tau^{\prime}+t^{\ast}\right)} &=& 
- \frac{\partial \ln\left[r^{\rm N}_A\left(\tau^{\prime}+t^{\ast}\right)-\ve{\sigma}\cdot\ve{r}^{\rm N}_A\left(\tau^{\prime}+t^{\ast}\right)\right]}
{\partial\,c \tau^{\prime}} 
\nonumber\\
&& + {\cal O}\left(\frac{v_A}{c}\right),   
\label{integration_by_parts_A}
\end{eqnarray}

\noindent
where the terms proportional to $v_A/c$ in (\ref{integration_by_parts_A}) will be given later; see Eq.~(\ref{Proof1_A}). 
The fact that the neglected terms (\ref{Integral_A_1}) are beyond 1PN approximation is evidenced in appendix \ref{Proof1}.

\section{Integrals ${\cal I}_{C}$, ${\cal I}_{D}$, ${\cal I}_{E}$, ${\cal I}_{F}$}\label{Appendix_Integral_C}

The four integrals in Eqs.~(\ref{Integral_I_C}) - (\ref{Integral_I_F}) will be determined;  
in what follows the time-arguments $\tau+t^{\ast}$ and $\tau_0+t^{\ast}$ of these integrals are omitted for simpler notation.
In the calculation of the integrals, all terms are neglected which are proportional to either $v_A/c$ or $\dot{M}_A^L/c$, 
because they are of higher-order beyond 1PN approximation. The proof for these assertions will not be given explicitly, because 
they go very similar as the example elaborated in appendix \ref{Proof1}.

\subsection{Integral ${\cal I}_{C}$}

The integral ${\cal I}_{C}$ reads:
\begin{eqnarray}
{\cal I}_{C} &=& \int\limits_{\tau_0}^{\tau}\,d c\tau^{\prime}\,
\frac{M_L^A\left(\tau^{\prime}+t^{\ast}\right)}{r^{\rm N}_A\left(\tau^{\prime}+t^{\ast}\right)}\,. 
\label{Appendix_Integral_C_5}
\end{eqnarray}

\noindent
This integral occurs in the first and fourth term of Eq.~(\ref{Second_Integration_10}).

\subsection{Integral ${\cal I}_{C}$ for the case $l=0$} 

Let us first consider the integral (\ref{Appendix_Integral_C_5}) for the case $l=0$, which occurs 
in the fourth term in (\ref{Second_Integration_10}). One obtains, by means of relation (\ref{integration_by_parts_A}), the following solution:   
\begin{eqnarray}
{\cal I}_{C}^{l=0} &=& \int\limits_{\tau_0}^{\tau}\,d c\tau^{\prime}\,
\frac{M_A}{r^{\rm N}_A\left(\tau^{\prime}+t^{\ast}\right)}
\nonumber\\
\nonumber\\
&=& - M_A\,\ln \frac{r^{\rm N}_A\left(\tau+t^{\ast}\right)-\ve{\sigma}\cdot\ve{r}^{\rm N}_A\left(\tau+t^{\ast}\right)}
{r^{\rm N}_A\left(\tau_0+t^{\ast}\right)-\ve{\sigma}\cdot\ve{r}^{\rm N}_A\left(\tau_0+t^{\ast}\right)}  
\nonumber\\
\nonumber\\
&& + {\cal O}\left(\frac{v_A}{c}\right). 
\label{Appendix_Integral_C_A}
\end{eqnarray}

\subsection{Integral ${\cal I}_{C}$ for the case $l \ge 1$} 

Now we consider the integral (\ref{Appendix_Integral_C_5}) for the case $l\ge 1$, which occurs in the 
first and fourth term in (\ref{Second_Integration_10}). In this case, we always have the differential operation 
$\displaystyle P^{ij}\frac{\partial}{\partial \xi^j}$ in front, 
\begin{eqnarray}
P^{ij}\frac{\partial}{\partial \xi^j}\,{\cal I}_{C} &=& 
P^{ij}\frac{\partial}{\partial \xi^j} \int\limits_{\tau_0}^{\tau}\,d c\tau^{\prime}\,
\frac{M_L^A\left(\tau^{\prime}+t^{\ast}\right)}{r^{\rm N}_A\left(\tau^{\prime}+t^{\ast}\right)}\,.
\label{Appendix_Integral_C_B}
\end{eqnarray}

\noindent
For evaluating this integral we can use the result in (\ref{Integral_A_1}), and obtain:  
\begin{widetext} 
\begin{eqnarray}
P^{ij}\frac{\partial}{\partial \xi^j}\,{\cal I}_{C}
&=& - M^A_L\left(\tau+t^{\ast}\right)
P^{ij}\,\frac{\partial \ln \left[r^{\rm N}_A\left(\tau+t^{\ast}\right) - \ve{\sigma}\cdot\ve{r}^{\rm N}_A\left(\tau+t^{\ast}\right)\right]}
{\partial \xi^j}
\nonumber\\
\nonumber\\
&& + M^A_L\left(\tau_0+t^{\ast}\right)
P^{ij}\,\frac{\partial \ln \left[r^{\rm N}_A\left(\tau_0+t^{\ast}\right) - \ve{\sigma}\cdot\ve{r}^{\rm N}_A\left(\tau_0+t^{\ast}\right)\right]}
{\partial \xi^j}
+ {\cal O} \left(\frac{\dot{M}^A_L}{c}\right) + {\cal O}\left(\frac{v_A}{c}\right). 
\label{Appendix_Integral_C_D}
\end{eqnarray}
\end{widetext}

\subsection{Integral ${\cal I}_{D}$}

According to expression (\ref{Second_Integration_10}), the differential operation $\displaystyle P^{ij}\frac{\partial}{\partial \xi^j}$ is always in front 
of the integral ${\cal I}_{D}$ ($p \ge 2$), so we may consider:  
\begin{widetext} 
\begin{eqnarray}
P^{ij}\frac{\partial}{\partial \xi^j}\,{\cal I}_{D} &=& P^{ij}\frac{\partial}{\partial \xi^j}\int\limits_{\tau_0}^{\tau}\,d c\tau^{\prime}\, 
M_L^A\left(\tau^{\prime}+t^{\ast}\right) \left(\frac{\partial}{\partial c\tau^{\prime}}\right)^{p-1} 
\frac{1}{r^{\rm N}_A\left(\tau^{\prime}+t^{\ast}\right)} 
\nonumber\\
&=& + M_L^A\left(\tau+t^{\ast}\right)\,\left(\frac{\partial}{\partial c\tau}\right)^{p-2}\,P^{ij}\frac{\partial}{\partial \xi^j}\, 
\frac{1}{r^{\rm N}_A\left(\tau+t^{\ast}\right)}  
- M_L^A\left(\tau_0+t^{\ast}\right)\,\left(\frac{\partial}{\partial c\tau_0}\right)^{p-2}\,P^{ij}\frac{\partial}{\partial \xi^j}\, 
\frac{1}{r^{\rm N}_A\left(\tau_0+t^{\ast}\right)} 
\nonumber\\
\nonumber\\
&& + {\cal O}\left(\frac{\dot{M}_L^A}{c}\right),   
\label{Appendix_Integral_C_10}
\end{eqnarray}
\end{widetext} 

\noindent
which has been solved using integration by parts. The proof that the correction terms are in fact of the order ${\cal O}\left(\dot{M}_L^A/c\right)$ 
is straightforward.

\subsection{Integral ${\cal I}_{E}$}

Now we consider the integral (\ref{Integral_I_E}). According to (\ref{Second_Integration_10}), the differential operation 
$\displaystyle P^{ij}\frac{\partial}{\partial \xi^j}$ is always in front of the integral ${\cal I}_{E}$ ($p \ge 1$), so we consider:
\begin{eqnarray}
P^{ij}\frac{\partial}{\partial \xi^j}\,{\cal I}_{E} &=& 
P^{ij}\frac{\partial}{\partial \xi^j}\,\int\limits_{\tau_0}^{\tau}\,d c\tau^{\prime}\,
M_L^A\left(\tau^{\prime}+t^{\ast}\right)\,
\nonumber\\
&& \hspace{-1.0cm} \times \ln \left[r^{\rm N}_A\left(\tau^{\prime}+t^{\ast}\right) - \ve{\sigma}\cdot\ve{r}^{\rm N}_A\left(\tau^{\prime}+t^{\ast}\right)\right]\,. 
\label{Appendix_Integral_E_5}
\end{eqnarray}

\noindent
In order to solve that integral, we may use the following relation: 
\begin{widetext}
\begin{eqnarray}
\ln \left(r^{\rm N}_A\left(\tau^{\prime}+t^{\ast}\right) - \ve{\sigma}\cdot\ve{r}^{\rm N}_A\left(\tau^{\prime}+t^{\ast}\right)\right)
&=& \frac{\partial \left[r^{\rm N}_A\left(\tau^{\prime}+t^{\ast}\right) + \ve{\sigma}\cdot\ve{r}^{\rm N}_A\left(\tau^{\prime}+t^{\ast}\right)
\ln \left(r^{\rm N}_A\left(\tau^{\prime}+t^{\ast}\right) - \ve{\sigma}\cdot\ve{r}^{\rm N}_A\left(\tau^{\prime}+t^{\ast}\right)\right)\right]}
{\partial c\tau^{\prime}}
\nonumber\\
&& + {\cal O}\left(\frac{v_A}{c}\right).
\label{Appendix_Integral_C_16}
\end{eqnarray}
\end{widetext}

\noindent
Like in relation (\ref{integration_by_parts_A}), the form of the expressions proportional to $v_A/c$ 
in (\ref{Appendix_Integral_C_16}) can easily be determined.
By inserting relation (\ref{Appendix_Integral_C_16}) into the integral (\ref{Appendix_Integral_E_5}), one obtains by integration by parts:  
\begin{widetext} 
\begin{eqnarray}
P^{ij}\frac{\partial}{\partial \xi^j}\,{\cal I}_{E} 
&=& + M_L^A\left(\tau+t^{\ast}\right)\,P^{ij}\frac{\partial}{\partial \xi^j}\, 
\left[r^{\rm N}_A\left(\tau+t^{\ast}\right) + \ve{\sigma}\cdot\ve{r}^{\rm N}_A\left(\tau+t^{\ast}\right) 
\ln \left(r^{\rm N}_A\left(\tau+t^{\ast}\right) - \ve{\sigma}\cdot\ve{r}^{\rm N}_A\left(\tau+t^{\ast}\right)\right) \right] 
\nonumber\\
\nonumber\\
&& - M_L^A\left(\tau_0+t^{\ast}\right)\,P^{ij}\frac{\partial}{\partial \xi^j}\, 
\left[r^{\rm N}_A\left(\tau_0+t^{\ast}\right) + \ve{\sigma}\cdot\ve{r}^{\rm N}_A\left(\tau_0+t^{\ast}\right)
\ln \left(r^{\rm N}_A\left(\tau_0+t^{\ast}\right) - \ve{\sigma}\cdot\ve{r}^{\rm N}_A\left(\tau_0+t^{\ast}\right)\right) \right]   
\nonumber\\
\nonumber\\
&& + {\cal O}\left(\frac{\dot{M}_L^A}{c}\right) + {\cal O}\left(\frac{v_A}{c}\right). 
\label{Appendix_Integral_C_15}
\end{eqnarray}
\end{widetext} 

\noindent
The proof that the neglected terms are in fact of the order $v_A/c$ goes very similar to the example elaborated in appendix \ref{Proof1}.  

\subsection{Integral ${\cal I}_{F}$}

Now we consider the integral (\ref{Integral_I_F}). According to (\ref{Second_Integration_10}), at least one differential operation
of the form $\displaystyle P^{ij}\frac{\partial}{\partial \xi^j}$ is always in front of the integral ${\cal I}_{F}$ ($p \ge 1$), so we consider:
\begin{widetext} 
\begin{eqnarray}
P^{ij}\frac{\partial}{\partial \xi^j}\,{\cal I}_{F} &=& \int\limits_{\tau_0}^{\tau}\,d c\tau^{\prime}\,
M_L^A\left(\tau^{\prime}+t^{\ast}\right) \left(\frac{\partial}{\partial c\tau^{\prime}}\right)^{p}
\frac{1}{r^{\rm N}_A\left(\tau^{\prime}+t^{\ast}\right)}
\nonumber\\
\nonumber\\
&=& + M_L^A\left(\tau+t^{\ast}\right)\,\left(\frac{\partial}{\partial c\tau}\right)^{p-1}\,P^{ij}\frac{\partial}{\partial \xi^j}\, 
\frac{1}{r^{\rm N}_A\left(\tau+t^{\ast}\right)}
- M_L^A\left(\tau_0+t^{\ast}\right)\,\left(\frac{\partial}{\partial c\tau_0}\right)^{p-1}\,P^{ij}\frac{\partial}{\partial \xi^j}\, 
\frac{1}{r^{\rm N}_A\left(\tau_0+t^{\ast}\right)}
\nonumber\\
\nonumber\\
&& + {\cal O}\left(\frac{\dot{M}_L^A}{c}\right), 
\label{Appendix_Integral_C_20}
\end{eqnarray}
\end{widetext} 

\noindent
which has been solved using integration by parts.

\section{Estimation of neglected terms: an example\label{Proof1}}

As a typical example, let us consider the neglected terms in the solution (\ref{Integral_A_1}), where
the relation (\ref{integration_by_parts_A}) has been used, which in its exact form reads  
(the variables $\tau^{\prime} + t^{\ast}$ will be suppressed for simpler notation):
\begin{eqnarray}
\frac{1}{r^{\rm N}_A} &=&
- \frac{\partial \ln \left[r^{\rm N}_A-\ve{\sigma}\cdot\ve{r}^{\rm N}_A\right]}{\partial\,c \tau^{\prime}}
+ \frac{1}{r^{\rm N}_A}\,\frac{\ve{v}_A}{c} \cdot \frac{r^{\rm N}_A\,\ve{\sigma} - \ve{r}_A^{\rm N}}{r^{\rm N}_A - \ve{\sigma} \cdot \ve{r}_A^{\rm N}}\,.
\nonumber\\
\label{Proof1_A}
\end{eqnarray}

\noindent
Inserting this relation into (\ref{Integral_A_1}) yields an additional integral proportional to $v_A/c$, namely:
\begin{eqnarray}
P^{ij}\frac{\partial}{\partial \xi^j} \int\limits_{-\infty}^{\tau} d c\tau^{\prime}
\frac{M^A_L}{r^{\rm N}_A}\,\frac{\ve{v}_A}{c} \cdot \left(\ve{\sigma} - \frac{\ve{d}_A}{r^{\rm N}_A - \ve{\sigma}\cdot \ve{r}_A^{\rm N}}\right),
\label{Proof1_B}
\end{eqnarray}

\noindent
where $\ve{r}_A^{\rm N} = \ve{d}_A + \ve{\sigma} \left(\ve{\sigma} \cdot \ve{r}_A^{\rm N}\right)$ has been used.
The first term of this integral is identical to the integral (\ref{Integral_A_1}), except the additional factor $\ve{\sigma}\cdot \ve{v}_A/c$.
So it remains to consider the second term in (\ref{Proof1_B}); the sign in front is not relevant here:
\begin{eqnarray}
{\cal I}_{G} &=& P^{ij}\,\frac{\partial}{\partial \xi^j} \int\limits_{-\infty}^{\tau} d c\tau^{\prime}
\frac{M^A_L}{r^{\rm N}_A}\,\frac{\ve{v}_A}{c} \cdot \frac{\ve{d}_A}{r^{\rm N}_A - \ve{\sigma}\cdot \ve{r}_A^{\rm N}}\,.
\label{Proof1_C}
\end{eqnarray}

\noindent
Using relation (\ref{dipole_35}), one can rewrite this integral in the following form:
\begin{eqnarray}
{\cal I}_{G} &=& P^{ij}\,\frac{\partial}{\partial \xi^j} P^{ab}\,\frac{\partial}{\partial \xi^b} \int\limits_{-\infty}^{\tau} d c\tau^{\prime}
M^A_L\,\frac{\ve{v}^a_A}{c}\,\ln \left[r^{\rm N}_A - \ve{\sigma}\cdot \ve{r}_A^{\rm N}\right].
\nonumber\\
\label{Proof1_D}
\end{eqnarray}

\noindent
Using relation (\ref{Appendix_Integral_C_16}), this integral can be integrated by parts:
\begin{eqnarray}
{\cal I}_{G} &=& M^A_L\,\frac{\ve{v}^a_A}{c}\,P^{ij}\,\frac{\partial}{\partial \xi^j} P^{ab}\,\frac{\partial}{\partial \xi^b}
\nonumber\\
&& \hspace{-0.5cm} \times \bigg[r^{\rm N}_A + \ve{\sigma}\cdot \ve{r}_A^{\rm N}\,\ln \left(r^{\rm N}_A - \ve{\sigma}\cdot \ve{r}_A^{\rm N}\right)\bigg]
\Bigg|_{- \infty}^{\tau} + {\cal O}\left(c^{-2}\right).
\nonumber\\
\label{Proof1_E}
\end{eqnarray}

\noindent
Performing the differentiations, one finally arrives at
\begin{eqnarray}
{\cal I}_{G} &=& M^A_L\,\frac{v^a_A}{c}\,\frac{1}{r^{\rm N}_A - \ve{\sigma}\cdot \ve{r}_A^{\rm N}}\,
\left(P^{ai} - \frac{d_A^a\,d_A^i}{r^{\rm N}_A\left(r^{\rm N}_A - \ve{\sigma}\cdot \ve{r}_A^{\rm N}\right)}\right),
\nonumber\\
\label{Proof1_F}
\end{eqnarray}

\noindent
up to terms of the order ${\cal O}\left(c^{-2}\right)$, and the absolute value can be estimated by
\begin{eqnarray}
\left| {\cal I}_{G} \right| &\le& 2\,M^A_L\,\frac{v_A}{c}\,\frac{1}{r^{\rm N}_A - \ve{\sigma}\cdot \ve{r}_A^{\rm N}}\,.
\label{Proof1_G}
\end{eqnarray}

\noindent
As stated in relation (\ref{Integral_A_1}), the expression in (\ref{Proof1_F}) is of the order $v_A/c$, hence beyond 1PN approximation. The fact
that in extreme astrometric configurations, $\ve{\sigma} \cdot \ve{r}_A^{\rm N} \rightarrow r^{\rm N}_A$, the expression
in (\ref{Proof1_F}) becomes formally large is not of much relevance, since there are many other terms of the order $v_A/c$ which presumably cancel
this term. The proof of such an assertion is, of course, beyond 1PN approximation and involves an exact consideration of all terms to that order.

\section{Partial derivatives for the first integration}\label{Appendix_Derivatives_1}

Throughout this section we will use the following abbreviatory notation: 
$\ve{r}^{\rm N}_A=\ve{r}^{\rm N}_A\left(\tau+t^{\ast}\right)$, $\ve{d}_A = \ve{d}_A\left(\tau+t^{\ast}\right)$, and 
corresponding notation for their absolute values.  

\subsection{Example} 

Let us consider an example on how the differentiation is meant within the formalism: 
\begin{eqnarray}
&& P^{i_1 j_1}\,\frac{\partial}{\partial \xi^{j_1}}\,\frac{1}{r^{\rm N}_A} 
\nonumber\\
&=& P^{i_1 j_1} \frac{\partial}{\partial \xi^{j_1}} 
\frac{1}{\sqrt{\ve{\xi}^2+c^2 \tau^2+\ve{x}_A^2-2 \ve{\xi}\cdot\ve{x}_A - 2 c \tau \ve{\sigma}\cdot\ve{x}_A}}, 
\nonumber\\
\label{Example_5}
\end{eqnarray}
 
\noindent
where (\ref{notation_15}) has been used; recall $\ve{\xi}\cdot\ve{\sigma}=0$ 
and $\ve{x}_A = \ve{x}_A\left(\tau+t^{\ast}\right)$.  
Inserting the projector (\ref{variable_3}), one finds 
\begin{eqnarray}
P^{i_1 j_1}\,\frac{\partial}{\partial \xi^{j_1}}\,\frac{1}{r^{\rm N}_A} &=&
- \frac{\xi^{i_1} - x_A^{i_1} + \sigma^{i_1} \left(\ve{\sigma}\cdot\ve{x}_A\right)}
{\left|\ve{\xi} + c\,\tau\,\ve{\sigma} - \ve{x}_A\right|^3} 
\nonumber\\
\nonumber\\
&& \hspace{-1.0cm} = - \frac{\xi^{i_1} + c\,\tau\,\sigma^{i_1} - x_A^{i_1} - \sigma^{i_1} \left(c\,\tau - \ve{\sigma}\cdot\ve{x}_A\right)}
{\left|\ve{\xi} + c\,\tau\,\ve{\sigma} - \ve{x}_A\right|^3}\,.  
\nonumber\\
\label{Example_10}
\end{eqnarray}

\noindent
In view of $\ve{\sigma}\cdot\ve{\xi} = 0$, the following term in the nominator can be rewritten as follows: 
$c\,\tau - \ve{\sigma}\cdot\ve{x}_A = \ve{\sigma}\cdot\left(\ve{\xi} + c\,\tau\,\ve{\sigma} - \ve{x}_A\right) = \ve{\sigma}\cdot\ve{r}_A^{\rm N}$. 
Then, by using the definition of impact vector (\ref{First_Integration_22}), we finally arrive at:  
\begin{eqnarray}
P^{i_1 j_1}\,\frac{\partial}{\partial \xi^{j_1}}\,\frac{1}{r^{\rm N}_A} &=&  
- \frac{d^{i_1}_A}{\left(r^{\rm N}_A\right)^3}\,.    
\label{Example_15}
\end{eqnarray}

\noindent
All subsequent derivatives have been determined in similar way.

\subsection{Partial derivatives for dipole-term}

In order to obtain the dipole-term in (\ref{dipole_5}) we need the following derivatives:
\begin{eqnarray}
P^{i_1 j_1}\,\frac{\partial}{\partial \xi^{j_1}}\,\frac{d_A^i}{r^{\rm N}_A - \ve{\sigma}\cdot\ve{r}^{\rm N}_A}\,\frac{1}{r^{\rm N}_A}
&=& \frac{1}{r^{\rm N}_A\left(r^{\rm N}_A - \ve{\sigma}\cdot\ve{r}^{\rm N}_A\right)}
\nonumber\\ 
&& \hspace{-4.0cm} \times \left(P^{i_1 i}-\frac{d_A^{i_1}\,d_A^{i}}{\left(r^{\rm N}_A\right)^2} 
- \frac{d_A^{i_1}\,d_A^{i}}{r^{\rm N}_A\left(r^{\rm N}_A - \ve{\sigma}\cdot\ve{r}^{\rm N}_A\right)}\right),
\label{dipole_10}
\end{eqnarray}

\noindent
and
\begin{eqnarray}
P^{i_1 j_1}\,\frac{\partial}{\partial \xi^{j_1}}\,\frac{1}{r^{\rm N}_A} 
&=& - \frac{d_A^{i_1}}{\left(r^{\rm N}_A\right)^3}\,,
\label{dipole_11}
\\  
\frac{\partial}{\partial c\tau}\,\frac{1}{r^{\rm N}_A} &=& - \frac{\ve{\sigma}\cdot\ve{r}^{\rm N}_A}{\left(r^{\rm N}_A\right)^3} 
+ {\cal O}\left(\frac{v_A}{c}\right).  
\label{dipole_12}
\end{eqnarray}

\subsection{Partial derivatives for quadrupole-term} 

In (\ref{Comparison_15}) the derivatives 
$\displaystyle \frac{\partial M_L^A}{\partial c \tau} = {\cal O}\left(\frac{\dot{M}_L^A}{c}\right)$ and   
$\displaystyle \frac{\partial d_A^i}{\partial c\tau} = {\cal O}\left(\frac{v_A}{c}\right)$ 
are beyond 1PN approximation. Hence, we are left with the following expressions:  
\begin{eqnarray}
P^{i_2 j_2}\,\frac{\partial}{\partial \xi^{j_2}}\,\frac{d_A^i}{\left(r^{\rm N}_A\right)^3}
&=& - 3\,\frac{d_A^i\,d_A^{i_2}}{\left(r^{\rm N}_A\right)^5} + \frac{P^{i i_2}}{\left(r^{\rm N}_A\right)^3}\,,  
\label{Comparison_20}
\end{eqnarray}

\noindent
and  
\begin{widetext}
\begin{eqnarray}
&& P^{i_1 j_1}\,P^{i_2 j_2}\,\frac{\partial}{\partial \xi^{j_1}}\,
\frac{\partial}{\partial \xi^{j_2}}\,\frac{d_A^i}{r^{\rm N}_A - \ve{\sigma}\cdot\ve{r}^{\rm N}_A}\,
\frac{1}{r^{\rm N}_A} 
\nonumber\\
\nonumber\\
&=& - P^{i_1 i_2}\,\frac{d_A^i}{r^{\rm N}_A-\ve{\sigma}\cdot\ve{r}^{\rm N}_A}\,
\frac{1}{\left(r^{\rm N}_A\right)^2} \left(\frac{1}{r^{\rm N}_A} + \frac{1}{r^{\rm N}_A-\ve{\sigma}\cdot\ve{r}^{\rm N}_A}\right)
- \frac{P^{i i_1}}{\left(r^{\rm N}_A\right)^2}\,\frac{d_A^{i_2}}{r^{\rm N}_A-\ve{\sigma}\cdot\ve{r}^{\rm N}_A}
\left(\frac{1}{r^{\rm N}_A} + \frac{1}{r^{\rm N}_A-\ve{\sigma}\cdot\ve{r}^{\rm N}_A}\right)
\nonumber\\
\nonumber\\
&& - \frac{P^{i i_2}}{\left(r^{\rm N}_A\right)^2}\,\frac{d_A^{i_1}}{r^{\rm N}_A-\ve{\sigma}\cdot\ve{r}^{\rm N}_A}
\left(\frac{1}{r^{\rm N}_A} + \frac{1}{r^{\rm N}_A-\ve{\sigma}\cdot\ve{r}^{\rm N}_A}\right)
+ \frac{d_A^i}{r^{\rm N}_A-\ve{\sigma}\cdot\ve{r}^{\rm N}_A}\,
\frac{d_A^{i_1}\,d_A^{i_2}}{\left(r^{\rm N}_A\right)^3}
\left(\frac{3}{\left(r^{\rm N}_A\right)^2}
+\frac{3}{r^{\rm N}_A\left(r^{\rm N}_A-\ve{\sigma}\cdot\ve{r}^{\rm N}_A\right)} + \frac{2}{\left(r^{\rm N}_A-\ve{\sigma}\cdot\ve{r}^{\rm N}_A\right)^2}\right),
\nonumber\\
\label{Comparison_25}
\end{eqnarray}
\end{widetext}

\begin{eqnarray}
P^{i_1 j_1}\,P^{i_2 j_2}\,\frac{\partial}{\partial \xi^{j_1}}\,
\frac{\partial}{\partial \xi^{j_2}}\,\frac{1}{r^{\rm N}_A}
&=& 3\,\frac{d_A^{i_1}\,d_A^{i_2}}{\left(r^{\rm N}_A\right)^5} - \frac{P^{i_1 i_2}}{\left(r^{\rm N}_A\right)^3}\,,
\nonumber\\
\label{Comparison_30}
\end{eqnarray}

\begin{eqnarray}
P^{i_2 j_2}\,\frac{\partial}{\partial \xi^{j_2}}\,\frac{\partial}{\partial c\tau}\,\frac{1}{r^{\rm N}_A}
&=& 3\,\frac{d_A^{i_2}}{\left(r^{\rm N}_A\right)^5} \,\left(\ve{\sigma}\cdot\ve{r}^{\rm N}_A\right) + {\cal O}\left(\frac{v_A}{c}\right), 
\nonumber\\
\label{Comparison_32}
\end{eqnarray}

\begin{eqnarray}
\frac{\partial}{\partial c\tau}\,\frac{\partial}{\partial c\tau}\,\frac{1}{r^{\rm N}_A} 
&=& - \frac{1}{\left(r^{\rm N}_A\right)^3} + 3\,\frac{\left(\ve{\sigma}\cdot\ve{r}^{\rm N}_A\right)^2}{\left(r^{\rm N}_A\right)^5} 
+ {\cal O}\left(\frac{v_A}{c}\right), 
\nonumber\\
\label{Comparison_34}
\end{eqnarray}

\begin{eqnarray}
\frac{\partial}{\partial c\tau}\,\frac{1}{\left(r^{\rm N}_A\right)^3} 
&=& - 3\,\frac{\left(\ve{\sigma}\cdot\ve{r}^{\rm N}_A\right)}{\left(r^{\rm N}_A\right)^5} + {\cal O}\left(\frac{v_A}{c}\right),
\label{Comparison_35}
\end{eqnarray}

\noindent
where  
$P^{i_2 j_2}\,\left(\xi^{j_2}-x_A^{j_2}\right)=d_A^{i_2}$  
has frequently been used; note that $\delta_{j_1 j_2}\,P^{i_1 j_1}\,P^{i_2 j_2}=P^{i_1 i_2}$.

\section{Partial derivatives for the second integration}\label{Appendix_Derivatives_2}

Throughout this section the abbreviatory notation is used:  
$\ve{r}^{\rm N}_A=\ve{r}^{\rm N}_A\left(\tau+t^{\ast}\right)$
and $r^{\rm N}_A = \left|\ve{r}^{\rm N}_A\left(\tau+t^{\ast}\right)\right|$.

\subsection{Partial derivatives for dipole-term}

In order to obtain the dipole-term in (\ref{dipole_20}) we need the following derivatives:
\begin{eqnarray}
P^{i_1 j_1}\,\frac{\partial}{\partial \xi^{j_1}}\,\frac{d_A^i}{r^{\rm N}_A - \ve{\sigma}\cdot\ve{r}^{\rm N}_A}   
&=& \frac{1}{r^{\rm N}_A - \ve{\sigma} \cdot\ve{r}^{\rm N}_A} 
\nonumber\\
&& \hspace{-3.0cm} \times \left(P^{i i_1} - \frac{d_A^i\,d_A^{i_1}}{r^{\rm N}_A \left(r^{\rm N}_A - \ve{\sigma}\cdot\ve{r}^{\rm N}_A\right)}\right),  
\label{dipole_30}
\\
\nonumber\\
\nonumber\\
P^{i_1 j_1} \frac{\partial}{\partial \xi^{j_1}} \ln \left(r^{\rm N}_A - \ve{\sigma}\cdot\ve{r}^{\rm N}_A\right)  
&=& \frac{d_A^{i_1}}{r^{\rm N}_A - \ve{\sigma} \cdot\ve{r}^{\rm N}_A} \frac{1}{r^{\rm N}_A}\,,  
\label{dipole_35}
\end{eqnarray}

\noindent
where the last relation has already been given by (\ref{First_Integration_20}).  

\subsection{Partial derivatives for quadrupole-term} 

In order to obtain the quadrupole-term in (\ref{Trajectory_Quadrupole_5}) we need the following derivatives:  
\begin{widetext}
\begin{eqnarray}
P^{i_1 j_1}\,P^{i_2 j_2}\,\frac{\partial}{\partial \xi^{j_1}}\,\frac{\partial}{\partial \xi^{j_2}}\, 
\frac{d_A^i}{r^{\rm N}_A - \ve{\sigma}\cdot\ve{r}^{\rm N}_A} &=&
- \frac{P^{i\,i_1}\,d_A^{i_2}}{r^{\rm N}_A\left(r^{\rm N}_A - \ve{\sigma}\cdot\ve{r}^{\rm N}_A\right)^2}  
- \frac{P^{i\,i_2}\,d_A^{i_1}}{r^{\rm N}_A\left(r^{\rm N}_A - \ve{\sigma}\cdot\ve{r}^{\rm N}_A\right)^2}  
- \frac{P^{i_1\,i_2}\,d_A^i}{r^{\rm N}_A \left(r^{\rm N}_A - \ve{\sigma}\cdot\ve{r}^{\rm N}_A\right)^2} 
\nonumber\\
\nonumber\\
&& + \frac{d_A^i\,d_A^{i_1}\,d_A^{i_2}}{\left(r^{\rm N}_A\right)^3 \left(r^{\rm N}_A - \ve{\sigma}\cdot\ve{r}^{\rm N}_A\right)^2} 
+ 2\,\frac{d_A^i\,d_A^{i_1}\,d_A^{i_2}}{\left(r^{\rm N}_A\right)^2 \left(r^{\rm N}_A - \ve{\sigma}\cdot\ve{r}^{\rm N}_A\right)^3}\,,
\nonumber\\
\label{quadrupole_40}
\\
\nonumber\\
P^{i_1 j_1}\,P^{i_2 j_2}\,\frac{\partial}{\partial \xi^{j_1}}\,\frac{\partial}{\partial \xi^{j_2}}\, 
\ln \left(r^{\rm N}_A - \ve{\sigma}\cdot\ve{r}^{\rm N}_A\right) &=&  
\frac{P^{i_1\,i_2}}{r^{\rm N}_A \left(r^{\rm N}_A - \ve{\sigma}\cdot\ve{r}^{\rm N}_A\right)} 
- \frac{d_A^{i_1}\,d_A^{i_2}}{\left(r^{\rm N}_A\right)^3 \left(r^{\rm N}_A - \ve{\sigma}\cdot\ve{r}^{\rm N}_A\right)}  
- \frac{d_A^{i_1}\,d_A^{i_2}}{\left(r^{\rm N}_A\right)^2 \left(r^{\rm N}_A - \ve{\sigma}\cdot\ve{r}^{\rm N}_A\right)^2}\,.  
\label{quadrupole_45}
\end{eqnarray}
\end{widetext}


\end{document}